\documentclass[12pt]{article}
\usepackage{geometry}                
\usepackage{graphicx}
\usepackage{amsmath}
\usepackage{xspace} 
\usepackage{amssymb}
\usepackage{epsfig}
\usepackage{siunitx}
\usepackage{booktabs}
\usepackage{epstopdf}
\usepackage{subfigure}
\usepackage{ifthen} 
\newboolean{inbibliography}
\setboolean{inbibliography}{false} 
\newboolean{articletitles}
\setboolean{articletitles}{true} 
\newboolean{uprightparticles}
\setboolean{uprightparticles}{false} 

\usepackage{xspace} 
\usepackage{upgreek}

\newcommand{\offsetoverline}[2][0.1em]{\kern #1\overline{\kern -#1 #2}}%

\def\lhcb   {\mbox{LHCb}\xspace}

\def\lhc    {\mbox{LHC}\xspace}

\def\velo   {VELO\xspace}

\def\MagUp {\mbox{\em Mag\kern -0.05em Up}\xspace}

\ifthenelse{\boolean{uprightparticles}}%
{

 \def\PDelta      {\ensuremath{\Delta}\xspace}                 
 \def\PXi         {\ensuremath{\Xi}\xspace}                 
 \def\PLambda     {\ensuremath{\Lambda}\xspace}                 
 \def\PSigma      {\ensuremath{\Sigma}\xspace}                 
 \def\POmega      {\ensuremath{\Omega}\xspace}                 
 \def\PUpsilon    {\ensuremath{\Upsilon}\xspace}

 \def\PB      {\ensuremath{\mathrm{B}}\xspace}                 
                  
 \def\PD      {\ensuremath{\mathrm{D}}\xspace}

 \def\PK      {\ensuremath{\mathrm{K}}\xspace}

 \def\Pi      {\ensuremath{\mathrm{i}}\xspace}

}
{

 \mathchardef\PDelta="7101
 \mathchardef\PXi="7104
 \mathchardef\PLambda="7103
 \mathchardef\PSigma="7106
 \mathchardef\POmega="710A
 \mathchardef\PUpsilon="7107
                  
 \def\PB      {\ensuremath{B}\xspace}                 
                  
 \def\PD      {\ensuremath{D}\xspace}

 \def\PK      {\ensuremath{K}\xspace}

 \def\Pi      {\ensuremath{i}\xspace}

}

\makeatletter
\ifcase \@ptsize \relax
  \newcommand{\miniscule}{\@setfontsize\miniscule{4}{5}}
\or
  \newcommand{\miniscule}{\@setfontsize\miniscule{5}{6}}
\or
  \newcommand{\miniscule}{\@setfontsize\miniscule{5}{6}}
\fi
\makeatother

\DeclareRobustCommand{\optbar}[1]{\shortstack{{\miniscule (\rule[.5ex]{1.25em}{.18mm})}
  \\ [-.7ex] $#1$}}

  \def\Kbar    {{\kern 0.2em\overline{\kern -0.2em \PK}{}}\xspace}

\def\KorKbar {\kern 0.18em\optbar{\kern -0.18em K}{}\xspace}

  \def\Dbar    {{\kern 0.2em\overline{\kern -0.2em \PD}{}}\xspace}

\def\DorDbar {\kern 0.18em\optbar{\kern -0.18em D}{}\xspace}

\def\Bbar    {{\ensuremath{\kern 0.18em\overline{\kern -0.18em \PB}{}}}\xspace}

\def\BorBbar    {\kern 0.18em\optbar{\kern -0.18em B}{}\xspace}

\def\Y#1S{\ensuremath{\PUpsilon{(#1S)}}\xspace}

\def\LorLbar     {\kern 0.18em\optbar{\kern -0.18em \PLambda}{}\xspace}

\def\to                 {\ensuremath{\rightarrow}\xspace}

\def\AT#1     {\ensuremath{A_{\mathrm{T}}^{#1}}\xspace}           

\def\C#1      {\ensuremath{\mathcal{C}_{#1}}\xspace}                       
\def\Cp#1     {\ensuremath{\mathcal{C}_{#1}^{'}}\xspace}                    
\def\Ceff#1   {\ensuremath{\mathcal{C}_{#1}^{\mathrm{(eff)}}}\xspace}        
\def\Cpeff#1  {\ensuremath{\mathcal{C}_{#1}^{'\mathrm{(eff)}}}\xspace}       
\def\Ope#1    {\ensuremath{\mathcal{O}_{#1}}\xspace}                       
\def\Opep#1   {\ensuremath{\mathcal{O}_{#1}^{'}}\xspace}                    

\newcommand{\tev}{\ifthenelse{\boolean{inbibliography}}{\ensuremath{~T\kern -0.05em eV}}{\ensuremath{\mathrm{\,Te\kern -0.1em V}}}\xspace}
\newcommand{\gev}{\ensuremath{\mathrm{\,Ge\kern -0.1em V}}\xspace}
\newcommand{\mev}{\ensuremath{\mathrm{\,Me\kern -0.1em V}}\xspace}
\newcommand{\kev}{\ensuremath{\mathrm{\,ke\kern -0.1em V}}\xspace}
\newcommand{\ev}{\ensuremath{\mathrm{\,e\kern -0.1em V}}\xspace}
\newcommand{\mevc}{\ensuremath{{\mathrm{\,Me\kern -0.1em V\!/}c}}\xspace}
\newcommand{\gevc}{\ensuremath{{\mathrm{\,Ge\kern -0.1em V\!/}c}}\xspace}
\newcommand{\mevcc}{\ensuremath{{\mathrm{\,Me\kern -0.1em V\!/}c^2}}\xspace}
\newcommand{\gevcc}{\ensuremath{{\mathrm{\,Ge\kern -0.1em V\!/}c^2}}\xspace}
\newcommand{\gevgevcc}{\ensuremath{{\mathrm{\,Ge\kern -0.1em V^2\!/}c^2}}\xspace} 
\newcommand{\gevgevcccc}{\ensuremath{{\mathrm{\,Ge\kern -0.1em V^2\!/}c^4}}\xspace} 

\def\cm   {\ensuremath{\mathrm{ \,cm}}\xspace}

\def\mm   {\ensuremath{\mathrm{ \,mm}}\xspace}

\def\gsim{{~\raise.15em\hbox{$>$}\kern-.85em
          \lower.35em\hbox{$\sim$}~}\xspace}
\def\lsim{{~\raise.15em\hbox{$<$}\kern-.85em
          \lower.35em\hbox{$\sim$}~}\xspace}

\def\geant      {\mbox{\textsc{Geant4}}\xspace}

\def\pythia     {\mbox{\textsc{Pythia}}\xspace}

\def\tell1  {TELL1\xspace}
\def\ukl1   {UKL1\xspace}

\newcommand{\eg}{\mbox{\itshape e.g.}\xspace}

\usepackage{cite} 
\usepackage{mciteplus}

\usepackage{rotating}
\usepackage{color}

\usepackage{hyperref}
\usepackage{hyperxmp}
\usepackage[all]{hypcap} 
\usepackage{comment}
\usepackage{lineno}

\begin{document}

\vspace*{-1.5cm}

\hspace*{-0.5cm}
\begin{center}

\begin{tabular*}{\linewidth}{lc@{\extracolsep{\fill}}r}
\vspace*{-1.3cm}
\mbox{\!\!\!\includegraphics[width=.12\textwidth]{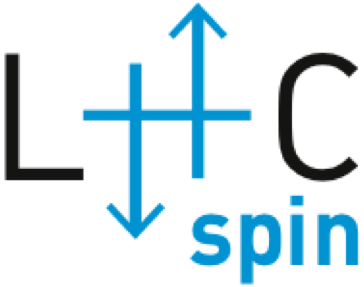}}
 \\
 & & \\
 \\
 & & \today \\
 & & \\
\hline
\end{tabular*}

\end{center}

\vspace*{0.3cm}

{\bf\boldmath\huge
\begin{center}
LHCspin: a Polarized Gas Target for LHC
\end{center}
}


\vspace*{0.5cm}

\begin{center}
A.~Accardi$^{1,2}$, 
A.~Bacchetta$^{3,4}$, 
L.~Barion$^5$, 
G.~Bedeschi$^{5,6}$, 
V.~Benesova$^{7}$, 
S.~Bertelli$^8$, 
V.~Bertone$^{9}$, 
C.~Bissolotti$^{10}$,
M.~Boglione$^{11,12}$,
G.~Bozzi$^{13,14}$,
N.~Bundaleski$^{15}$, 
V.~Carassiti$^5$, 
F.G.~Celiberto$^{16}$, 
Z.~Chen$^{17}$, 
G.~Ciullo$^{5,6}$, 
M.~Constantinou$^{18}$,
P.~Costa Pinto$^{19}$,
A.~Courtoy$^{20}$,
U.~D’Alesio$^{13,14}$,
C.~De Angelis$^{13,14}$,
E.~De Lucia$^8$, 
I.~Denisenko$^{21}$,
P.~Di Nezza$^{8,*}$,
M.~Diehl$^{22}$,
F.~Donato$^{11,12}$,
N.~Doshita$^{23}$, 
O.M.N.~Duarte Teodoro$^{15}$, 
M.G.~Echevarria$^{24}$,
T.~El-Kordy$^{25}$,
R.~Engels$^{26,27}$, 
F.~Fabiano$^{13,14}$,
I.P.~Fernando$^{28}$,
M.~Ferro-Luzzi$^{19}$, 
C.~Flore$^{13,14}$,
L.~Gamberg$^{29}$, 
G.R.~Goldstein$^{30}$,
J.O.~Gonzalez-Hernandez$^{11,12}$,
B.~Gou$^{31,32}$, 
A.~Gridin$^{21}$,
A.~Guskov$^{19}$,
C.~Hadjidakis$^{33}$, 
V.~Hejny$^{26,27}$, 
T.~Iwata$^{23}$, 
D.~Keller$^{28}$,
N.~Koch$^{34}$, 
A.~Kotzinian$^11$, 
J.P.~Lansberg$^{33}$, 
P.~Lenisa$^{5,6}$, 
X.~Li$^{17}$,
H.W.Lin$^{35}$,
S.~Liuti$^{28}$,
R.~Longo$^{36}$,
M.~Maggiora$^{11,12}$,
G.~Manca$^{13,14}$,
S.~Mariani$^{19}$, 
J.~Matousek$^{7}$,
T.~Matsuda$^{37}$,
A.~Metz$^{38}$,
M.~Mirazita$^8$, 
Y.~Miyachi$^{23}$,
A.~Movsisyan$^{39}$, 
F.~Murgia$^{14}$,
A.~Nass$^{26,27}$, 
E.R.~Nocera$^{11,12}$,
C.~Oppedisano$^11$, 
L.L.~Pappalardo$^{5,6}$, 
B.~Parsamyan$^{11,19,23,39}$, 
B.~Pasquini$^{3,4}$, 
M.~Pesek$^{7}$, 
A.~Piccoli$^{5,6}$, 
C.~Pisano$^{13,14}$,
D.~Pitonyak$^{40}$,
J.~Pretz$^{26,41}$,
A.~Prokudin$^{2,42}$,
M.~Radici$^{4}$,
F.~Rathmann$^{43}$, 
M.~Rotondo$^8$, 
M.~Santimaria$^8$,
G.~Schnell$^{24}$, 
R.~Shankar$^{5,6}$, 
A.~Signori$^{11,12}$,
D.~Sivers$^{44}$,
S.~Squerzanti$^5$, 
M.~Stancari$^{45}$,
E.~Steffens$^{46}$, 
L.~Sun$^{31,32}$,
H.~Suzuki$^{47}$,
G.~Tagliente$^{48}$,
F.~Tessarotto$^{49}$,
C.~Van Hulse$^{16}$,
Q.~Xu$^{17}$,
Z.~Ye$^{50}$,
J.~Zhang$^{17}$

\bigskip

{\normalfont\itshape\footnotesize 
$^{1}$ Christopher Newport University, Newport News, Virginia, 23606, USA \\
$^{2}$ Jefferson Lab, Newport News, Virginia 23606, USA \\
$^{3}$ Dipartimento di Fisica, Università degli Studi di Pavia, 27100 Pavia, Italy \\
$^{4}$ INFN Sezione di Pavia, 27100 Pavia, Italy \\
$^5$ INFN Sezione di Ferrara, Ferrara, Italy \\
$^{6}$ University of Ferrara, Ferrara, Italy \\
$^{7}$ Faculty of Mathematics and Physics, Charles University, 12000 Praha, Czechia \\
$^8$ INFN Laboratori Nazionali di Frascati, Frascati (Rome), Italy \\
$^{9}$ CEA Paris-Saclay, France \\
$^{10}$ Argonne National Laboratory, PHY Division, Lemont, IL, USA \\
$^{11}$ INFN Sezione di Torino, Torino, Italy \\
$^{12}$ Department of Physics, University of Turin, via Pietro Giuria 1, I-10125 Torino, Italy \\
$^{13}$ Dipartimento di Fisica, Università di Cagliari, I-09042 Monserrato (CA), Italy \\
$^{14}$ INFN Sezione di Cagliari, Monserrato (CA), Italy \\
$^{15}$ CEFITEC, Nova University Lisbon, Caparica P-2829-516, Portugal \\
$^{16}$ Universidad de Alcal\'a (UAH), E-28805 Alcal\'a de Henares, Madrid, Spain \\
$^{17}$ Institute of Frontier and Interdisciplinary Science, Shandong University,
Binhai Road 72, Qingdao, Shandong 266237, China \\
$^{18}$ Physics Department, Temple University, 1925 N. 12th Street Philadelphia, PA 19122-1801, USA \\
$^{19}$ European Organization for Nuclear Research (CERN), Geneva, Switzerland \\
$^{20}$ Instituto de Fìsica, Universidad Nacional Autònoma de Mèxico,
Apartado Postal 20-364, 01000 Ciudad de Mèxico, Mexico \\
$^{21}$ International laboratory covered by a cooperation agreement with CERN \\
$^{22}$ Deutsches Elektronen-Synchrotron DESY, Notkestraße 85, 22607 Hamburg, Germany \\
$^{23}$ University of Yamamata, Yamagata 990-8560, Japan \\
$^{24}$ Department of Physics \& EHU Quantum Center, University of the Basque Country UPV/EHU, and IKERBASQUE, Bilbao, Spain \\
$^{25}$ FH Aachen – University of Applied Sciences, Aachen, Germany \\ 
$^{26}$ Institut für Kernphysik, Forschungszentrum Jülich, Wilhelm-Johnen-Straße, Jülich, 52428, NRW, Germany \\
$^{27}$ GSI Helmholtzzentrum für Schwerionenforschung GmbH, Planckstr, Darmstadt (Germany) \\
$^{28}$ University of Virginia, Charlottesville, Virginia 22901, USA \\
$^{29}$ Division of Science, Penn State University Berks, Reading, Pennsylvania 19610, USA \\
$^{30}$ Department of Physics and Astronomy Tufts University Medford, MA 02155, USA \\
$^{31}$ Institute of Modern Physics, Chinese Academy of Sciences, Lanzhou 730000, China \\
$^{32}$ University of Chinese Academy of Sciences, Beijing 100049, China \\
$^{33}$ Universit\'e Paris-Saclay, CNRS, IJCLab, 91405 Orsay, France \\
$^{34}$ TH Nuremberg, Germany \\
$^{35}$ Department of Physics and Astronomy, Michigan State University, East Lansing, MI 48824, USA \\
$^{36}$ Department of Physics, University of Illinois at Urbana-Champaign, Urbana IL, USA \\
$^{37}$ University of Miyazaki, Miyazaki 889-2192, Japan \\
$^{38}$ Department of Physics, Barton Hall, Temple University, Philadelphia, PA 19122, USA \\
$^{39}$ AANL - A. Alikhanyan National Laboratory, 2 Alikhanyan Brothers Street, 0036, Yerevan, Armenia \\
$^{40}$ Department of Physics, Lebanon Valley College, Annville, Pennsylvania 17003, USA \\
$^{41}$ RWTH Aachen University, Germany \\
$^{42}$ Division of Science, Penn State University Berks, Reading, Pennsylvania 19610, USA \\
$^{43}$ Brookhaven National Laboratory, Upton, 11793, New York, United States \\
$^{44}$ Portland Physics Institute Portland, OR 97239 \\
$^{45}$ Fermi National Accelerator Laboratory, Batavia IL USA \\
$^{46}$ Friedrich-Alexander-Universit{\"a}t Erlangen-N{\"u}rnberg (FAU), Erlangen-N{\"u}rnberg, Germany \\
$^{47}$ Chubu University, Kasugai 487-8501, Japan \\
$^{48}$ INFN Sezione di Bari, Bari, Italy \\
$^{49}$ INFN Sezione di Trieste, Trieste, Italy \\
$^{50}$ Tsinghua University, Haidian District, Beijing 100084 \\ 
}
\end{center}

{\normalfont\itshape\footnotesize $^*$corresponding author: Pasquale.DiNezza@lnf.infn.it}


\begin{abstract}
\noindent
 
The goal of the LHCspin project is to develop innovative solutions for measuring the 3D structure of nucleons in high-energy polarized fixed-target collisions at LHC, exploring new processes and exploiting new probes in a unique, previously unexplored, kinematic regime. 
A precise multi-dimensional description of the hadron structure has, in fact, the potential to deepen our understanding of the strong interactions and to provide a much more precise framework for measuring both Standard Model and Beyond Standard Model observables. 
This ambitious task poses its basis on the recent experience with the successful installation and operation of the SMOG2 unpolarized gas target in front of the LHCb spectrometer. Besides allowing for interesting physics studies ranging from astrophysics to heavy-ion physics, SMOG2 provides an ideal benchmark for studying beam-target dynamics at the LHC and demonstrates the feasibility of simultaneous operation with beam-beam collisions.
With the installation of the proposed polarized target system, LHCb will become the first experiment to simultaneously collect data from unpolarized beam-beam collisions at $\sqrt{s}$=14 TeV and polarized and unpolarized beam-target collisions at $\sqrt{s_{NN}}\sim$100 GeV. LHCspin has the potential to open new frontiers in physics by exploiting the capabilities of the world's most powerful collider and one of the most advanced spectrometers.
This document also highlights the need to perform an R\&D campaign and the commissioning of the apparatus at the LHC Interaction Region 4 during the Run 4, before its final installation in LHCb. This opportunity could also allow to undertake preliminary physics measurements with unprecedented conditions. 

\end{abstract}

{\it Editorial Committee: V.~Carassiti, G.~Ciullo, P.~Di Nezza, R.~Engels, P.~Lenisa, A.~Nass, L.L.~Pappalardo, B.~Parsamyan, M.~Santimaria, E.~Steffens}
 
\cleardoublepage
\tableofcontents
\cleardoublepage
\newcommand{\minidaqone}{MiniDAQ1\xspace}
\newcommand{\minidaqtwo}{MiniDAQ2\xspace}

%
\section{Introduction}

\label{sec:introduction}

To understand the complex structure of protons and neutrons in terms of quarks
and gluons is an outstanding task in the field of particle physics~\cite{Kuhn_2009,Leader_2014,aidala2020probingnucleonsnucleihigh}.
Despite impressive theoretical progress, making precise predictions for nucleon
structure from first principles remains a formidable challenge, and detailed
experimental measurements are crucial for progress in this field.
This challenge has been instrumental in spurring a series of past experiments worldwide (for recent reviews see \textit{e.g.} Refs.~\cite{Anselmino:2020vlp,Avakian:2019drf}), and it serves as a foundational basis for proposals of new-generation experimental facilities, such as the EIC (US)~\cite{Abdul_Khalek_2022}, LHeC (CERN)~\cite{osti_1177771}, and JLab22 (US)~\cite{accardi2023stronginteractionphysicsluminosity}. In order to understand the microscopic composition of the nucleon, the particle physics community has historically placed considerable reliance on two distinct types of physical quantities: the spatial distribution of electric charge and magnetic moment, as probed through elastic lepton-nucleon scattering and described by the Electric and Magnetic Form Factors (FFs) and the longitudinal momentum distributions of partons, described by the collinear parton distribution functions (PDFs). The PDFs, in particular, have been measured in deep inelastic scattering (DIS) experiments in which high-energy leptons scatter off individual quarks by exchanging a virtual photon~\cite{Collins_2011}. Although FFs and PDFs have contributed significantly to shape our physical picture of the nucleon, they provide only partial information about its internal structure. The FFs lack dynamic information regarding the nucleon's constituents, such as their linear momenta, while collinear PDFs offer no insight into their spatial distributions or transverse motion.

A significant advancement in the field is represented by the measurements accessing the 3-Dimensional structure of the nucleon, thereby accounting for transverse degrees of freedom in momentum and spatial distributions~\cite{Boglione_2016,Bacchetta_2016,Hadjidakis_2021} through the connection to specific observables sensitive to either Generalized Parton Distributions (GPDs)~\cite{Diehl_2016} or Transverse Momentum Dependent parton distribution functions (TMDs)~\cite{Angeles_Martinez_2015}. Particularly important and relevant in the contemporary context is the measurement of heavy-quarks observables sensitive to the largely unknown gluon TMDs, such as, e.g., the unmeasured gluon Sivers function~\cite{Sivers:1989cc}, which provides valuable information about the spin-orbit correlations of gluons inside the nucleon and is sensitive to the unknown gluon orbital angular momentum~\cite{ji2020protonspin30years}.

The nucleon and parton spin orientations represent a pivotal instrument in accessing this novel domain of 3D structure of the nucleon. The spin-dependence of the cross section for processes such as Semi-Inclusive DIS (SIDIS), Drell-Yan or inclusive production of hadrons in polarized hadron-hadron collisions gives rise to distinct azimuthal distributions (asymmetries) of the final-state particles. These asymmetries are expressed as ratios of convolution integrals over parton intrinsic momentum (structure functions), incorporating specific combinations of TMDs. Therefore, measuring these asymmetries provides access to the encoded information about the TMDs.

The proposed LHCspin project foresees the possibility to perform a range of unique and relevant measurements at the LHC exploiting polarized fixed-target proton-proton and proton-deuterium collisions. Furthermore, the possibility of merging the LHC heavy-ion program with spin physics will allow, for the first time, to study polarized lead-proton and lead-deuterium collisions at $\sqrt{s_{NN}}\sim72$ GeV. 

A realistic possibility for exploiting this access door in a relatively short time and at relatively low costs is to implement and install a polarized target of gaseous hydrogen (protons) or deuterium (protons and neutrons) inside an already existing and highly performing accelerator, such as the LHC. 
Polarized $^3$He (acting essentially as a neutron target) could be achievable in a hypothetical future phase with an upgraded setup. The study of fixed-target collisions with a polarized gaseous target~\cite{Hadjidakis_2021,LHCSpin_3,LHCSpin_4,LHCFT} offers several unique advantages:

\begin{itemize}
\item a very high polarization degree (up to 85\%);
\item absence of dilution effects due to the presence of unpolarized materials in the target;
\item possibility to invert very quickly the polarization direction (to reduce systematics); 
\item possibility to achieve relatively high luminosities with sufficiently dense targets;
\item precise determination of the beam-gas luminosity;
\item negligible effects on the beam life-time;
\item possibility to also inject unpolarized gases;
\item negligible impact on the LHCb beam-beam physics program and performances.
\end{itemize}

Furthermore, beams at the TeV scale in conjunction with a forward spectrometer allow to access a unique and largely unexplored kinematic regime characterized by the coverage of the large negative Feynman $x_F$ and the large Bjorken-$x$ regions at intermediate Q$^2$ with beam-gas collisions at $\sqrt{s_{NN}}\sim$100 GeV;    

The project benefits from the existence of a polarized gas target system developed for the HERMES experiment \cite{Nass:2003mk} but requires modifications to ensure that all components of this complex system comply with LHC requirements. In addition, as detailed in Section~\ref{sec:abs_pol} of this document, a new absolute polarimeter is necessary and needs to be developed and commissioned. 

Furthermore, part of this technology (\textit{e.g.} an openable storage cell for the target gas) has already been implemented at the LHC, in the framework of the SMOG2 project \cite{Garcia_2024}. Specifically, the SMOG2 cell was installed upstream of the LHCb detector, within the LHC primary vacuum, during the LS2 and, is being successfully employed to collect fixed-target data since the beginning of the Run 3. In these years of operation, SMOG2 has demonstrated a negligible impact on the LHCb beam-beam collisions performances and data taking while being operated simultaneously with the collider mode.

The LHCspin apparatus can be developed and commissioned at the LHC Interaction Region 4 (IR4), in a section presently occupied by a Beam Gas Vertex (BGV) device \cite{Vlachos:2018lfe}, out of operation since years. It is worth noting that our proposed apparatus could in principle be also used to provide beam parameter measurements, should the LHC decide to pursue this aspect further.

LHCb, in its scoping document for the Upgrade II \cite{LHCbcollaboration:2903094}, clarifies the importance to perform this specific R\&D in the optics of the future upgrade of the SMOG2 system with the LHCspin one.

This document is organized as follows: {\bf Chapter 2} focuses on the physics perspectives with a polarized target at LHCb; {\bf Chapter 3} discusses the current experience of implementing and running an unpolarized target system, including a storage cell, at LHC and LHCb; {\bf Chapter 4} describes the proposed polarized target system, its main components, and its implementation in LHCb; {\bf Chapter 5} presents the simulations and expected performances of polarized fixed-target collisions with this system implemented in LHCb; {\bf Chapter 6} addresses the issue of atomic recombination connected to the storage cell coating and explores the possibility to develop a polarized molecular target; {\bf Chapter 7} illustrates the main features of an absolute polarimeter, necessary for molecular polarization measurements; {\bf Chapter 8} outlines the experimental setup proposed for installation and commissioning at the Interaction Region 4 and related physics opportunities; 
{\bf Chapter 9} summarizes the working group organization; 
finally {\bf Chapter 10} presents the conclusions.
\section{Spin physics perspectives at LHCb}

\label{sec:physics}

The primary physics objective of the LHCspin project is to study the spin-dependent dynamical structure of the nucleon in terms of quarks and gluons degrees of freedom, within the unique kinematic domain covered by the setup.

\begin{figure}[h!]
\centering
\includegraphics[width=0.7\textwidth]{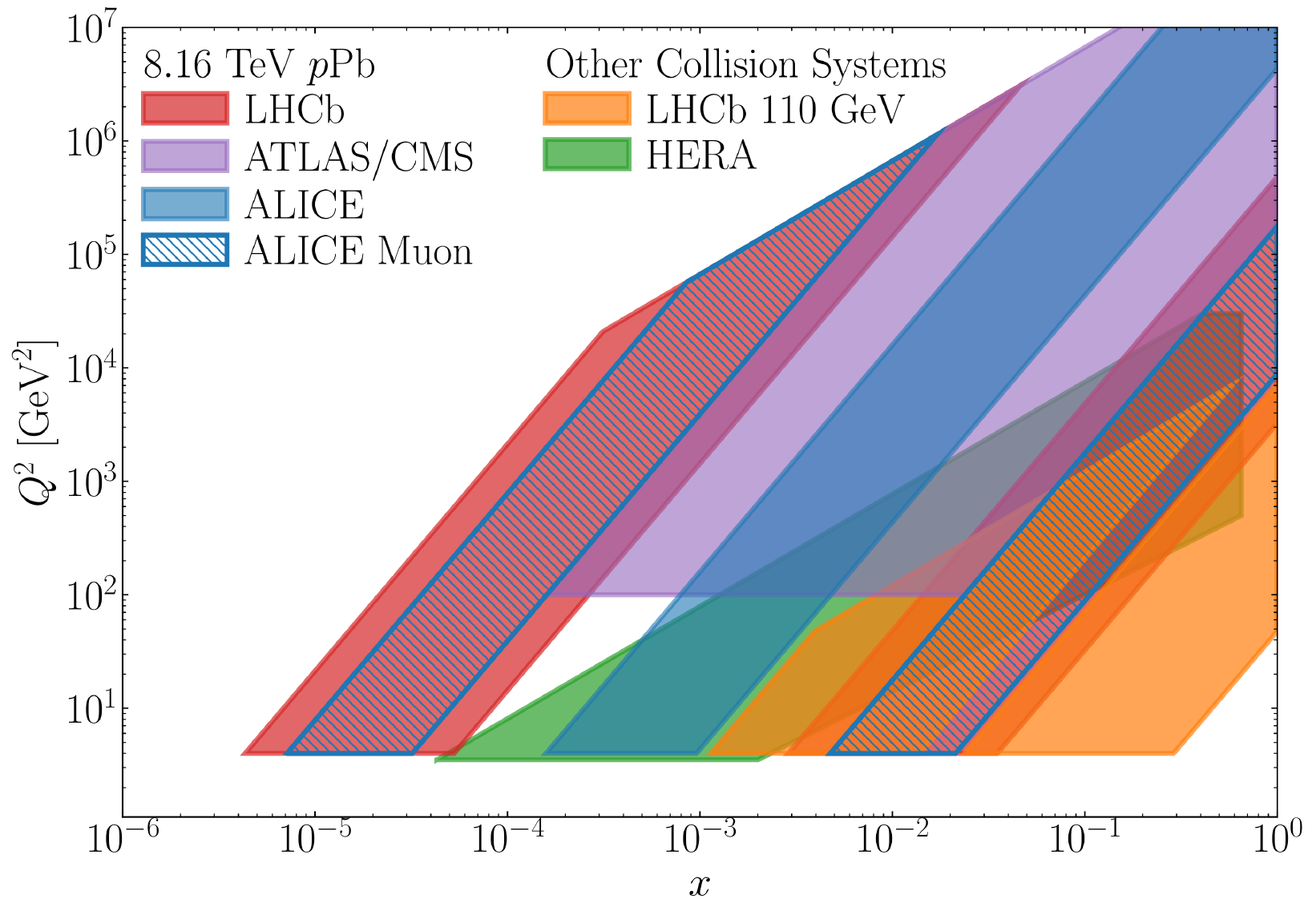}
\caption{Kinematic coverage of LHC fixed-target experiments (orange) compared to other experiments.}
\label{fig:kin}
\end{figure}

Fixed-target beam-gas collisions with a $7~\rm{TeV}$ proton beam occur at a centre-of-mass energy per nucleon of $\sqrt{s_{NN}}=115~\rm{GeV}$ and allow to cover backward and central rapidities in the centre-of-mass frame ($-3<y_{\rm{CM}}<0$), offering an unprecedented opportunity to investigate partons carrying a large fraction of the target nucleon momentum, i.e.~large Bjorken$-x$ values, and a large-negative Feynman-$x$ ($x_F$) across a wide $Q^2$ range\footnote{Here $Q^2$ can be approximated to the squared transverse energy of the produced particle. For example, for inclusive $J/\psi$ production, it is assumed that $Q^2 \sim E_T^2=m^2_{J/\psi}+p^2_{T,J/\psi}$.} (see Fig.~\ref{fig:kin}). 

The proposed research will be pursued by extracting, from the collected polarized fixed-target collisions data, experimental observables that provide access to both quarks and gluons transverse-momentum-dependent PDFs (TMDs) and generalized parton distributions functions (GPDs) (see \eg~\cite{DIEHL_2023} for a recent review on the subject). Specifically, polarized quark and gluon distributions can be probed by LHCspin through proton collisions on polarized hydrogen and deuterium targets.

\begin{figure}[h!]
\centering
\includegraphics[width=0.99\textwidth]{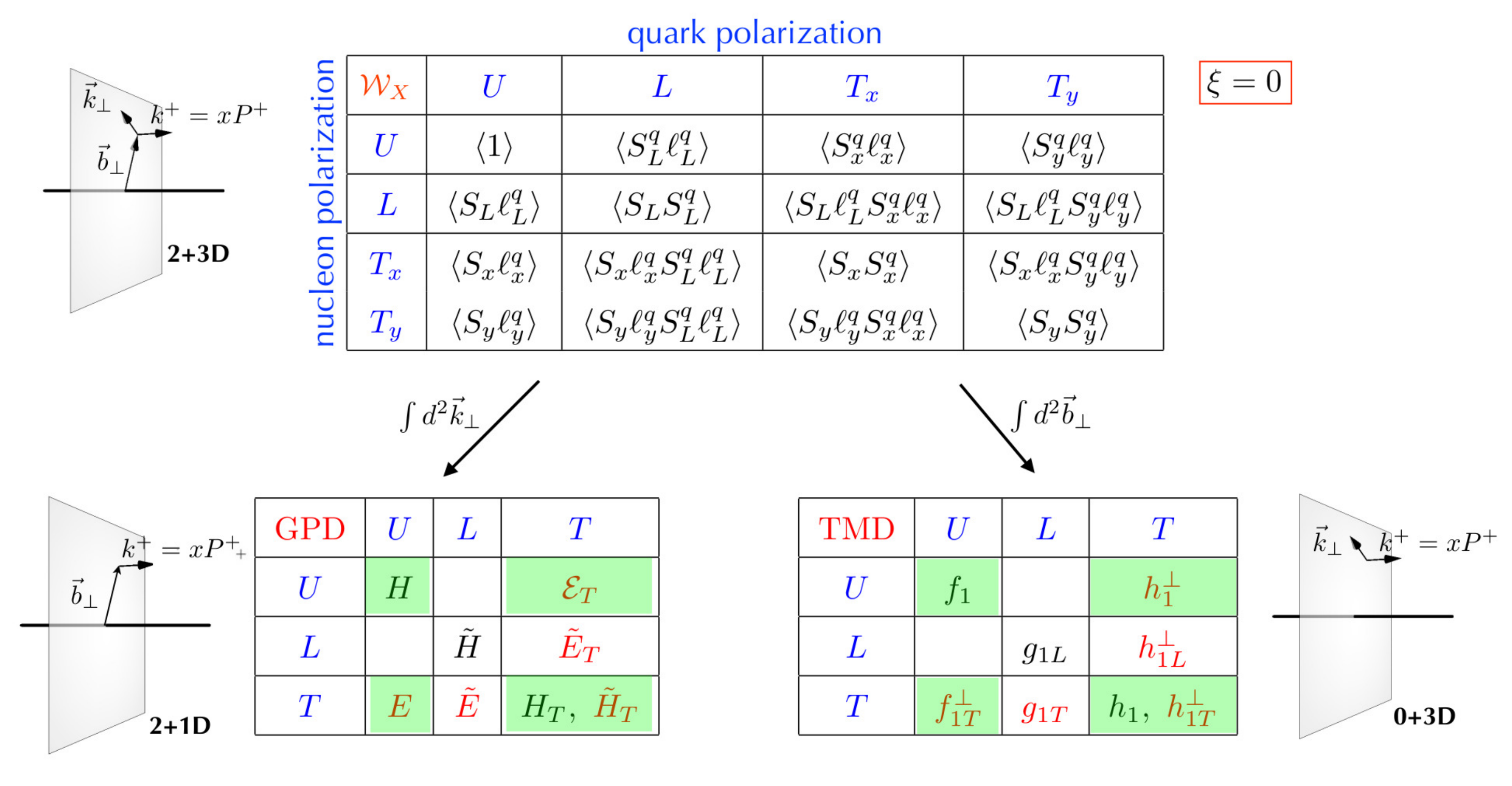}
\caption{Wigner distributions (top) and leading-twist GPDs (bottom, left) and TMDs (bottom, right) for different combinations of quark and nucleon polarization states~\cite{Pasquini}. Distributions marked in red vanish for no orbital angular momentum contribution to the nucleon spin, while the quantities highlighted in green can be accessed at LHCspin.}
\label{fig:wigner}
\end{figure}

Figure~\ref{fig:wigner} shows how the 5D Wigner distributions~\cite{BHATTACHARYA_2017} connect to the GPD and TMD functions. Integration over the transverse momentum of Wigner distributions gives rise to the GPDs, while integration over the transverse coordinate (impact parameter) leads to the TMDs. Leading-twist distributions, probed using either unpolarized or transversely polarized targets, including Boer-Mulders ($h_{1}^{\perp,q}(x,p_T^2)$)~\cite{Boer:1997nt}, transversity ($h_1^q(x,p_T^2)$), Sivers ($f_{1T}^{\perp,q}(x,p_T^2)$)~\cite{Sivers:1989cc} and Kotzinian-Mulders(worm-gear-L, $h_{1L}^{\perp,q}(x,p_T^2)$)~\cite{K-M_96} PDFs, provide independent fundamental information on different quark-nucleon spin and quark transverse momentum correlations and the spin structure of the nucleon.

\subsection{Quark TMDs}
\label{sec:qTMDs}

\begin{figure}[ht]
\centering
\includegraphics[width=0.46\textwidth]{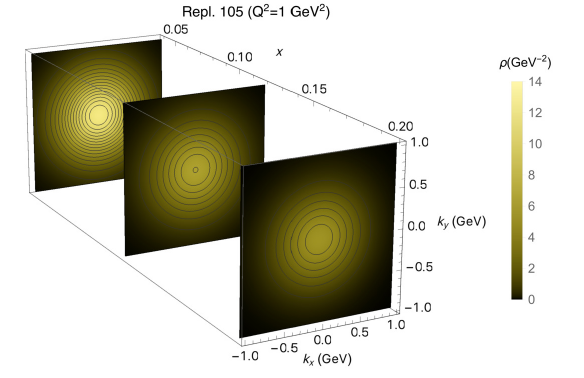}
\hfill
\includegraphics[width=0.53\textwidth]{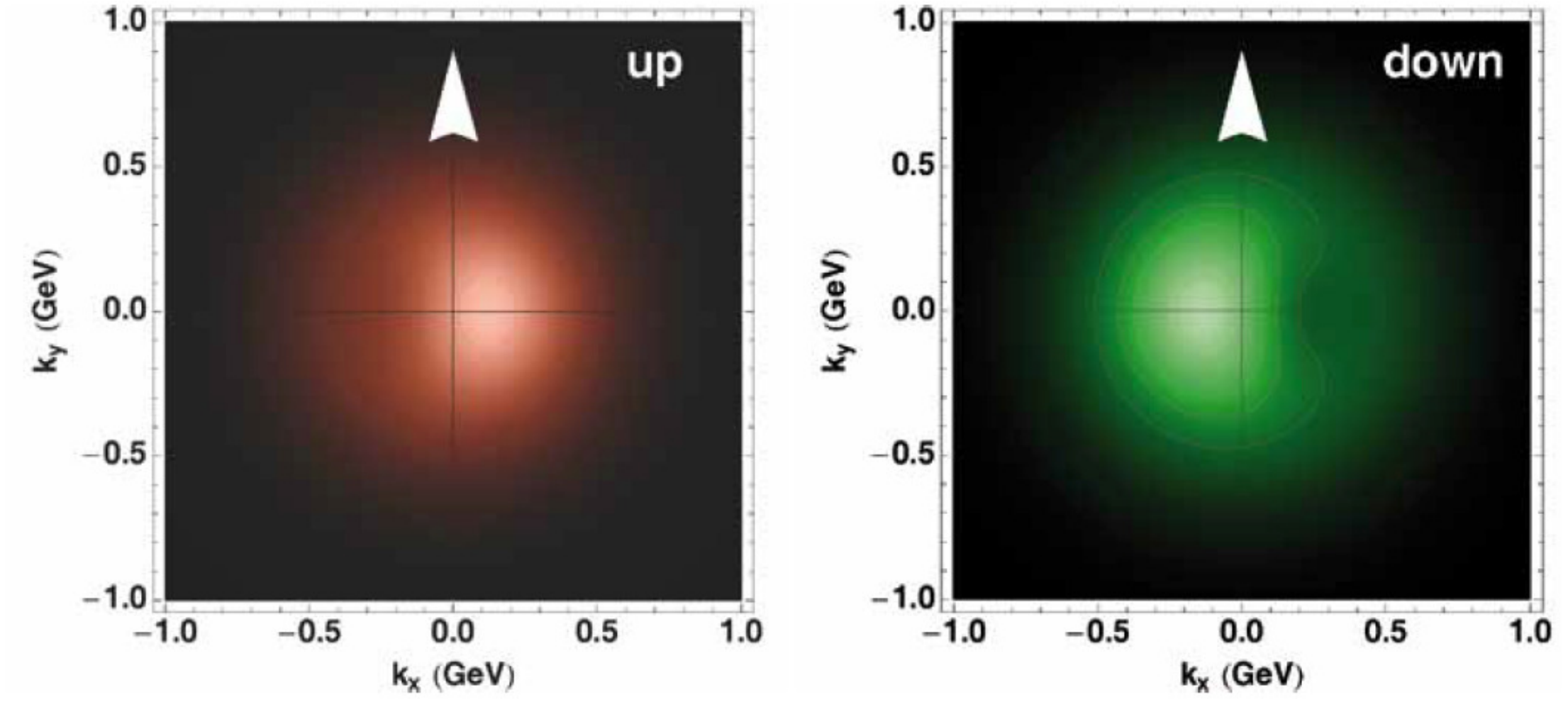}
\caption{Left: Up quark densities in momentum space~\cite{Bacchetta:2017gcc}. Right: Distortion of the up and down quark distributions in the momentum space when spin is taken into account~\cite{saggiatore}. These images are elaborated starting from real data and show that the distortion for up- and down-quarks is opposite.}
\label{fig:tmds}
\end{figure}

The study of quark TMDs is among the main physics goals of LHCspin. Quark TMDs encode different correlations between the spin of the nucleon, the spin of the quarks and their intrinsic transverse momentum. A comprehensive knowledge of the TMDs allows for the construction of 3D maps of the nucleon structure in the momentum space (often referred to as {\it nucleon tomography}), as shown in Fig.~\ref{fig:tmds}. 

The golden channel for accessing the quark TMDs in hadronic collisions is the Drell-Yan (DY) process, in which a quark and an anti-quark annihilate to produce a charged lepton pair (e.g. $\mu^+\mu^-$) in the final state~\cite{Arnold_2009}. At the LHC fixed-target kinematic conditions the dominant contribution to the process is the one where the anti-quark from the proton beam is probed at small-$x$, and the quark from the target proton is probed at large-$x$. By injecting unpolarized hydrogen in the target one can get sensitivity to the Boer-Mulders function, $h_1^\perp(x,p_T^2)$, through the azimuthal dependence ($\rm cos(2\phi)$ modulation) of the DY cross section 

\begin{equation}
d\sigma_{UU}^{DY} \propto f_1^{\bar{q}} \otimes f_1^{{q}}+{\rm cos}(2\phi)~h_1^{\perp,\bar{q}} \otimes h_1^{\perp,q}~,
\end{equation}

\noindent
where $f_1^{q(\bar{q})}$ denotes the unpolarized quark TMD, the symbol $\otimes$ denotes a convolution integral over the incoming quarks transverse momenta, $\phi$ is the azimuthal orientation of the lepton pair in the di-lepton centre-of-mass frame and the subscript $UU$ indicates unpolarized beam (first index) and target (second index). 
Employing a transversely polarized hydrogen (or deuterium) target enables sensitivity to the transverse-nucleon-spin-dependent quark TMDs, including the Sivers function, $f_{1T}^{\perp,q}(x,p_T^2)$, and the transversity PDF, $h_1^q(x,p_T^2)$, through a measurement of the corresponding Transverse Single-Spin Asymmetries (TSSA):

\begin{equation}
   \frac{1}{\it{P}}\frac{\sigma^{\uparrow}-\sigma^{\downarrow}}{\sigma^{\uparrow}+\sigma^{\downarrow}} \sim A_{UT}^{{\rm sin}(\phi_s)}{\rm sin}(\phi_s) + A_{UT}^{{\rm sin}(2\phi-\phi_s)}{\rm sin}(2\phi-\phi_s) +\cdots~, 
\end{equation}

\noindent
where $P$ denotes the effective target polarization degree (e.g. $80\%$) and $\phi_s$ the azimuthal angle of the target transverse polarization with respect to the reaction plane. The azimuthal amplitudes, denoted as $A_{UT}^{{\rm sin}(\omega(\phi,\phi_s))}$, with $\omega(\phi,\phi_s)$ indicating the relevant combinations of the azimuthal angles $\phi$ and $\phi_s$, provide direct access to the combinations of quark TMDs, e.g.:

\begin{equation}
  A_{UT}^{{\rm sin}(\phi_s)} \sim \frac{f_1^{\bar{q}} \otimes f_{1T}^{\perp,q}}{f_1^{\bar{q}} \otimes f_1^q},~~~A_{UT}^{{\rm sin}(2\phi-\phi_s)} \sim \frac{h_1^{\perp,\bar{q}} \otimes h_1^q}{f_1^{\bar{q}} \otimes f_1^q},~~~{\rm etc}. 
\end{equation}

\begin{figure}[h!]
\centering
\includegraphics[width=0.50\textwidth]{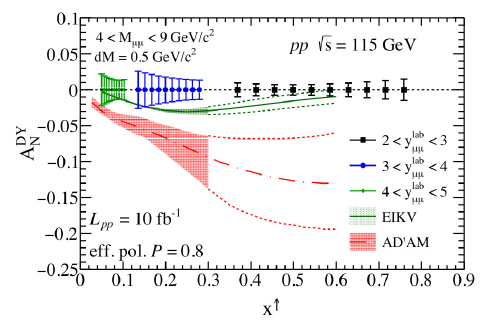}
\hfill
\includegraphics[width=0.36\textwidth]{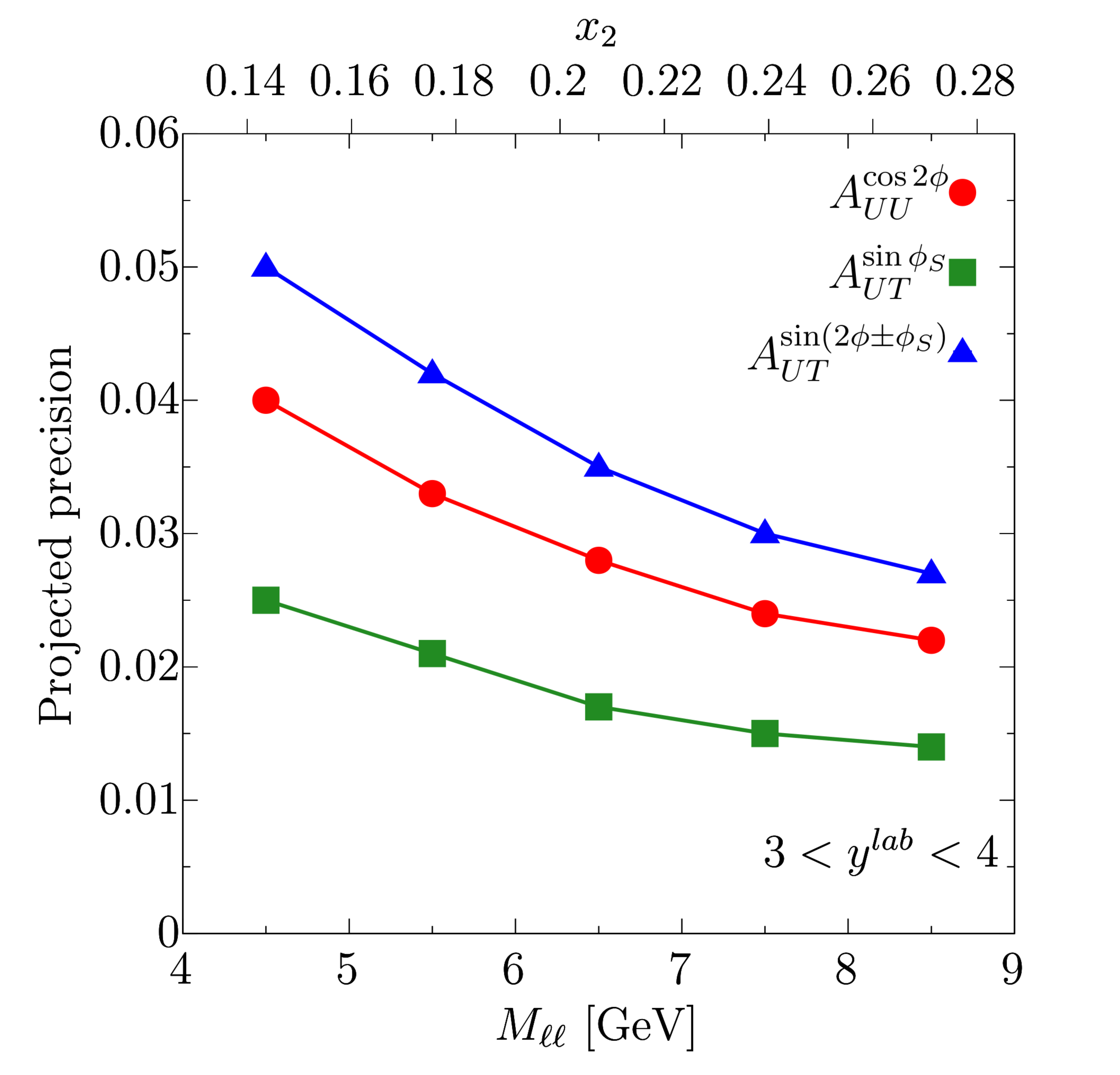}
\caption{Left: projections of TSSAs as a function of $x$ for DY events at the LHCb fixed-target kinematics compared to theoretical predictions. Right: projected precision for selected azimuthal asymmetry amplitudes with DY data in a specific rapidity interval, as a function of the di-lepton invariant mass~\cite{Hadjidakis:2018ifr}.}
\label{fig:dy}
\end{figure}

\noindent
The accumulated knowledge on transversity PDF is based on existing SIDIS measurements from HERMES~\cite{Airapetian:2009ae,HERMES:2020ifk} and COMPASS~\cite{Alekseev:2008aa,Airapetian:2009ae,Adolph:2012sp,Adolph:2014fjw,Adolph:2014zba,Adolph:2016dvl,Parsamyan:2013ug} experiments, and is mainly restricted to valence quarks and to a relatively limited $x$ region. The PDF is of considerable interest for two principal reasons. Firstly, it is one of the three TMDs that survive integration over quark transverse momentum (along with the $f_1$ and $g_1$ distributions). Secondly, a precise determination of its first moment, the tensor charge, could set stringent constraints on physics beyond the Standard Model~\cite{Courtoy:2015haa}. The two time-reversal-odd (T-odd) TMDs, i.e.~the Sivers and the Boer-Mulders functions, are expected to have an opposite sign when extracted from DY data with respect to the same quantities extracted in SIDIS~\cite{Collins:2002kn}. 
This fundamental QCD prediction is being addressed by several experiments operating at different energies (COMPASS~\cite{COMPASS:2017jbv,COMPASS:2023vqt}, STAR~\cite{Adamczyk:2015gyk}, SPINquest~\cite{Keller:2022abm}). In this respect, the primary advantages of LHCspin in comparison with COMPASS and SPINquest fixed-target measurements are primarily attributable to the absence of a dilution factor, resulting from the intrinsically pure nature of the target, and the significantly enhanced mass resolution, attributable to the absence of an absorber in front of the spectrometer.
This, in conjunction with the different and unique kinematic coverage, makes future LHCspin measurements highly competitive, yet complementary to those conducted at COMPASS and SPINquest.
In addition to the sign-change studies for T-odd PDFs, 
isospin effects can be investigated by comparing results from ${\rm p}$-$\rm{H}$ and ${\rm p}$-$\rm{D}$ collisions. Projections for DY measurements with a transversely polarized target evaluated at the LHCb fixed-target kinematics and based on an integrated luminosity of $10~\rm{fb}^{-1}$ are shown in Fig.~\ref{fig:dy}~\cite{Hadjidakis:2018ifr}.

A recently published study~\cite{Fernando_Keller_2023} has introduced an innovative data-driven methodology employing deep neural networks for a minimally biased extraction of the SU(3) flavor-dependent Sivers function, enabling a rigorous assessment of the extraction quality, precise uncertainty propagation, and systematic error quantification as a function of kinematics. These new approaches will ensure an enhanced reliability in the extraction of TMDs from future measurements, including those to be performed with LHCspin.

In the context of LHCspin measurements involving a longitudinally polarized target, access to the Kotzinian-Mulders $h_{1L} ^{\perp,q}$ TMD, also referred to as Worm-gear-L PDF, becomes a viable prospect. 
The corresponding DY asymmetry is given by the following ratio of convolutions:
\begin{equation}
  ~A_{UL}^{{\rm sin}(2\phi)}\sim \frac{h_{1}^{\perp,\bar{q}} \otimes h_{1L}^{\perp,q}}{f_1^{\bar{q}} \otimes f_1^q},
\end{equation}
\noindent
This PDF provides a detailed description of transversely polarized quarks within a longitudinally polarized nucleon. Experimental efforts have been made to measure the corresponding effect in SIDIS by HERMES, JLab, and COMPASS experiments. However, the available data is rather limited in precision, leaving significant gaps in our understanding of this particular PDF. 
In principle, with a longitudinally polarized target (and an unpolarized beam) one could also have access to the $g_1$ function (collinear or TMD) by exploiting the charged-current DY process, where a $W$ boson is exchanged~\cite{Huang_2016}. This process could also provide sensitivity to the $g_{1T}$ TMD function if, instead, a transversely polarized target is used. However, both measurements are expected to be difficult to achieve at fixed-target energies.

\subsection{Gluon TMDs}
\label{sec:gTMDs}
While first phenomenological extractions of quark TMDs have been performed in recent years, mainly using available SIDIS data, gluon TMDs are presently largely unexplored. Measurements of observables sensitive to gluon TMDs, such as, e.g. the gluon Sivers function~\cite{Boer:2015}, represent nowadays the new frontier of this research field. Given that heavy quarks are predominantly produced via gluon-gluon fusion at the LHC, the production of quarkonia and open heavy-flavour states represents the most efficient method of studying gluon dynamics inside nucleons and probing gluon TMDs~\cite{PhysRevD.86.094007,PhysRevD.110.034038}. In particular, the LHCb detector is optimized for measurements of heavy-meson production, such as ${\rm J/\psi}$, ${\rm \psi}'$, ${\rm D}^0$, $\eta_c$, $\chi_c$, $\chi_b$, etc., and thus has potential to become a unique and highly efficient facility for these studies.

\begin{figure}[ht]
\centering
\includegraphics[width=0.50\textwidth]{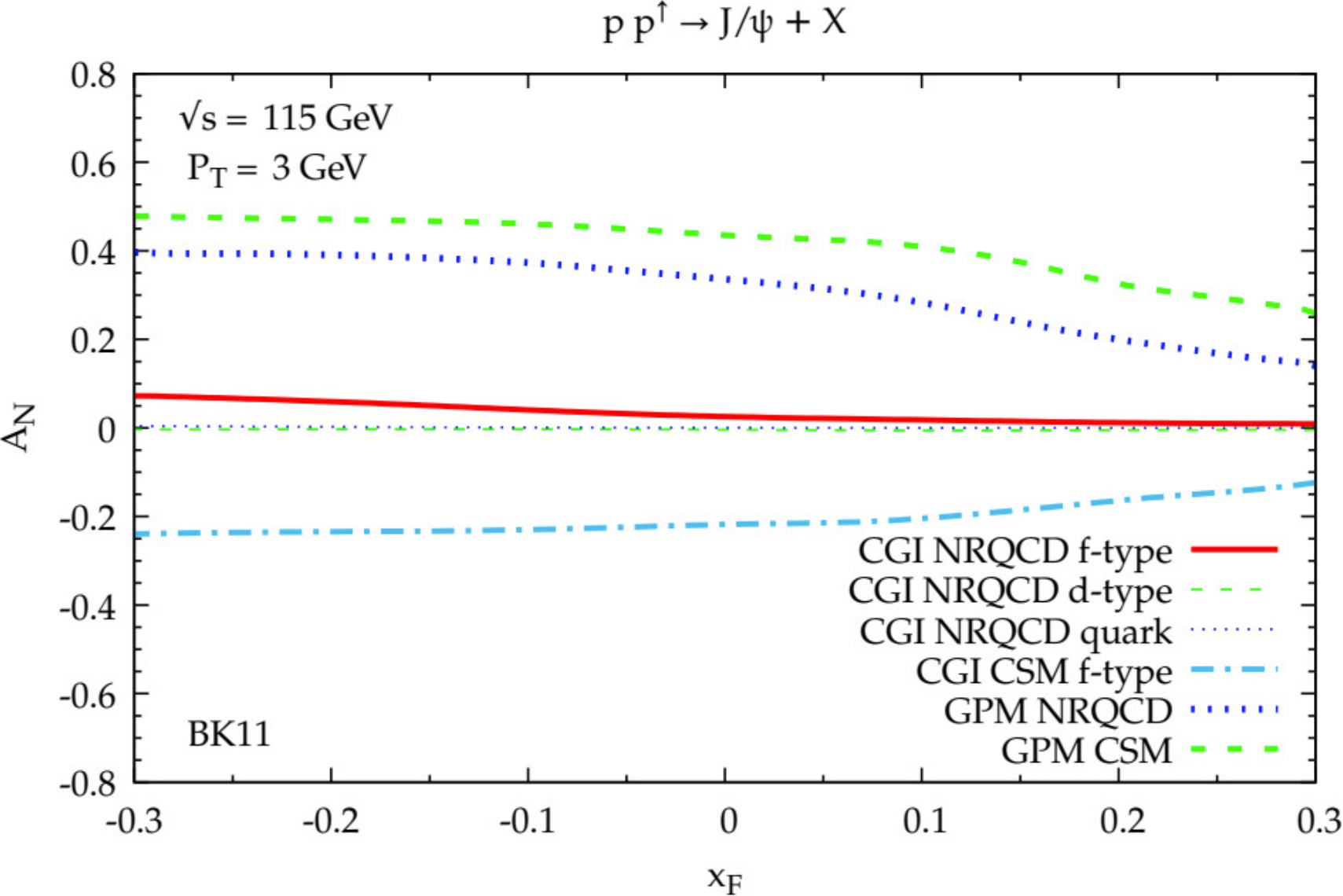}
\includegraphics[width=0.40\textwidth]{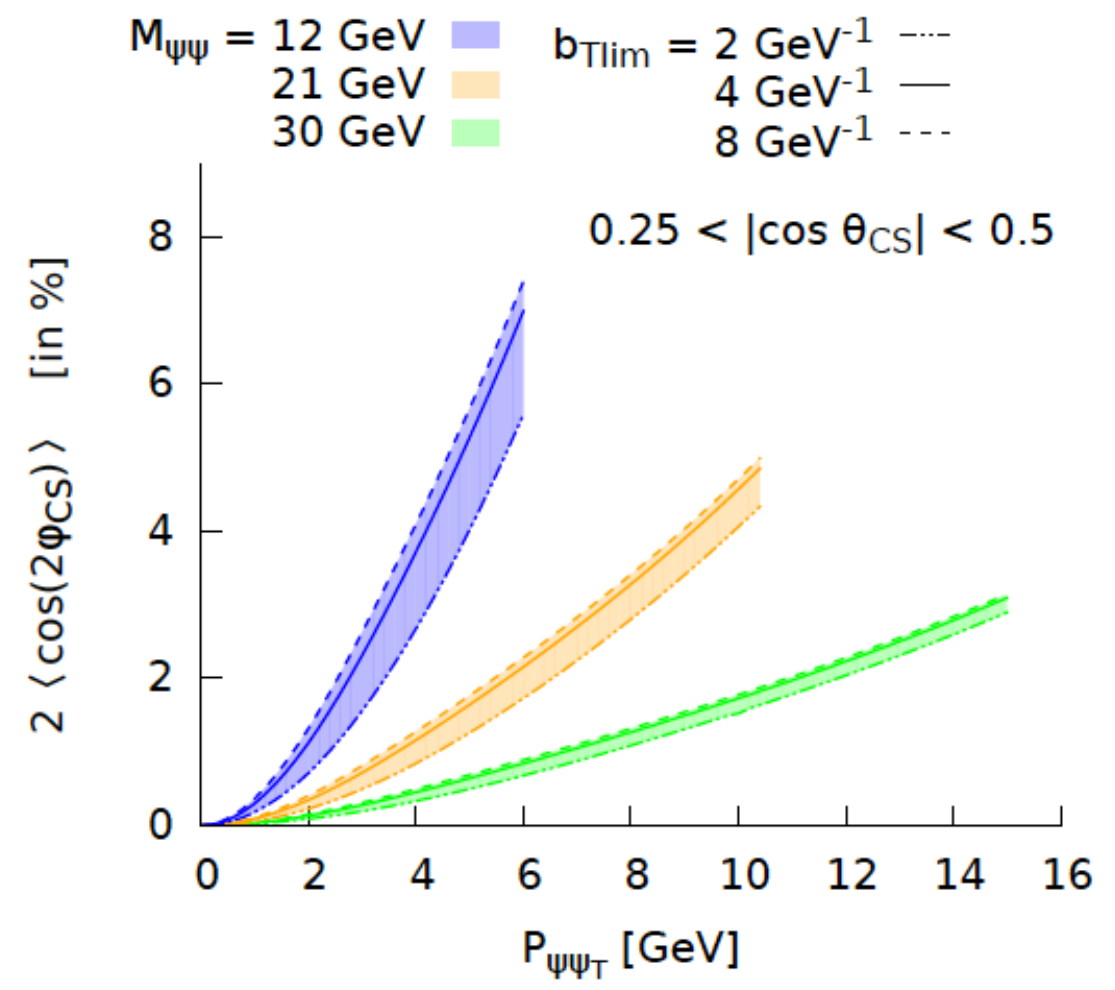}
\caption{Theoretical predictions for (left) TSSAs in inclusive ${\rm J/\psi}$ production~\cite{DAlesio:2020}; (right) ${\rm cos}(2\phi)$ asymmetry amplitudes of the unpolarized cross section for di-${\rm J/\psi}$ production as a function of the relative transverse momentum~\cite{Scarpa:2020}.}
\label{fig:gtmds}
\end{figure}

While the unpolarized $f_1^g$ and the linearly polarized (Boer-Mulders) $h_1^{\perp,g}$ gluon TMDs can be accessed through the study of the azimuthal dependence of the spin-independent cross section, the gluon TMDs that require a transversely polarized nucleon, such as the gluon Sivers function $f_{1T}^{\perp,g}$, can be probed through a measurement of corresponding target transverse-spin asymmetries:

\begin{equation}
   \frac{1}{\it{P}}\frac{\sigma^{\uparrow}-\sigma^{\downarrow}}{\sigma^{\uparrow}+\sigma^{\downarrow}} \propto [f_1^g(x_b,p_{T,b}) \otimes f_{1T}^{\perp,g}(x_t,p_{T,t})]{\rm sin}(\phi_s)
   +\cdots~, 
\end{equation}

\noindent
where 
the indices $b$ and $t$ denote the beam and the target proton, respectively. Figure~\ref{fig:gtmds} (left) shows the $x_F$ dependence of several model predictions for these TSSAs in inclusive ${\rm J/\psi}$ events~\cite{DAlesio:2020}. Asymmetries as large as $30$-$40~\%$ could be expected in the negative $x_F$ region, where the LHCspin sensitivity is highest.

Since transverse-momentum-dependent QCD factorization requires $p_T(Q) \ll M_Q$, where $Q$ denotes a heavy quark, the ideal inclusive process to be studied with a polarized hydrogen target is associated quarkonium production, e.g.:

\begin{equation}
{\rm pp^\uparrow \rightarrow J/\psi+J/\psi+X,~~~pp^\uparrow \rightarrow J/\psi+\psi'+X,~~~pp^\uparrow \rightarrow \Upsilon+\Upsilon+X,~~~etc.,} 
\end{equation} 

\noindent
where only the relative $p_T$ has to be small compared to $M_Q$. Asymmetries as large as $5~\%$ are predicted as a function of the relative $p_T$ for the ${\rm cos}(2\phi)$ and ${\rm cos}(4\phi)$ modulations of the unpolarized cross-section for quarkonium-pair production~\cite{Scarpa:2020}, as shown in Fig.~\ref{fig:gtmds} (right).    

Comparisons between forthcoming data to be collected at the LHCspin and model-dependent calculations of polarized time-reversal even~\cite{Bacchetta:2020vty} and odd~\cite{Bacchetta:2024fci} gluon TMDs are expected to offer invaluable guidance in unveiling the full three-dimensional dynamics of gluons within unpolarized and polarized protons.

\subsection{GPDs}
\label{sec:GPDs}
\begin{figure}[ht]
\centering
\includegraphics[width=0.64\textwidth]{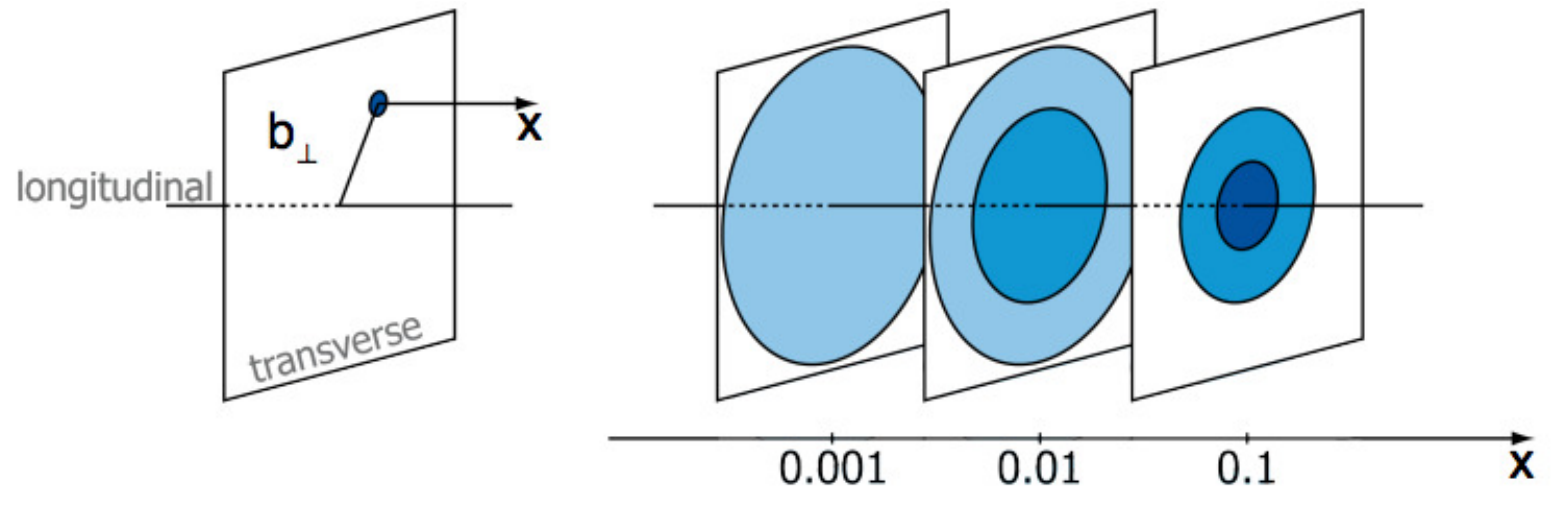}
\hfill
\includegraphics[width=0.34\textwidth]{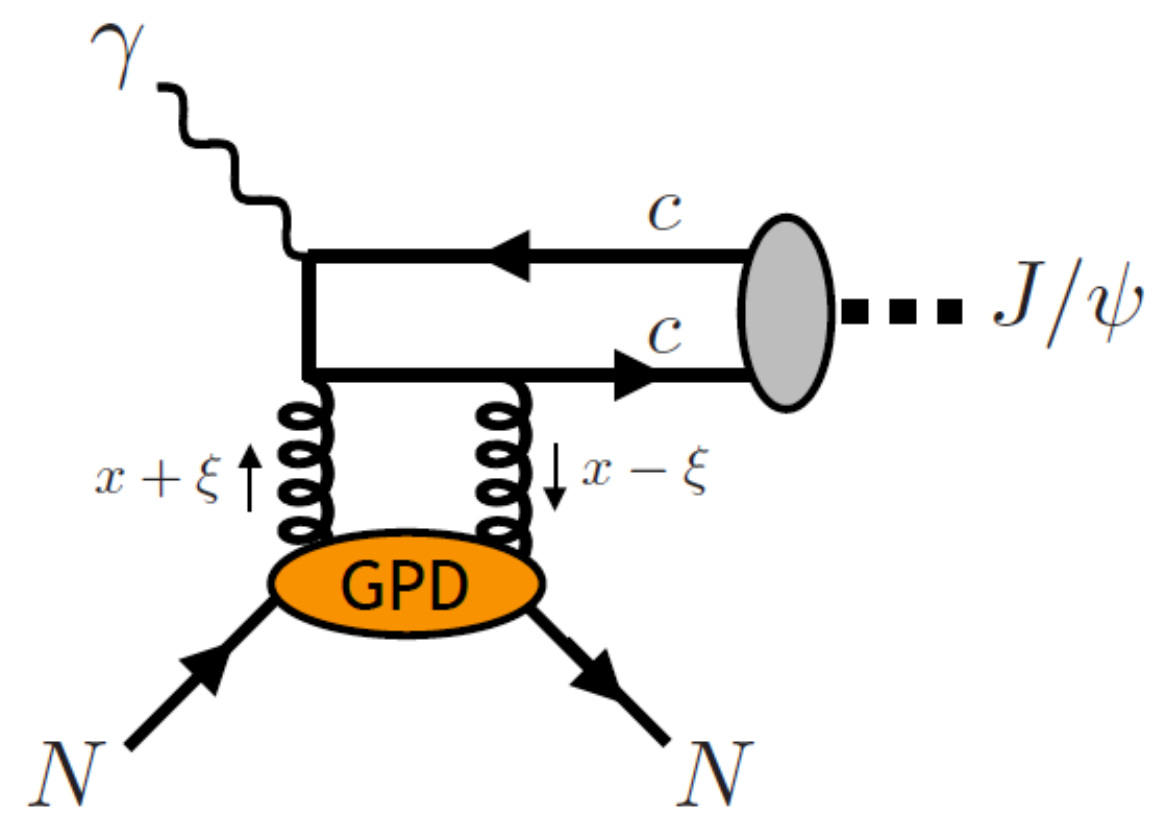}
\caption{Left: Nucleon tomography in coordinate space. Right: Access to gluon GPDs in UPC.}
\label{fig:GPDs}
\end{figure}

While TMDs provide a ``tomography'' of the nucleon in momentum space (see Fig.~\ref{fig:tmds}), complementary 3D maps can be obtained in the spatial coordinate space by measuring GPDs (see Fig.~\ref{fig:GPDs}, left panel). Correlating transverse position and longitudinal momentum, GPDs provide an access to the parton orbital angular momentum, whose contribution to the total nucleon spin can be inferred via the Ji sum rule~\cite{Ji:1996ek}. The essentially unknown gluon GPDs can be experimentally probed at LHC in exclusive quarkonia production in Ultra-Peripheral Collisions (UPCs), which are dominated by the electromagnetic interaction, in events where a pomeron exchange with the target nucleon occurs\cite{Koempel:2012} (Fig.~\ref{fig:GPDs}, right). First measurements of ${\rm J/\psi}$ and $\psi(2S)$ production in UPC in ${\rm PbPb}$ collisions have recently been reported by the LHCb collaboration\cite{Aaij:2022, Aaij:2022_2, Aaij:2023}. The LHCspin polarized target opens the possibility to measure
TSSAs in UPCs, and therefore potentially access the $E_g$ GPD~\cite{Koempel_2012}, which has never been measured so far and represents a key element of the proton spin puzzle.

\subsection{Heavy ion collisions}
\label{sec:heavy_ion}
\begin{figure}[h]
\centering
\includegraphics[width=0.47\textwidth]{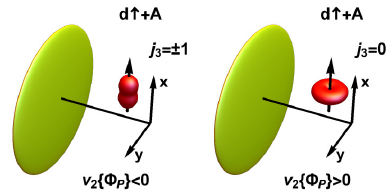}
\hspace{1cm}
\includegraphics[width=0.42\textwidth]{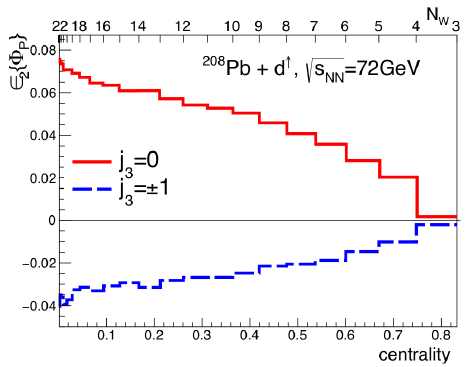}
\caption{Left: sketch of a ultra-relativistic collision of a lead nucleus against a transversely polarized deuteron in two different angular momentum projections. Right: ellipticity with respect to the polarization axis as a function of the collision centrality at LHCspin kinematics~\cite{Broniowski:2019kjo}.}
\label{fig:deuteron}
\end{figure}

Finally, the opportunity to use a polarized target in conjunction with the LHC heavy-ion beams paves the way for unique measurements that combine heavy-ion physics and spin physics. 
One such possibility
concerns the study of collective phenomena in heavy-light systems through ultrarelativistic collisions of heavy nuclei with transversely polarized deuterons \cite{PhysRevLett.121.202301,Broniowski:2019kjo}. 
Polarized deuteron targets provide a unique opportunity to control the orientation of the formed fireball by measuring the elliptic flow relative to the polarization axis (ellipticity). The spin-$1$ deuteron nucleus is prolate (oblate) in the $j_3=\pm1$ ($j_3=0$) configuration, where $j_3$ is the projection of the spin along the polarization axis. The deformation of the target deuteron can influence the orientation of the fireball in the transverse plane, as shown in Fig.~\ref{fig:deuteron}.

Another interesting topic consists in studying inclusive hadron production in ultra-peripheral proton-nucleus collisions as a new channel to investigate the assumed dominance of the contribution from twist-three fragmentation functions to the single spin asymmetries~\cite{Beni__2018}. 

LHCspin is the only facility where these measurements can be performed in the near future, thanks to the availability of the high-intensity LHC heavy-ion beams and the possibility of using a transversely polarized hydrogen and deuterium target.

\section{The present LHCb fixed-target system}

\label{sec:lhc}

\begin{figure}[h!]
    \begin{minipage}{13pc}
      \includegraphics[width=16.5pc]{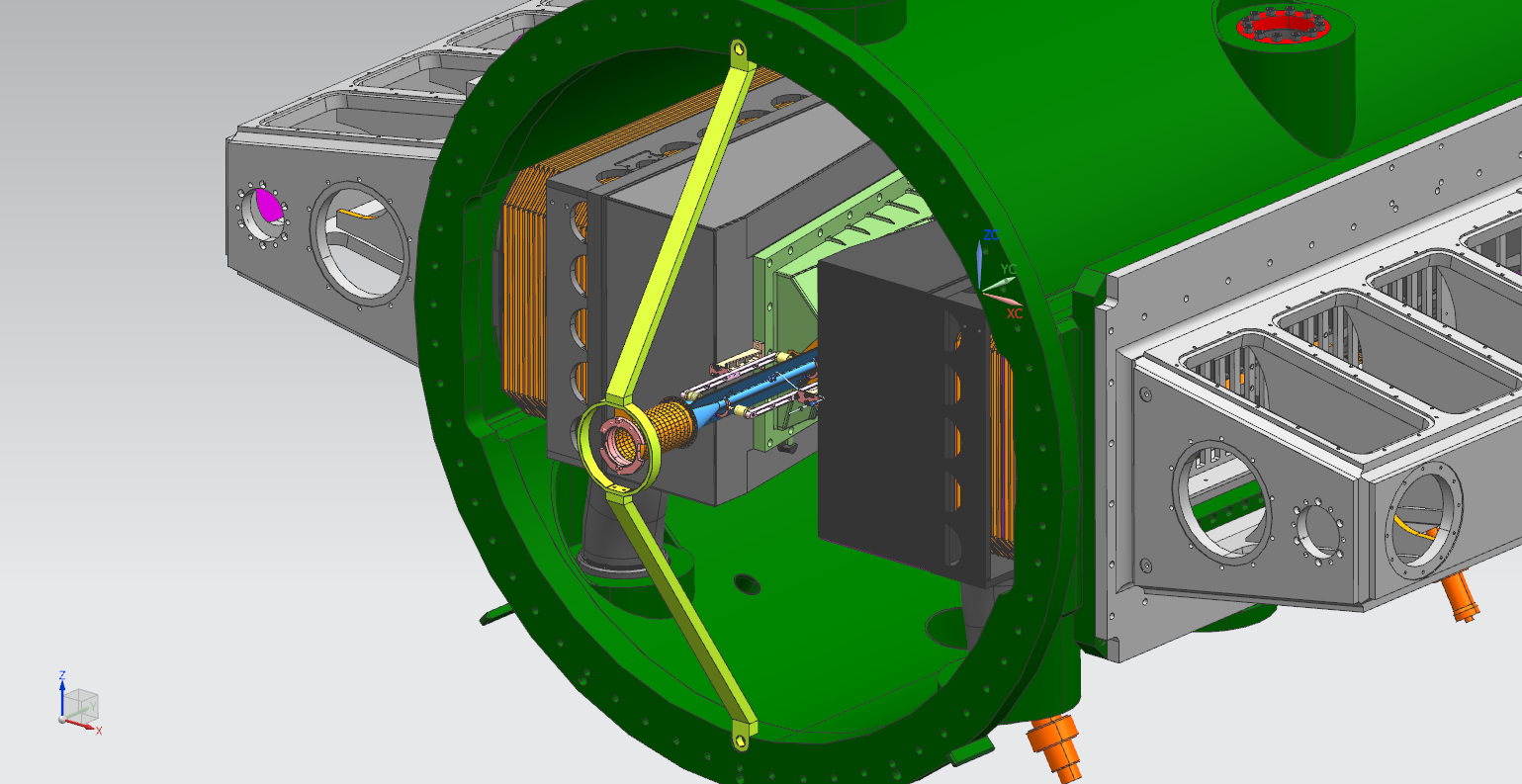}
    \end{minipage}
    \hspace{4 pc}%
    \begin{minipage}{13pc}
      \includegraphics[width=17.0pc]{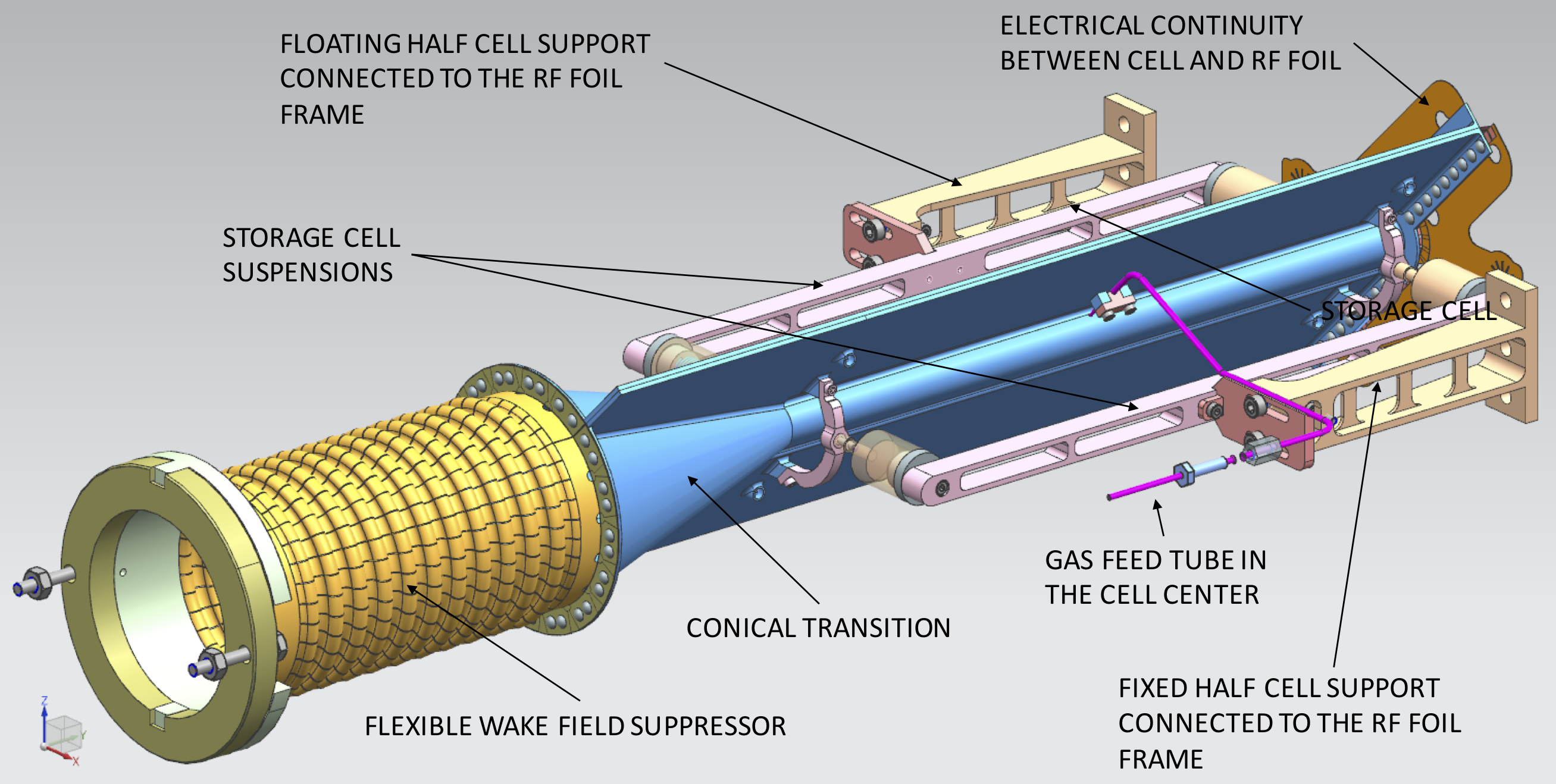}
    \end{minipage}
    \hspace{4pc}%
    \caption{\label{cell} Left: SMOG2 storage cell mounted inside LHCb, in front of the VELO detector. Right: details of the storage cell.}
\end{figure}

Among the LHC experiments, LHCb is the only one that can run both in collider and fixed-target mode. 
During the LHC Long Shutdown 2, the SMOG2 system has been installed in LHCb~\cite{Barshel:2014, SMOG2_paper}. It uses a storage cell for the target gas, installed at the upstream edge of the LHCb VErtex LOcator (VELO) detector and a gas feed system with multiple injection lines, which allows for precise density measurements as well as for the possibility to inject more gas species, including hydrogen and deuterium. 
The design of the storage cell and its arrangement inside the VELO vessel are shown in Fig.~\ref{cell}.

\subsection{The SMOG2 experience}
\label{sec:sc}

The SMOG2 system was commissioned using the early Run 3 LHC beams and operated continuously during the 2024 data taking period. The system operated by injecting H$_2$, D$_2$, He, Ne, and Ar gases during proton beam operations, and Ne and Ar when circulating lead beam,
producing 0.3 fb$^{-1}$ beam-gas collision data. It was clearly observed that gas injection does not affect LHC beam lifetimes, and essentially no impact on the spectrometer performance was observed.

\begin{figure}[h!]
  \begin{center}
       \includegraphics[width=26pc, angle=0]{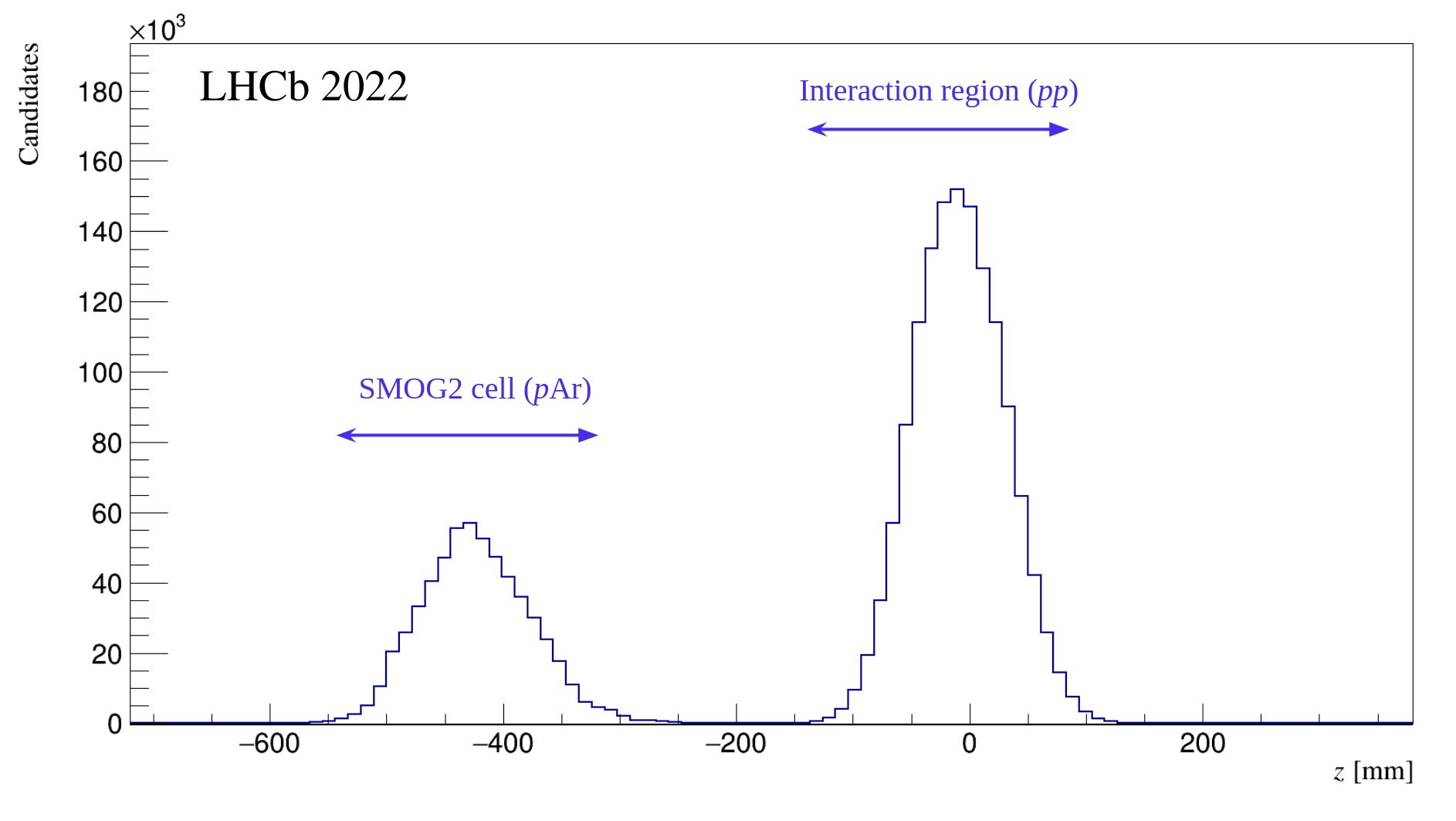}
    \caption{\label{PVz} Distributions of primary vertices along the beam direction acquired during a simultaneous collider (pp) and fixed-target (pAr) data taking.}
    \end{center}
\end{figure}

Moreover, thanks to the well separated beam-beam and beam-gas interaction regions achievable with the use of the storage cell, simultaneous collider and fixed-target data takings were performed, as shown in Fig.~\ref{PVz}, where the p-p (collider) and p-Ar (fixed-target) primary vertex regions along the beam direction are very well separated.

\begin{figure}[h!]
    \centering
    \includegraphics [scale=0.2] {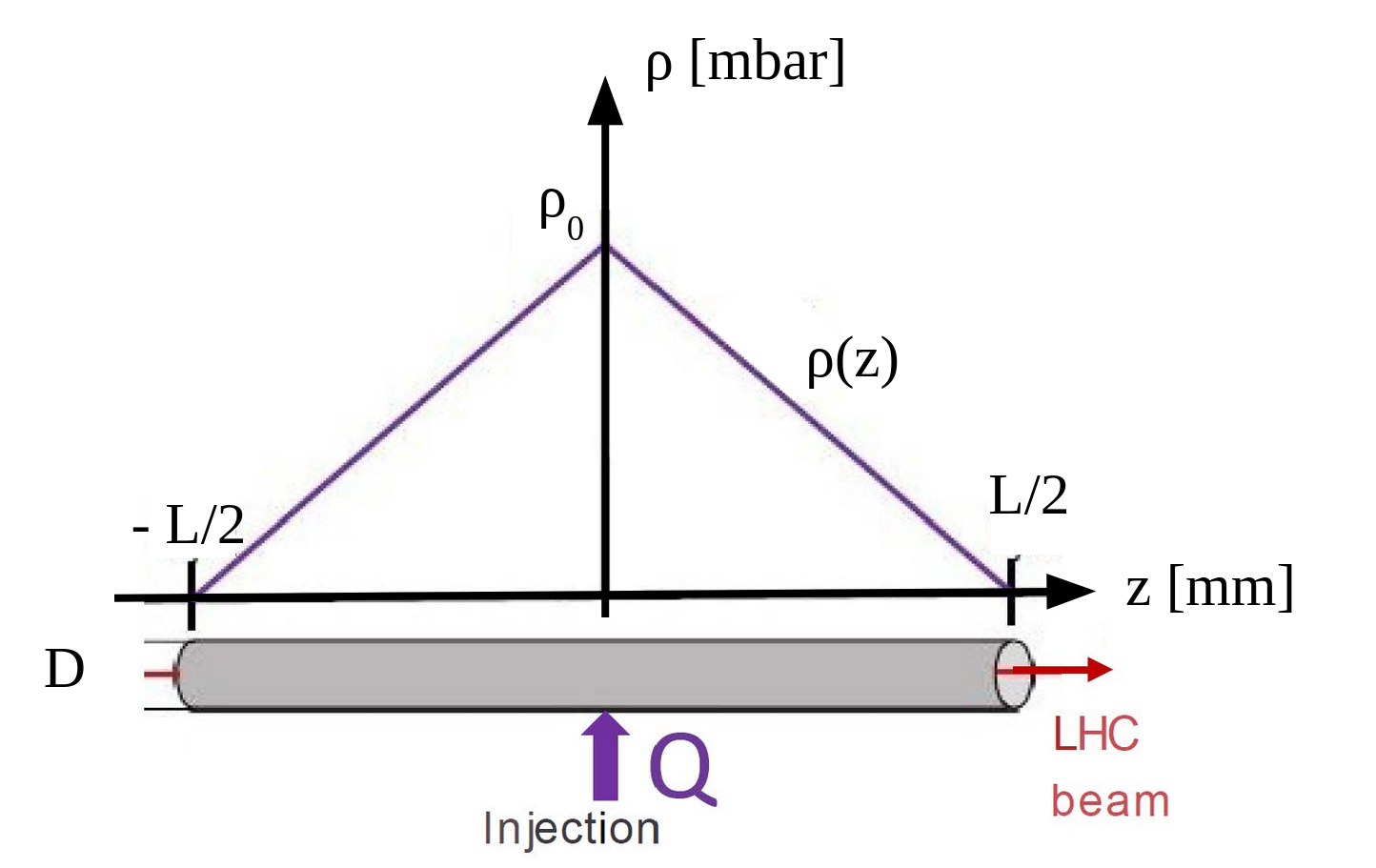}
    \caption{Scheme of the SMOG2 tubular storage cell with length $L$ and inner diameter $D$. Injection occurs in the center with flow rate $Q$, resulting in a triangular density distribution $\rho(z)$ with maximum $\rho_0$ at the center.}
    \label{fig:sc}
\end{figure}

The SMOG2 storage cell \cite{Haeberli1, article-steffens} consists of an open-ended tube positioned around the beam path, as schematically shown in Fig.~\ref{fig:sc}. Gas is injected at the center of the tube of length $L$ from where the molecules or atoms diffuse towards both ends. 

In the cylindrical storage cell of length $L$, the gas forms a triangular pressure profile with maximal density $\rho_0$ at the center and an average target areal density $\theta = \rho_0 \cdot L/2$. At the typical densities used in the SMOG2 storage cell, gas diffusion occurs in the molecular flow regime, where wall collisions dominate and re-emissions angles follow the Knudsen's cosine law~\cite{Knudsen}.
The flow rate and the corresponding volume density can be determined through (i) the Analytic Method (AM) employing parameters such as the cell geometry, the gas molecular mass, and the wall temperature, or with (ii) Numerical Simulations, such as the \textit{Molflow+} program~\cite{MolFlow}.

\subsubsection{Mechanical design and construction}
\begin{figure}
    \centering
    \includegraphics [scale=.35] {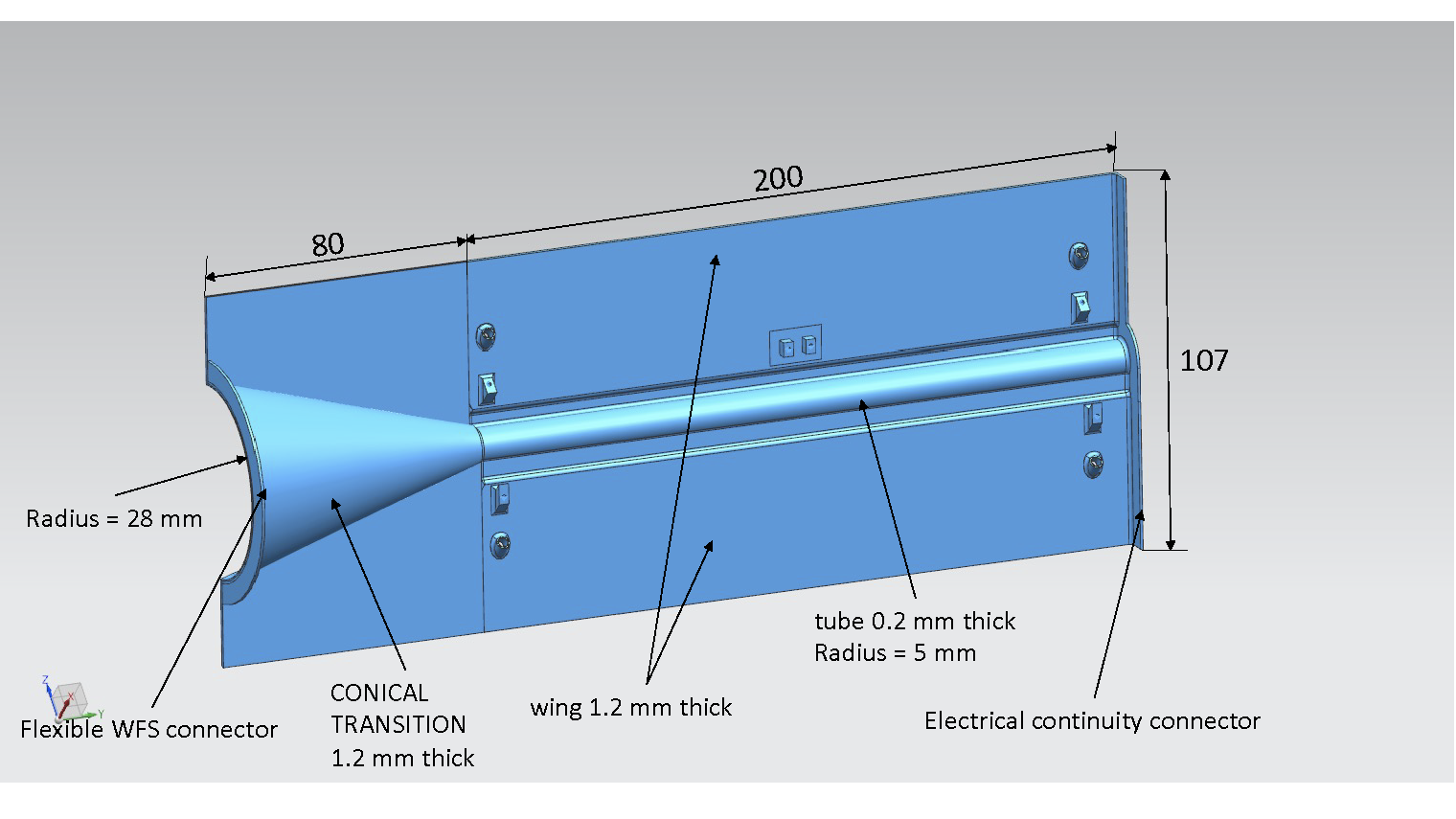}
    \caption{Dimensions of one half of the cell and its transition cone pointing to the upstream side of the \velo.}
    \label{fig:halfcell}
\end{figure}

\begin{figure}
    \centering
    \includegraphics [width = 1.0\textwidth] {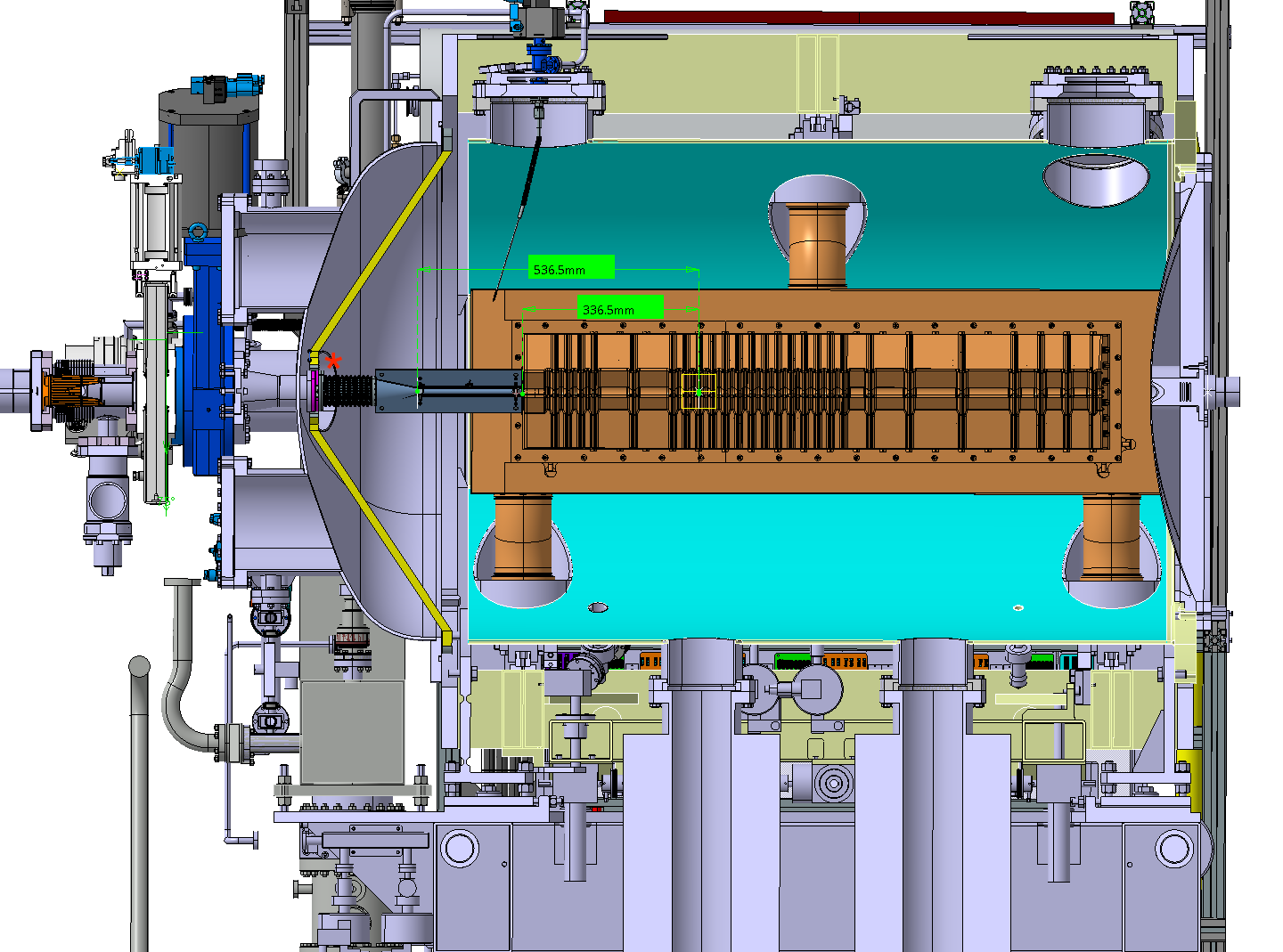}    
    \caption{Overall view of the \velo vessel with the storage cell (in dark blue) positioned just upstream of the RF boxes (light green). The distances of the cell edges from the beam-beam interaction point are indicated in yellow,  covering 200 mm from -536.5 mm to -336.5. The red star indicates the injection position when the injection type is chosen to be as for the previous SMOG system.}
    \label{fig:cell1b}
\end{figure}

\begin{figure}
    \centering
    \includegraphics [width = 0.8\textwidth] {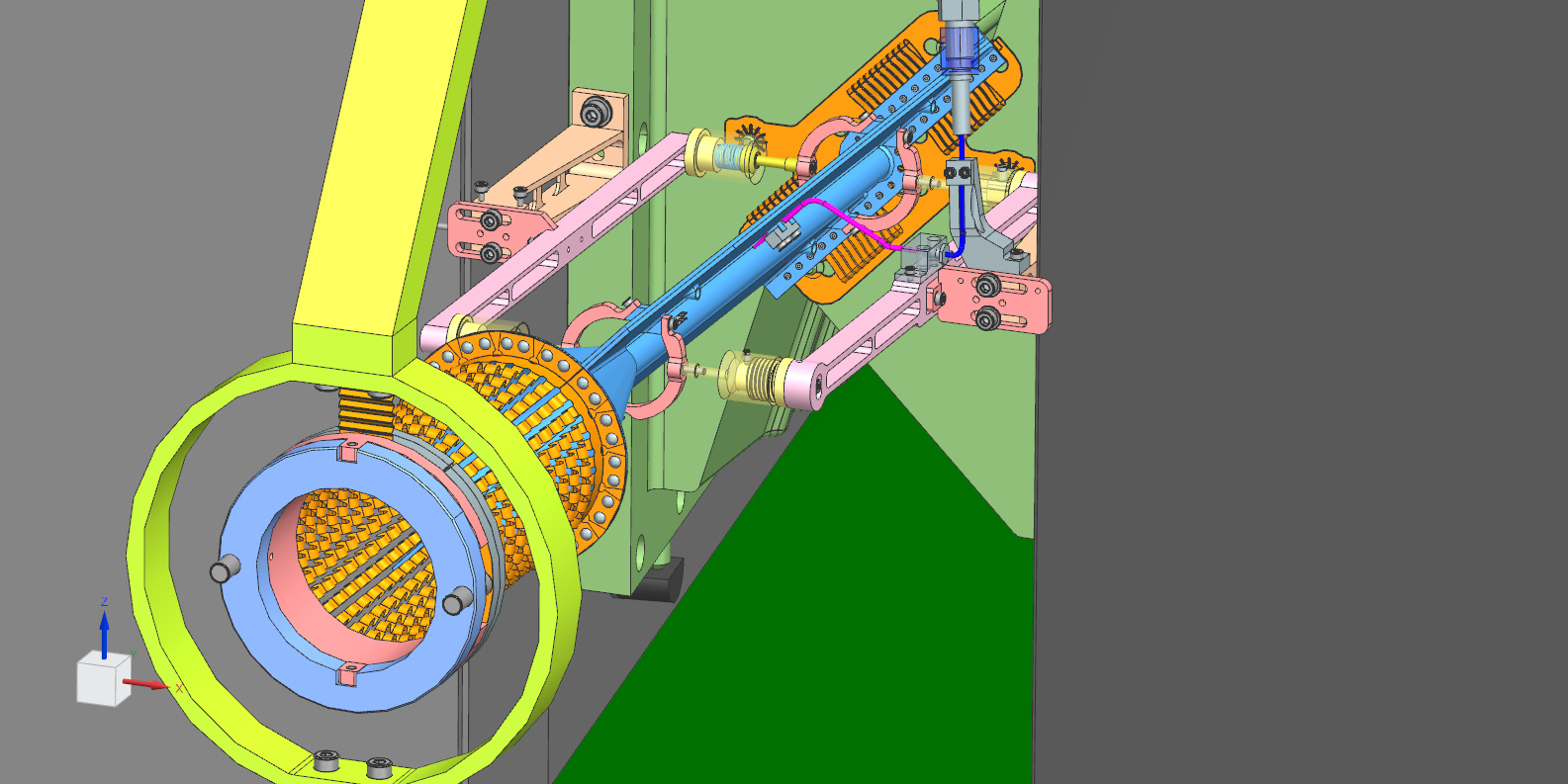}
    \caption{Zoom on the storage cell to show the supports and attachment to the \velo RF boxes and upstream beam pipe ring (light blue) via wake field suppressors (in gold).}
    \label{fig:cell1a}
\end{figure}

The SMOG2 cell consists of two halves, rigidly connected to the two \velo detector box. Due to the large transverse size of the \lhc beam at the injection energy of 450 GeV, the cell is kept open together with the \velo boxes during beam injection and tuning, and closed once the stable beam condition is reached.
The core of the storage cell consists of a tube connected on one side to the upstream beam pipe and on the other side to the \velo Radio-Frequency (RF) box~\cite{PCollins_VELO}. The tube has a length $L$ of 20\cm, an inner diameter $D$ of 1\cm (in the closed position), and a wall thickness of 200 $\mu$m. It is followed by a short conical extension, made out of the same piece of aluminum, allowing the diameter to be adapted to the one of the upstream beam pipe. Two 5\cm wide side wings provide a lateral sealing. 
In Fig.~\ref{fig:halfcell}, the main dimensions of the half cell are reported. Figure~\ref{fig:cell1b} shows the CAD transverse view of the cell installed in the \velo vessel, whereas Fig.~\ref{fig:cell1a} shows the cell system in its closed position.

The cell is rigidly mounted to the \velo boxes by two cantilevers screwed to the flange of the \velo RF boxes.
One half of the cell is rigidly fixed to the detector box, while the other one is mounted on a spring system that allows for an adequate flexibility when reaching the closed position.
During the installation phase, the alignment system of the fixed half-cell enabled the centering of the cell axis with respect to the \velo detector axis.

\begin{figure}
    \centering
    \includegraphics [scale=1.2] {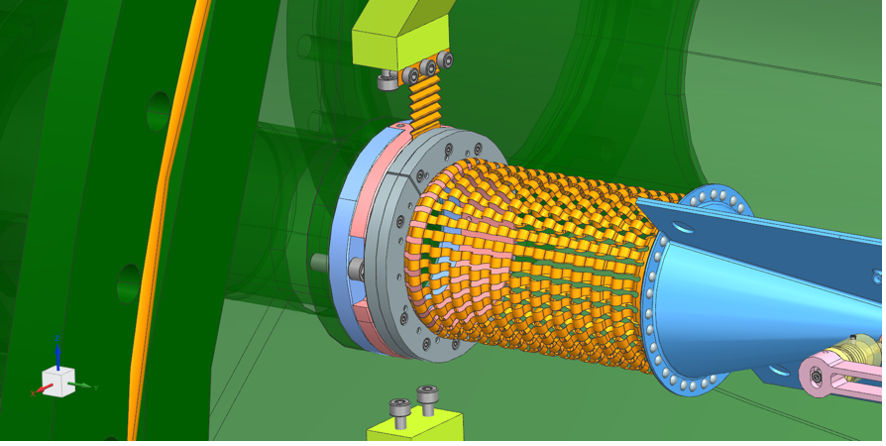} 
    \caption{Details of the upstream WFS and its connection to the beam pipe flange.}
    \label{fig:flange}
\end{figure}

Two Cu-Be2 Wake Field Suppressors (WFS) positioned at the upstream and downstream ends of the cell ensure electrical continuity, Fig. \ref{fig:flange}. 

\begin{figure}
    \centering
    \includegraphics [scale=0.19] {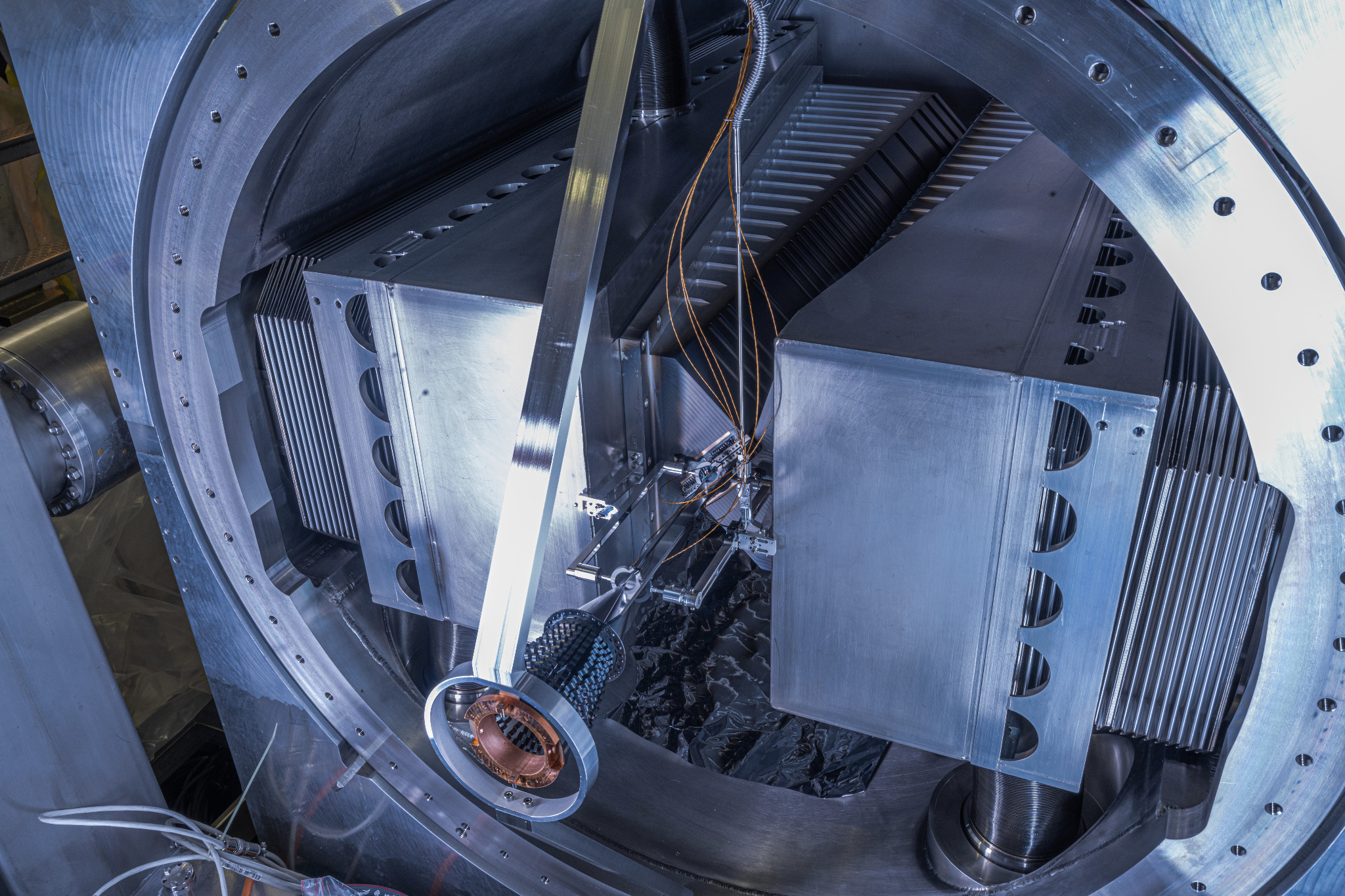}
    \caption{Picture of the storage cell, in closed position, installed in front of the \velo RF foil inside the \velo vessel.}
    \label{fig:photo}
\end{figure}

Figure \ref{fig:photo} shows a picture of the storage cell in front of the \velo RF foil, within the \velo vessel, during the SMOG2 installation in 2020.

The temperature of the SMOG2 cell is monitored using five K-Type twisted pair thermocouple wires. 
The values read by the sensors provide the temperature profile along the cell, affecting the conductance of the injected gas and allowing for the estimate of the integrated areal density of the cell and the beam-gas luminosity.

\subsection{Interference with LHC}
\label{sec:beam}

\begin{figure}[h!]
    \centering
    \includegraphics [scale=0.3] {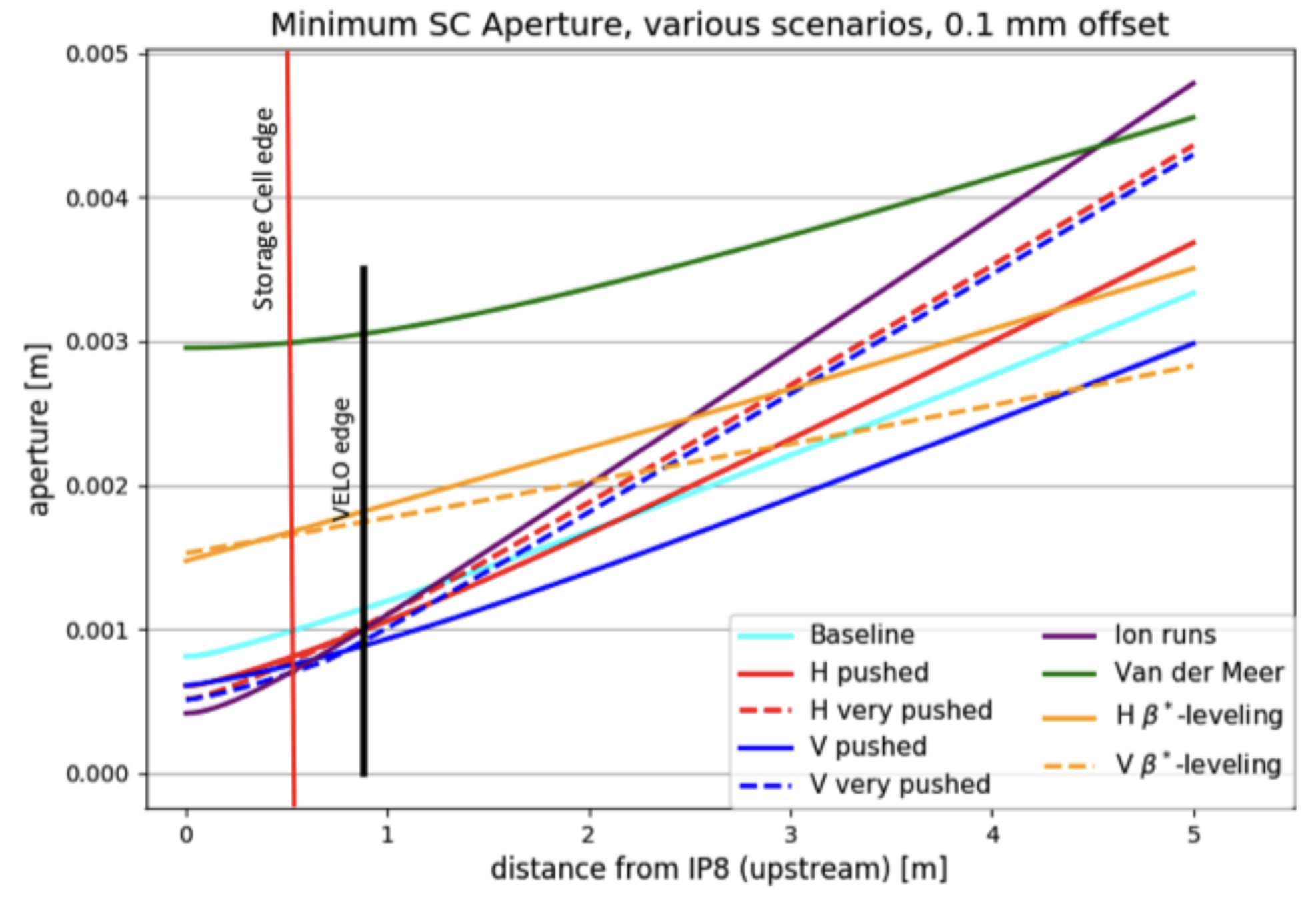}
    \caption{Minimum aperture for all studied scenarios for the SMOG2 cell. A 0.1 mm offset due to orbit drifts is assumed. The figure is from Ref.~\cite{Aperture_studies}  while the vertical red line, representing the SMOG2 storage cell edge, has been added by the authors of this document.}
    \label{fig:pedro1}
\end{figure}

To maximize the gas density within the storage cell, such to obtain the highest possible beam-gas luminosity for a given injected flux, the conductance of the storage cell has to be as small as possible. This condition can be obtained by either enhancing the length of the cell or by reducing its internal diameter. The minimal diameter of the cell is limited by the beam size in correspondence of the cell position. A minimal aperture of the cell has to be ensured to avoid any direct interaction of the beam with the cell material. The beam size depends on different factors and varies a lot along the LHC ring. Therefore, a careful inspection of the beam size at the cell position had to be performed in order to ensure an adequate and safe aperture.\\
At \lhcb, the upgraded \velo detector has, in its closed position, a minimal nominal distance of 3.5\mm from the beam axis, an aperture that is considered safe in the expected (HL-)LHC conditions of Run~3 and Run~4~\cite{LHCb-PUB-2012-018}. In nominal conditions the aperture is always limited by the downstream part of the RF boxes, Fig.~\ref{fig:cell1b}.
However, for a precise determination of the minimal allowed aperture, several effects must be taken into account, including the transverse offset imposed by the beam crossing configuration, waist shift, beta-beating, and the expected orbit shift during the physics fill.
Furthermore, several machine configurations need to be studied, with baseline optics as well as smaller values of $\beta^*$, both horizontal and vertical crossing configurations, and also special runs like $\beta^*$-leveling, ion runs and van der Meer scans. These studies have been performed for the case of the SMOG2 cell and have shown that the minimum allowed aperture over the longitudinal range of the SMOG2 storage cell is imposed by the van der Meer scan configuration, and amounts to 3\mm (assuming that the storage cell is centered around the closed orbit at every fill), see Fig.~\ref{fig:pedro1}. Based on these studies, a 5\mm storage cell inner radius, which accommodates these tolerances with a sufficient margin, was considered the best compromise between aperture and luminosity requirements.

Bunched beams with 40~MHz bunch frequency and high bunch charge represent strong sources of electromagnetic fields. The general rules for guiding these beams safely are: (i) to surround them with conducting surfaces that vary as smoothly as possible in cross-section in order to keep the RF field close to the beams, and (ii) to avoid excitation of cavity-like structures or other resonating systems. 
Electromagnetic simulations were used to clarify the impact of the WFS system of the SMOG2 cell on the \lhc. This consisted of eigenmode calculations, frequency domain wire simulations, and time-domain wakefield simulations.
The simulations revealed that the \lhc longitudinal and transverse beam stability is not altered significantly by the addition of the SMOG2 storage cell. Additionally, no evidence has been found that the SMOG2 setup modifies longitudinal and transverse resonant modes in both open and closed positions~\cite{CERN-PBC-Notes-2018-008}.

The electro-mechanical features of the storage cell of the proposed (LHCspin) polarized target system (geometry, materials, electric contacts, WFS, etc.) will be very similar to those of the SMOG2 cell, so a similarly negligible impact on the beam stability is expected. Nonetheless, dedicated simulation studies are in order to quantify the impact of the WFS system.\\

 \subsubsection{Impact on beam lifetime}
\begin{figure}[h!]
    \centering
    \includegraphics [scale=0.5] {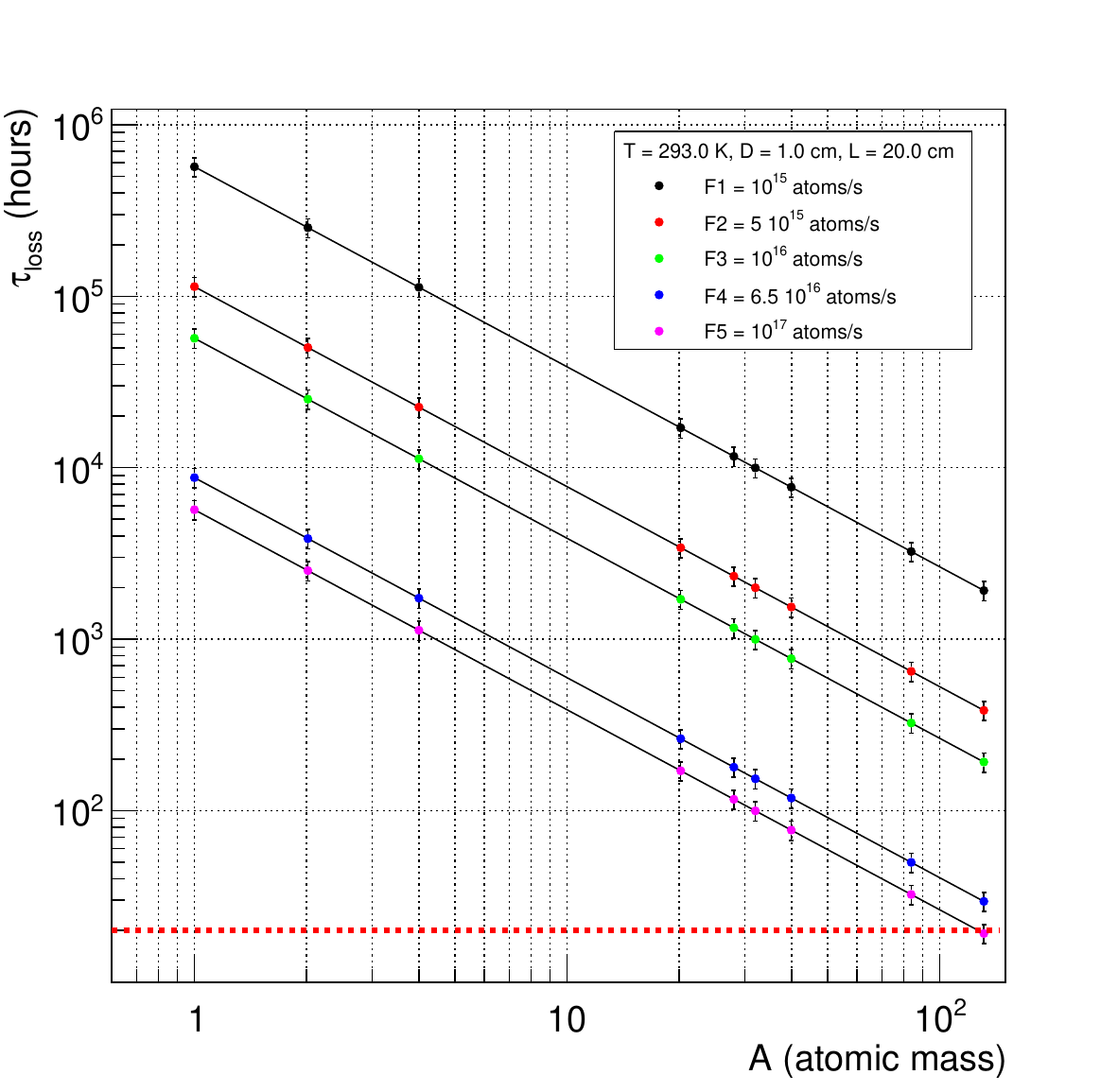}
    \caption{Beam lifetime as a function of the target gas atomic mass for different hypoteses of injected fluxes. The horizontal dotted line at 20 hours represents the maximal duration of a typical LHC fill.}
    \label{fig:tloss}
\end{figure}

With the injection of gas into the storage cell, the circulating beams undergo additional collisions which contribute to decrease their intensity. Defining the lifetime $\tau_{loss}$ of the beams as the time interval in which the intensity of the beams is reduced to $1/e$ of the initial one due (solely) to collisions with the target gas, one has: $N_{b}(t)=N_{b}(0)\cdot\exp(-t/\tau_{loss})$. The lifetime $\tau_{loss}$ depends on the total beam-gas interaction cross section $\sigma_{pA}$ and on the instantaneous luminosity $\mathcal{L}$ via the following equation

\begin{equation}
   \tau_{loss}=\frac{N_{b}}{\mathcal{L} \cdot \sigma_{pA}(100~GeV)}=\frac{1}{f_{rev} \cdot \theta \cdot \sigma_{pA}(100~GeV)}~,
\end{equation}

\noindent
where $f_{rev} = 11245$ Hz is the beam revolution frequency and $\theta$ is the areal density of the target gas, expressed in ${\rm atoms}/{\rm cm}^2$. The total beam-gas cross section is estimated in terms of the proton-proton cross section at $\sqrt{s}=100$ GeV, $\sigma_{pp}(100~GeV) \simeq 0.050~b$ through the approximated relation

\begin{equation}
  \sigma_{pA}\simeq A^{2/3}\sigma_{pp},
\end{equation}

\noindent
where $A$ denotes the atomic/molecular mass of the target gas species.
To ensure a marginal impact on the beam intensity, the lifetime $\tau_{loss}$ must be much larger than the typical duration of a LHC fill ($\approx 10-15$ hours). Considering as a reference the geometry of the SMOG2 storage cell ($L=20~cm, D=1.0~cm$) and different injection fluxes at room temperature, ranging from $10^{15}$ to $10^{17} {\rm atoms}/s$, the resulting beam lifetimes for light gases (e.g. H, D) is several orders of magnitude above this limit, as shown in Fig.~\ref{fig:tloss}.

\noindent
Even assuming a relatively high injected flux of $10^{17}$ hydrogen atoms per second, a beam lifetime of the order of $\sim 6 \times 10^3$ hours is expected under normal conditions, which largely exceeds the duration of a typical fill.

\subsubsection{Impact of H$_2$ injection}

\label{sec:molflow}

Electron multipacting has been observed in particle accelerators with positively charged beams, leading to the formation of electron clouds that may cause beam instabilities, pressure rise, and heat loads. To suppress this effect, two types of coating materials are used for the LHC beam pipes: Ti-Zr-V (NEG)~\cite{CHIGGIATO2006382, HENRIST200195} and amorphous carbon (a-C)~\cite{PhysRevSTAB.14.071001, vollenberg:ipac2021-wepab338}. The low secondary electron yield (SEY) of the NEG coating is achieved after reducing the surface oxides by heating in vacuum at a temperature above 180 $^\circ$C for a few hours (a process called activation). However, after activation, the Ti-Zr-V film pumps hydrogen and other reactive species by gettering effect. As hydrogen is one of the gases to be injected as a target, the gettering effect of the NEG film could compromise the stability and reproducibility of the gas density in the storage cell. Differently from the NEG, the a-C coating does not require any activation process and is inert with respect to the injected gases. For these reasons it was decided to use a-C as a coating material for the internal surface of the SMOG2 storage cell. However, while this solution is optimal for the case of unpolarized gases, in a polarized gas target the cell coating has also to preserve as much as possible the atomic polarization. Dedicated studies on this specific aspect are reported in Section~\ref{sec:coating}.

The injection of molecular (H$_2$) and, especially, of atomic (H) hydrogen thus requires dedicated studies to address possible issues related to potentially detrimental effects on the NEG coating and to set realistic fluxes and injection time limits to the operation of the gas feed system. Dedicated  \textit{Molflow+} simulations and laboratory measurements have been performed for the case of (unpolarized) molecular hydrogen, as detailed in~\cite{SMOG2_paper}.

\begin{figure}
	\centering
\includegraphics[height=0.335\textwidth, clip]{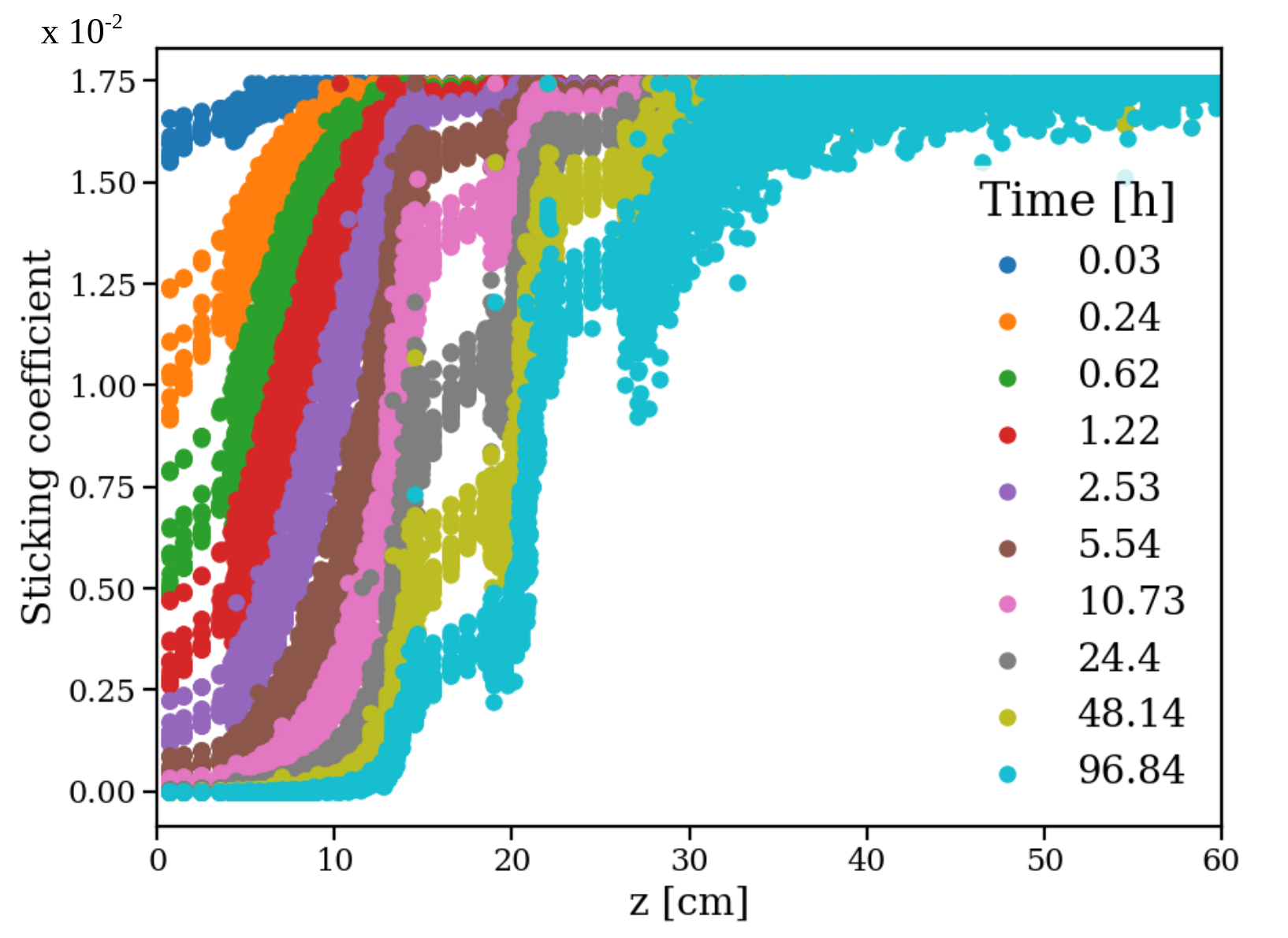}	\includegraphics[height=0.325\textwidth, clip]{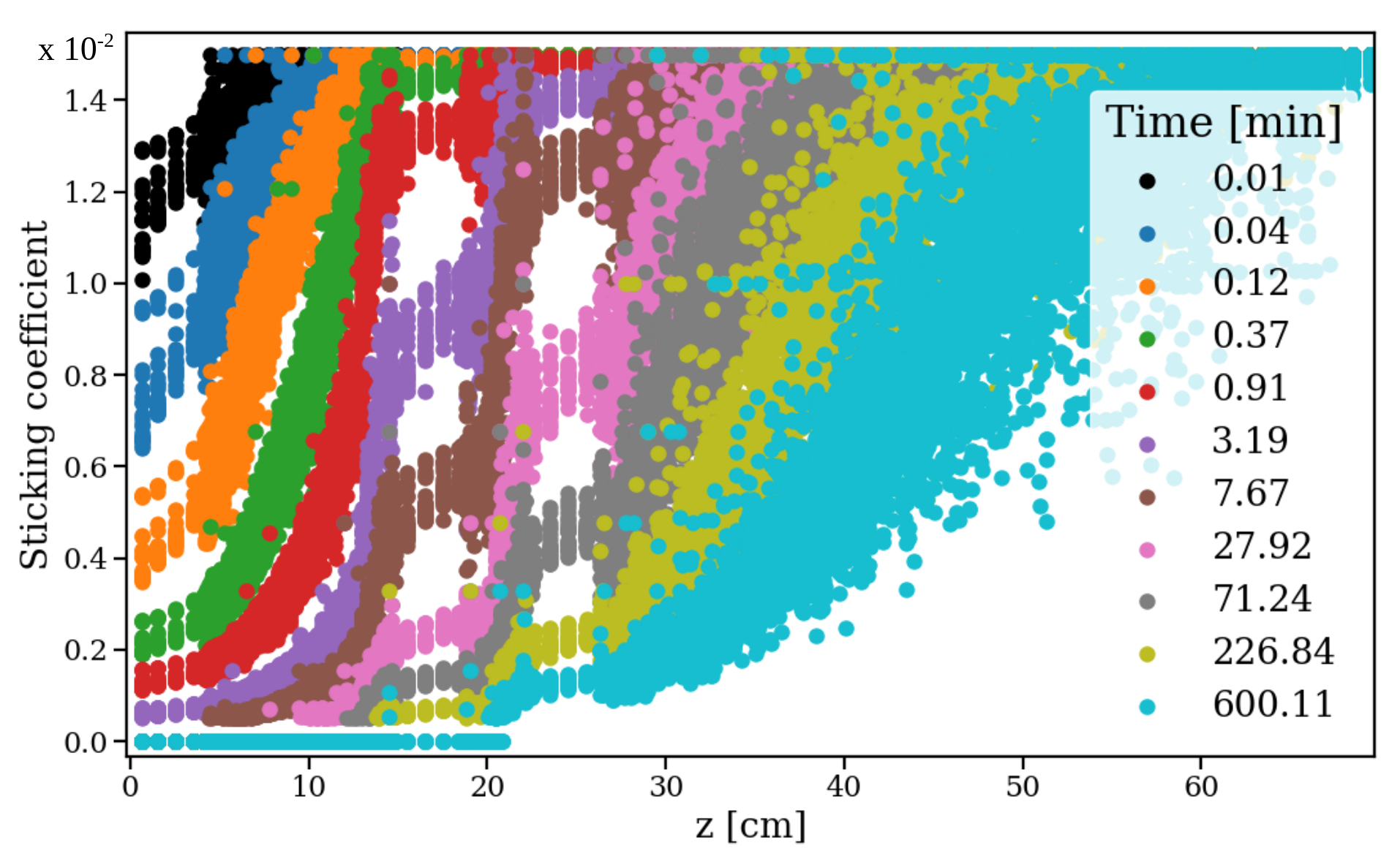}	
\caption{RF foil sticking coefficient as a function of the longitudinal position $z$ for different periods of injected H$_{2}$ (left) or N$_2$ (right). Each dot represents the sticking coefficient of a facet, with $z$ indicating the longitudinal coordinate of its center and being $z=0$ the upstream boundary of the RF foil.}
	\label{fig:molH2}
\end{figure}

Figure ~\ref{fig:molH2} presents the results of the simulated sticking coefficient evolution with the \textit{z} coordinate and the injection time. On the left, for H$_{2}$, the onset of saturation begins after around $20$ hours of injection and, after $100$~hours, the saturation only reaches the central region of the first 100\mm of RF foil, corresponding to around $2\%$ of the total area. On the right, for N$_{2}$, the saturation onsets after the first minutes from the injection and progresses much faster, as expected, reaching up to $200$~mm, corresponding to more than $15\%$ of the total area, after $10$~hours.\\
The results of these simulations, together with direct measurements on NEG samples in laboratory, assured that the level of saturation prospected in the RF box during Run 3 operation causes no safety issue to the \lhc operations.

\section{A polarized internal gas target for LHCb}

\label{sec:pgt}

\begin{figure}[h!]
\centering
\includegraphics[width=0.75\textwidth]{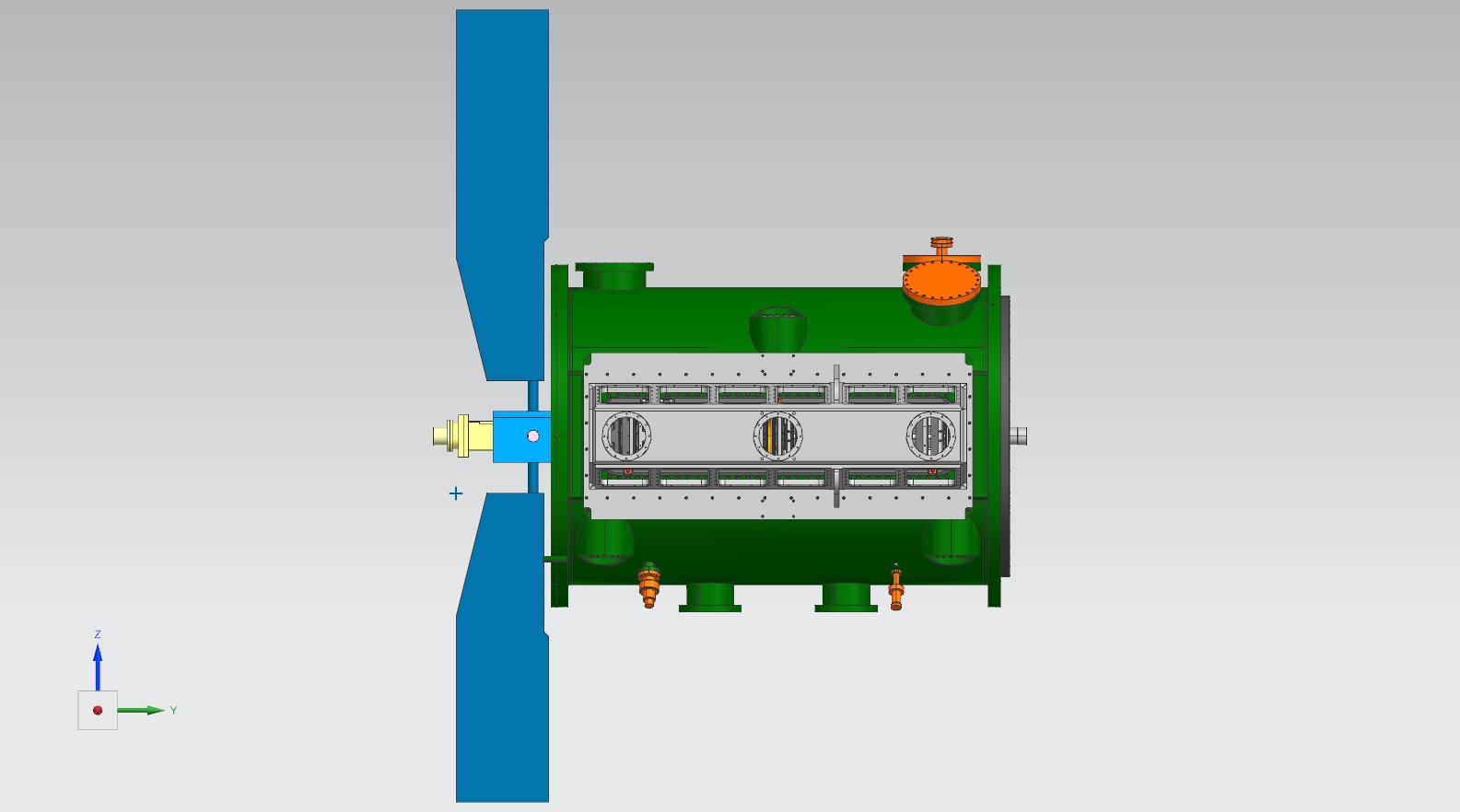}
\caption{Conceptual design of the full setup installed in the VELO alcove. On the left side of the VELO vessel, the vacuum chamber is visible, along with the ABS on the top and the BRP on the bottom. The absolute polarimeter is not shown.}
\label{fig:alcove}
\end{figure}

The apparatus of the polarized target comprises four main components: an Atomic Beam Source (ABS), a vacuum chamber hosting the storage cell and surrounded by the dipole magnetic field, an absolute polarimeter and a Breit-Rabi polarimeter (BRP). It is proposed to be installed at the IP8 during the LHC Long Shutdown 4, in the VELO alcove upstream of the \lhcb spectrometer, as shown in Fig.~\ref{fig:alcove}. 

\begin{itemize}

\item{The ABS}

The ABS (see e.g.~Ref.~\cite{Nass:2003mk}) consists of a dissociator to produce an atomic hydrogen gas, a beam forming system (nozzle, skimmer, collimator) to create a supersonic atomic beam, a sextupole system to focus atoms with z-component of the electron spin of +1/2 and defocus those with -1/2 based on the Stern-Gerlach principle, and high frequency transitions to interchange hyperfine state populations (Fig.~\ref{ABS-schematic}). This way the H-atoms are nuclear-spin polarized and are then injected into the storage cell.
\begin{figure}[ht!]
\begin{center}
    \includegraphics[width=10.5pc]{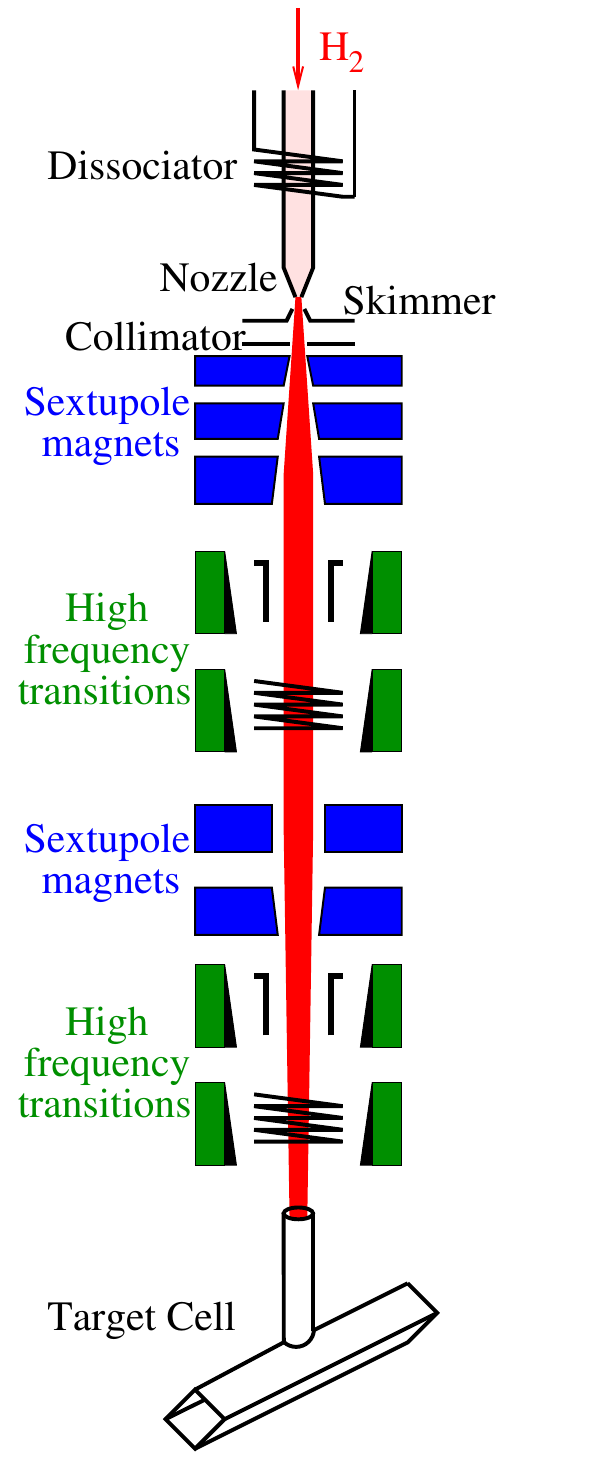}
    \hspace{4pc}%
    \caption{\label{ABS-schematic} Schematic drawing of an ABS.}
\end{center}
\end{figure}

A new dedicated ABS should be designed for LHCb in order to use up-to-date technology.
High remanence sextupole magnets will allow for higher atomic beam acceptance and for larger gap spaces between the magnets. This reduces rest gas attenuation in the magnet system and thus increases the ABS output intensity. New turbomolecular pumps with compression ratios for hydrogen as high as $10^6$, will lead to a drastically reduced forepumping system compared with previous ABSs. This allows for a more compact set-up which facilitates integration in the experiment. The design of a new ABS would need dedicated calculations to optimize the intensity, new tools are available\cite{Nass:2008th} and will be employed and developed further.

\item{The vacuum chamber}

The ABS injects a beam of polarized hydrogen or deuterium into the storage cell, to be located in the LHC primary vacuum along the beam pipe section upstream of the VELO. The storage cell, based on the same concept of the SMOG2 one, is placed inside a vacuum chamber and surrounded by a dipole magnet, as shown in Fig.~\ref{fig:pgt}. The chamber will be fabricated from low-carbon AISI 316L stainless steel. The magnet generates a $300~\rm{mT}$ static transverse field with a homogeneity of 10\% over the full volume of the cell, which is necessary to maintain the transverse polarization of the gas inside the cell, and to avoid beam-induced depolarization~\cite{Steffens:2019kgb}. 
A key difference with SMOG2 is the need to use proper storage cell coating which maintains a high level of nuclear polarization (see Section \ref{sec:coating}). 

\begin{figure}[h!]
\centering
\includegraphics[width=0.72\textwidth]{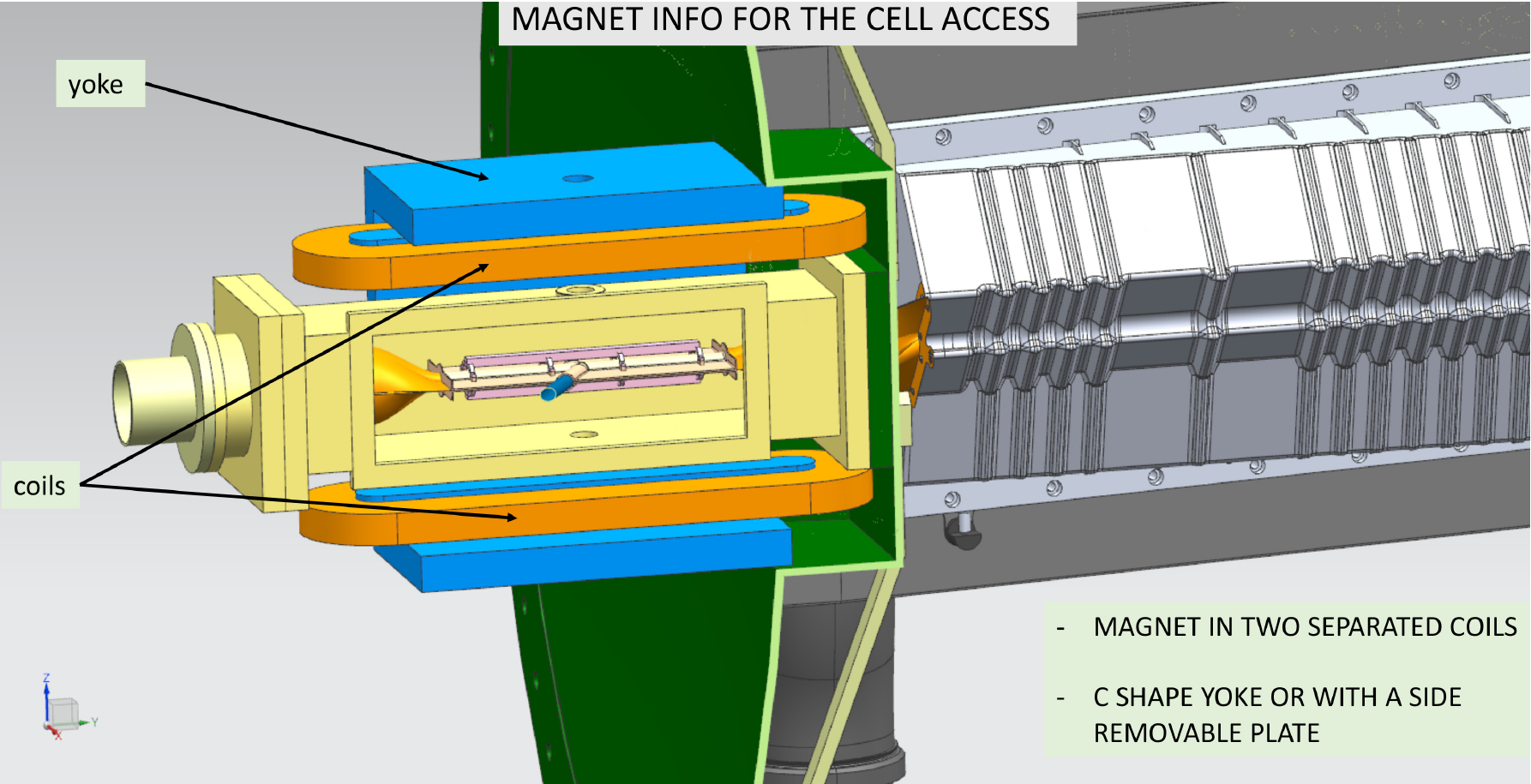}
\caption{A drawing of the LHCspin vacuum chamber (yellow) hosting the storage cell. The chamber is inserted between the coils of the magnet (orange) and the iron return yoke (blue). The VELO vessel and RF box are shown in green and grey, respectively.}
\label{fig:pgt}
\end{figure}

\item{Absolute polarimeter}

The studies conducted during the R\&D phase have revealed that, based on the cell coating material (Section~\ref{sec:coating}), only the molecular polarization of the gas survives the wall collisions (while the atomic polarization is largely lost). An absolute polarimeter will be therefore required. The calibration and commissioning of the absolute polarimeter will be conducted at the IR4, as described in Section~\ref{sec:abs_pol}. 

\item{Breit-Rabi polarimeter}

The BRP will be used for proper calibration of the ABS RF-transition units. The working principle and performance of a Breit-Rabi polarimeter are described in Ref.~\cite{HERMES:BRP}.

\item {The safety system}

The entire apparatus will be designed and realized in accordance with the LHC safety system. Two gate valves will separate the ABS and BRP from the LHC primary vacuum.

\end{itemize}

With this apparatus installed upstream of the upgraded \lhcb spectrometer, high-precision polarized fixed-target data will be collected for the first time at LHC during the Run 5, in parallel with collider data, as discussed in Section~\ref{sec:performance}.

\section{Performance and physics measurements}

\label{sec:performance}
In this section, expected LHCspin event rates and performance for p-H collisions are provided for several physics channels of interest. 

\subsection{Expected event rates}
\label{sec:lumi}
The expected yields are computed by projecting early SMOG2 results and assume the current (Run 3) reconstruction and selection efficiency, i.e. the expected performance increase of the upgraded LHCb detector~\cite{CERN-LHCC-2021-012} is conservatively neglected. 

\begin{figure}[ht!]
\begin{center}
\includegraphics[width=.7\textwidth]{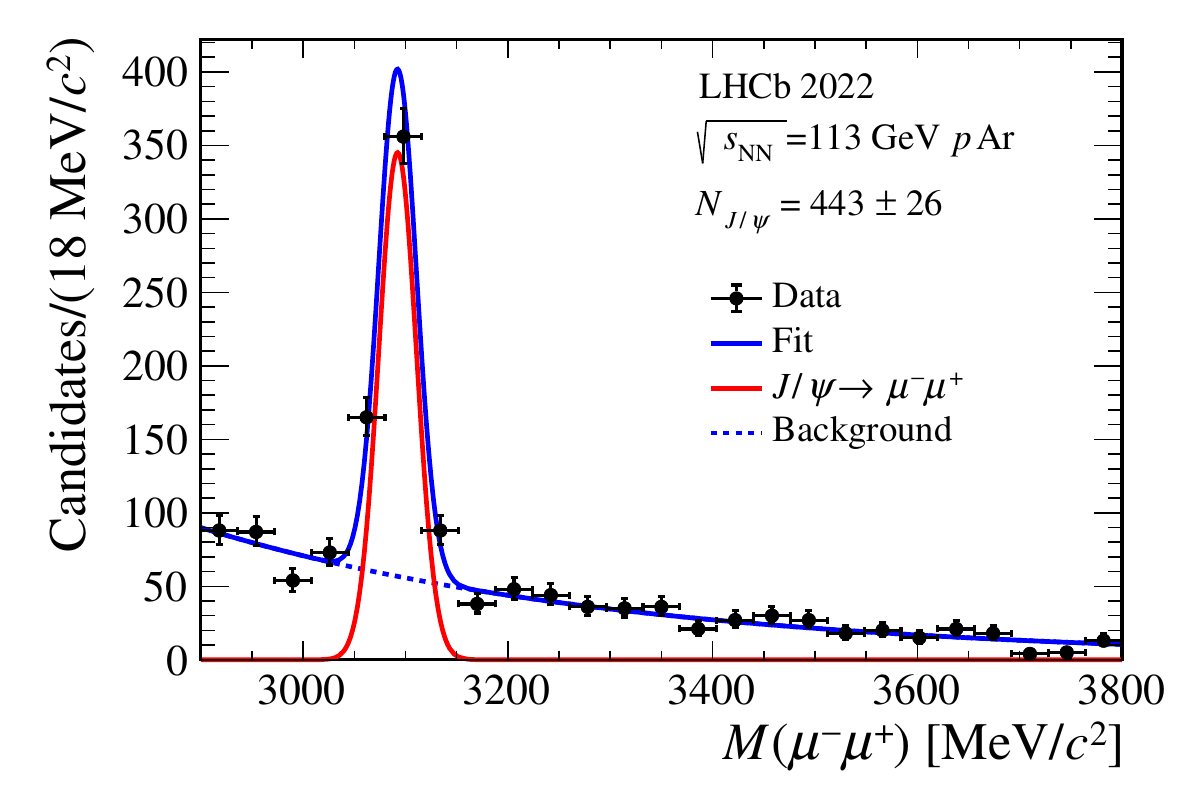}
\end{center}
\caption{Reconstructed $J/\psi \to \mu^+\mu^-$ decays from p-Ar collision collected with SMOG2~\cite{LHCB-FIGURE-2023-008}.}
\label{fig:jpsimumu_smog2}
\end{figure}

The SMOG2 data considered here have been collected in 2022 in only 18 minutes of p-Ar collisions with an injection pressure of $\rm{p} = 9.7 \times 10^{-8}~\rm{mbar}$. The clean invariant mass of $J/\psi \to \mu^+\mu^-$ events reconstructed with this dataset is shown in Fig.~\ref{fig:jpsimumu_smog2}.

The nucleon areal density for a triangular density profile~\cite{BoenteGarcia:2024kba} can be computed as: 
\begin{equation}
    \theta_{\rm{SMOG2}} = \frac{\rm{p}}{k_B T} \times \frac{L}{2} \times  \times A_{Ar} =  1.88 \times 10^{12} ~ \rm{nucleons/cm}^2,
\end{equation}
where a cell length of $L=20$ cm, a gas temperature $T=300~\rm{K}$ and $A_{Ar}=40$ have been used. Each nucleon is assumed to contribute incoherently to the production cross section, i.e. nuclear effects are neglected.

The areal density for the LHCspin cell filled with atomic hydrogen can instead be estimated as:
\begin{equation}
\label{eq:theta}
    \theta_{\rm{LHCspin}} = \frac{1}{2} \frac{\phi}{C} L = 3.7 \times 10^{13} ~\rm{atoms/cm}^2,
\end{equation}
where $\phi = 6.5 \times 10^{16}$ atoms/s is the expected flux from the atomic beam source (Sec.~\ref{sec:pgt}), $L_{cell}=20~\rm{cm}$ the length of the LHCspin cell and $C=17.5$ l/s the expected conductance of the cell (two open-ended half-length tubes) plus the feed tube, computed according to~\cite{conductance}:
\begin{equation}
    C=C_{\rm{cell}} + C_{\rm{tube}} = 
    2 \times 3.81\sqrt{\frac{T}{M}} \frac{D^3}{L_{\rm{cell}}/2 + 4/3 D}
    + 3.81\sqrt{\frac{T}{M}} \frac{D^3}{L_{\rm{tube}} + 4/3 D}~,
\end{equation}
where $T=300~\rm{K}$ is the gas temperature, $M=1$ the molecular mass, $D=1~\rm{cm}$ the cell and tube diameter and $L_{\rm{tube}}=10~\rm{cm}$ the length of the feed tube, respectively.
A triangular density profile is again assumed, giving the factor $1/2$ in Eq.~\ref{eq:theta}.
The areal density of a jet target (i.e. without a storage cell) would be around 40 times lower in case of hydrogen injection.

The ratio of LHCspin/SMOG2 event rates can therefore be obtained via the scaling factor:
\begin{align}
 f &= \frac{\theta_{\rm{LHCspin}}}{\theta_{\rm{SMOG2}}} 
 \times \frac{\epsilon}{\epsilon_{\rm{SMOG2}}}
 \times \frac{N_b^{\rm{Run 3}}}{{N_b^{\rm{SMOG2}}}}
 = \frac{3.7 \times 10^{13}}{9.36 \times 10^{11}}
 \times 3.24
 \times \frac{2808}{1735}
 = 208~,
 \label{eq:from_smog2}
\end{align}
where the number of colliding bunches has been scaled from the 1735 used in SMOG2 to the nominal 2808, and the efficiency correction accounts for known detector inefficiencies during the commissioning phase of LHCb.
The beam current, i.e. the total amount of circulating protons in the LHC ring is assumed to remain at the Run 3 value of $1.4\times10^{11}$ though it is likely to increase in Run 4 and Run 5.

The results of the projections are given in Tab.~\ref{tab:stat_from_smog2}, which reports the expected weekly rate of fully-reconstructed and selected events as well as a total yield obtained across 120 weeks of data-taking, corresponding to a full run.

\begin{table}[ht]
\centering
\begin{tabular}{c|c|c} \toprule
\bf{Channel} & \bf{Events / week} & \bf{Total yield} \\\midrule
$J/\psi \to \mu^+ \mu^-$ & \num{2.6E+07} & \num{3.1E+09} \\
$D^0 \to K^- \pi^+$ & \num{1.3E+08} & \num{1.6E+10} \\
$\psi(2S) \to \mu^+ \mu^-$ & \num{4.6E+05} & \num{5.5E+07} \\
$J/\psi J/\psi \to \mu^+ \mu^- \mu^+ \mu^-$ (DPS) & \num{1.7E+01} & \num{2.1E+03} \\
$J/\psi J/\psi \to \mu^+ \mu^- \mu^+ \mu^-$ (SPS) & \num{5.1E+01} & \num{6.1E+03} \\
Drell Yan ($5<M_{\mu\mu}<9$ GeV) & \num{1.5E+04} & \num{1.8E+06} \\
$\Upsilon \to \mu^+ \mu^-$ & \num{1.1E+04} & \num{1.3E+06} \\
$\Lambda_c^+ \to p K^- \pi^+$ & \num{2.6E+06} & \num{3.1E+08} \\
\bottomrule
\end{tabular}
\label{tab:stat_from_smog2}
\caption{Estimated weekly reconstructed and selected event rates at LHCspin and the total yield over a Run (120 weeks) for various channels.}
\end{table}

Estimates for several channels are provided by scaling their expected yield relatively to $J/\psi \to \mu^+\mu^-$ as provided  in~\cite{Citron:2018lsq,Bursche:2649878,Santimaria:2023xdy}. Di-$J/\psi$ events produced via Double Parton Scattering (DPS) are estimated via the pocket formula~\cite{Belyaev:2017sws}:
\begin{equation}
    \sigma(J/\psi J/\psi)_{\rm{DPS}} = \frac{1}{2}\frac{\sigma(J/\psi)^2}{\sigma_{\rm{eff}}}
    ~~~~
    \to
    ~~~~
    \frac{\sigma(J/\psi J/\psi)_{\rm{DPS}}}{\sigma(J/\psi)} \approx 6\times10^{-5}~,
\end{equation}
having used $\sigma_{\rm{eff}}\approx 10~\rm{mb}$ from~\cite{Lansberg:2015lva} and $\sigma(J/\psi)\approx 1226~\rm{nb}$ per nucleon from~\cite{LHCb:2018jry}. According to Ref.~\cite{Lansberg:2015lva}, single parton scattering (SPS) is expected to be about three times more frequent.

\paragraph{Luminosity at Run 4}
The above estimates do not rely on the knowledge of the instantaneous luminosity.
This can anyway be computed as
\begin{equation}
    \mathcal{L} = \frac{dN_p}{dt} \times \theta =
    f_{rev} \times N_b \times N_p \times \theta =
    1.6\times10^{32}~\rm{cm^{-2}s^{-1}},
\end{equation}
where $f_{rev} = 11245~\rm{Hz}$ is the LHC revolution frequency, $N_b=2808$ is the number of colliding bunches, $N_p=1.4\times10^{11}$
is the number of protons in each bunch and the areal density is given by Eq.~\ref{eq:theta}.
With 120 weeks of data-taking this corresponds to an integrated luminosity of $5.8~\rm{fb}^{-1}$ at the end of the Run 4.

\subsection{Expected precision on TSSAs}
\begin{figure}[ht!]
\begin{center}
\includegraphics[width=.49\textwidth]{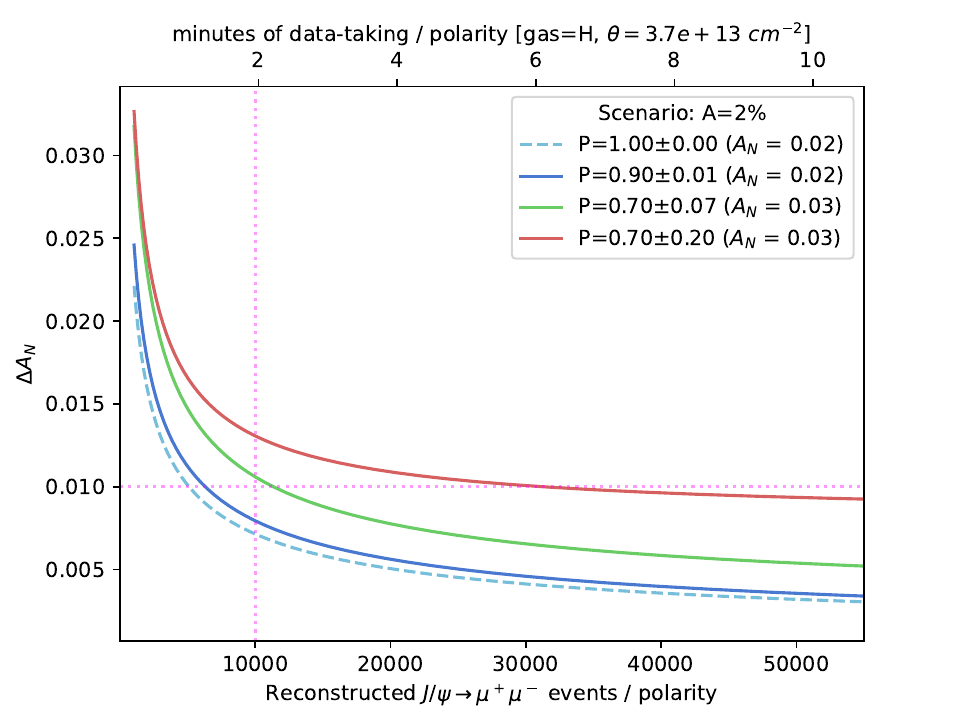}
\hfill
\includegraphics[width=.49\textwidth]{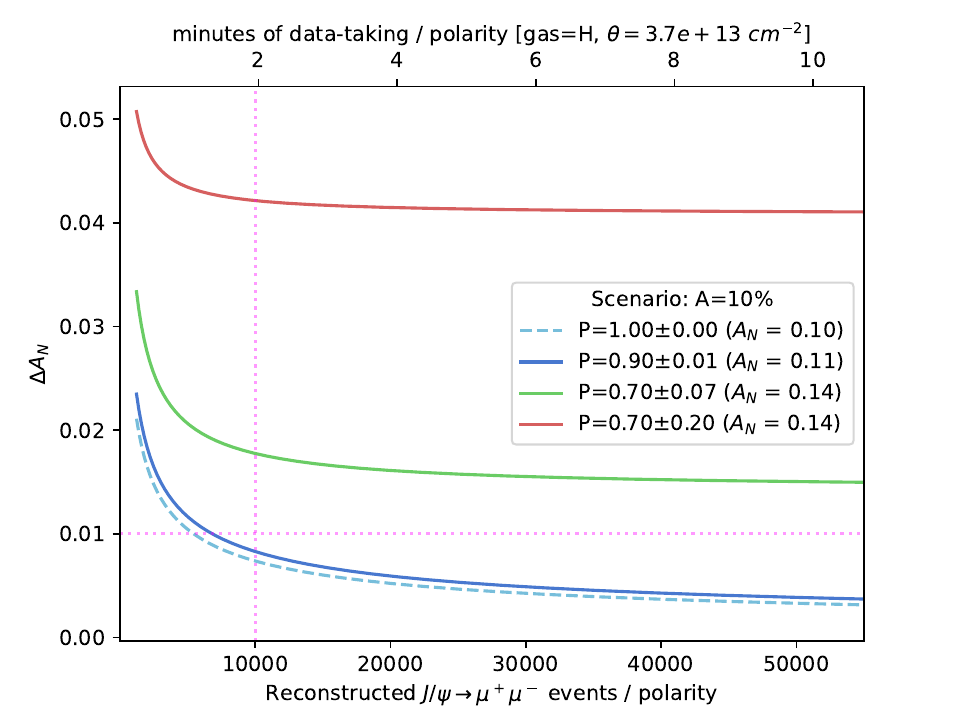}
\end{center}
\caption{Number of fully-reconstructed events and data-taking time to reach a given precision on a spin asymmetry at LHCspin with four different polarization degrees and related uncertainties.}
\label{fig:lumi}
\end{figure}

Fig.~\ref{fig:lumi} shows the data-taking time needed for each polarity state to reach a given precision on a transverse-target single spin asymmetry (TSSA). 
The TSSA is defined as
\begin{equation}
\label{eq:tssa}
    A_N = \frac{1}{P}\frac{N^{\uparrow}-N^{\downarrow}}{N^{\uparrow}+N^{\downarrow}} \equiv \frac{A}{P},
\end{equation}
where $P$ is the polarization degree and $N^{\uparrow(\downarrow)}$ denote the particle yields per each target polarization state.

The four curves represent the uncertainty on $A_N$ coming from both the statistical uncertainty and the systematic uncertainty associated with the knowledge of the polarization degree.
It is remarkable that an absolute precision better than 0.01 can be attained in just few minutes of data-taking on $J/\psi \to \mu^+\mu^-$ decays and similar high-statistics channels.

As an example, if $P = 0.70 \pm 0.07$ and $A = 5\%$, with 10000 collected events one has $A_N = 0.0714 \pm 0.016$, i.e. a relative precision of about 22\%. If instead $P = 0.90 \pm 0.01$, then $A_N = 0.0714 \pm 0.011$, i.e. a relative precision of about 20\%. 

Of course the interplay between statistical and systematic errors also depends on the kinematic binning choice: a good knowledge of the polarization degree is mandatory for a precise $A_N$ measurement on high-statistics channels, while a low-statistic channel, or a poorly populated kinematic region of a high-statistics channel, does not benefit from it since $A_N$ will be statistically limited.

\subsection{Kinematic coverage and efficiencies}
\label{sec:perf}
The LHCb performance for LHCspin data are analyzed in this section by means of full LHCb simulations.

Simulated samples of several decays are generated with \pythia~\cite{Sjostrand:2007gs}, configured specifically for \lhcb~\cite{LHCb-PROC-2010-056}, with the colliding proton beam momentum set to match the momentum per nucleon of both the beam and the target in the centre-of-mass frame. 
The four-momentum of the decay products is then embedded into p-H minimum bias events generated by the EPOS-LHC event generator~\cite{Pierog_2015}. The interaction of the generated particles with the detector, along with its response, are simulated using the \geant toolkit~\cite{GEANT4:2002zbu, Allison:2006ve}, as described in Ref.~\cite{Clemencic:2011zza}.
The simulated samples are reconstructed and analyzed using the same software tools employed for data processing. 

p-H interaction vertices are simulated with a flat distribution over a broad region, covering $\rm{PV}_z \in [-800,200]~\rm{mm}$ where $\rm{PV}_z$ is the $z$ coordinate (along the beam line) of the Primary Vertex (PV). A triangular density distribution has shown to yield similar kinematic coverage.

\begin{figure}[ht!]
\begin{center}
\includegraphics[width=.49\textwidth]{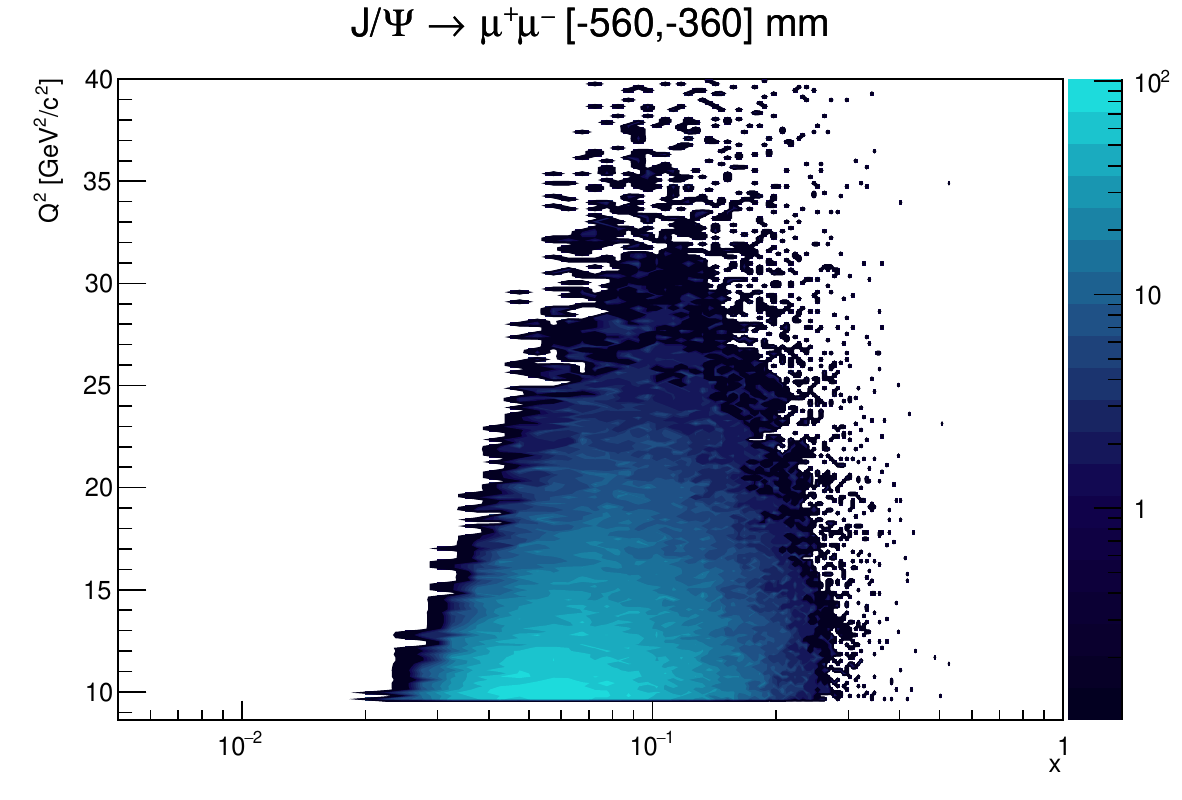}
\hfill
\includegraphics[width=.49\textwidth]{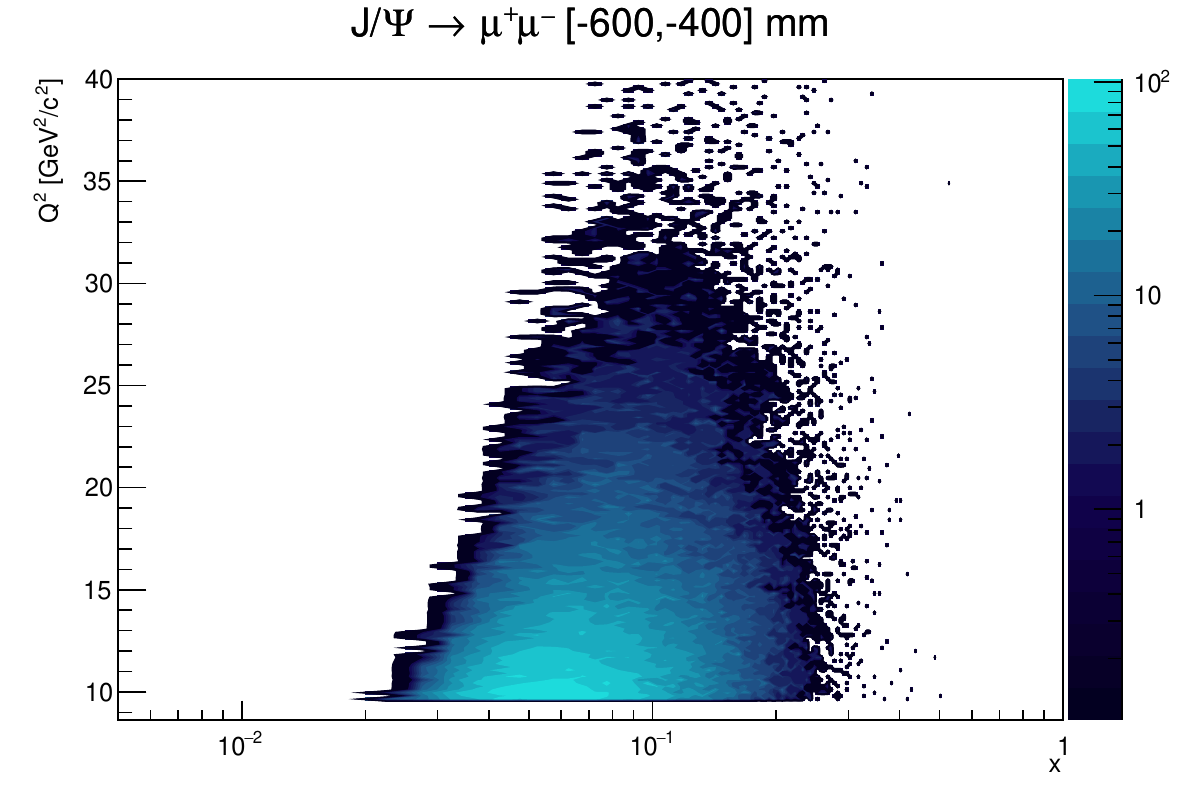}
\newline
\includegraphics[width=.49\textwidth]{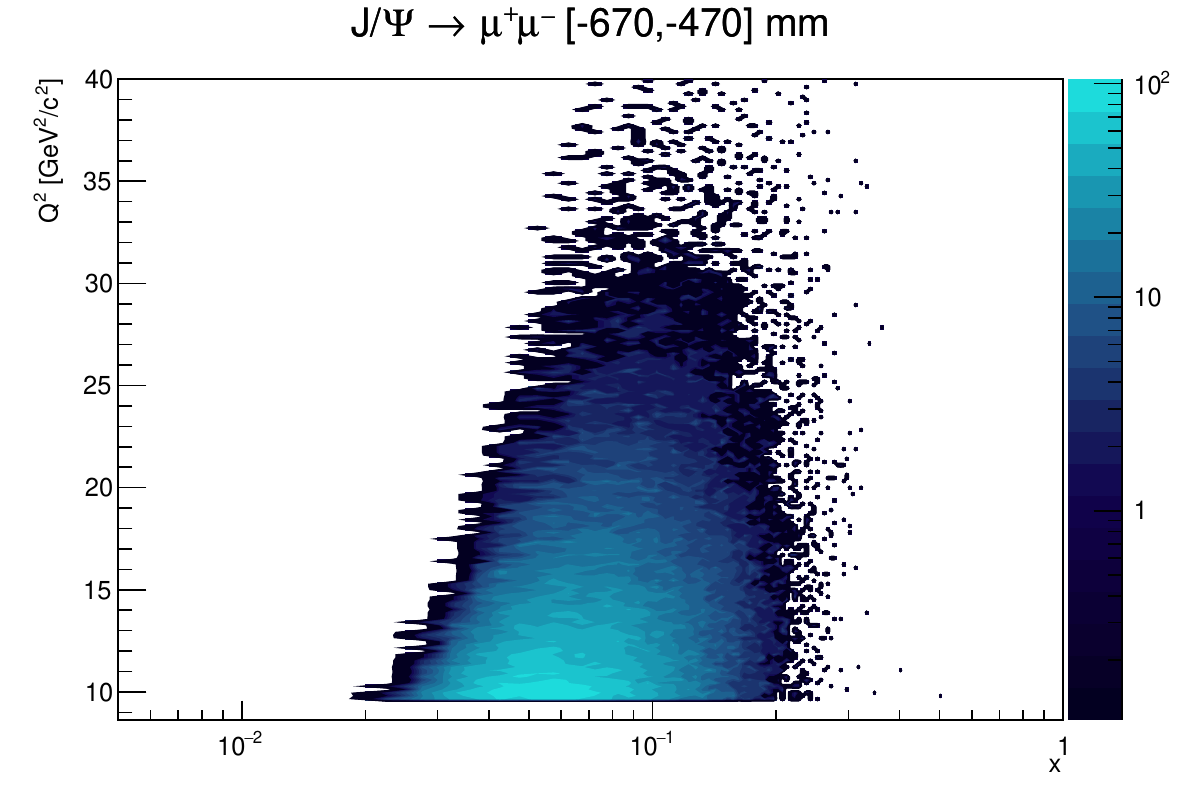}
\hfill
\includegraphics[width=.49\textwidth]{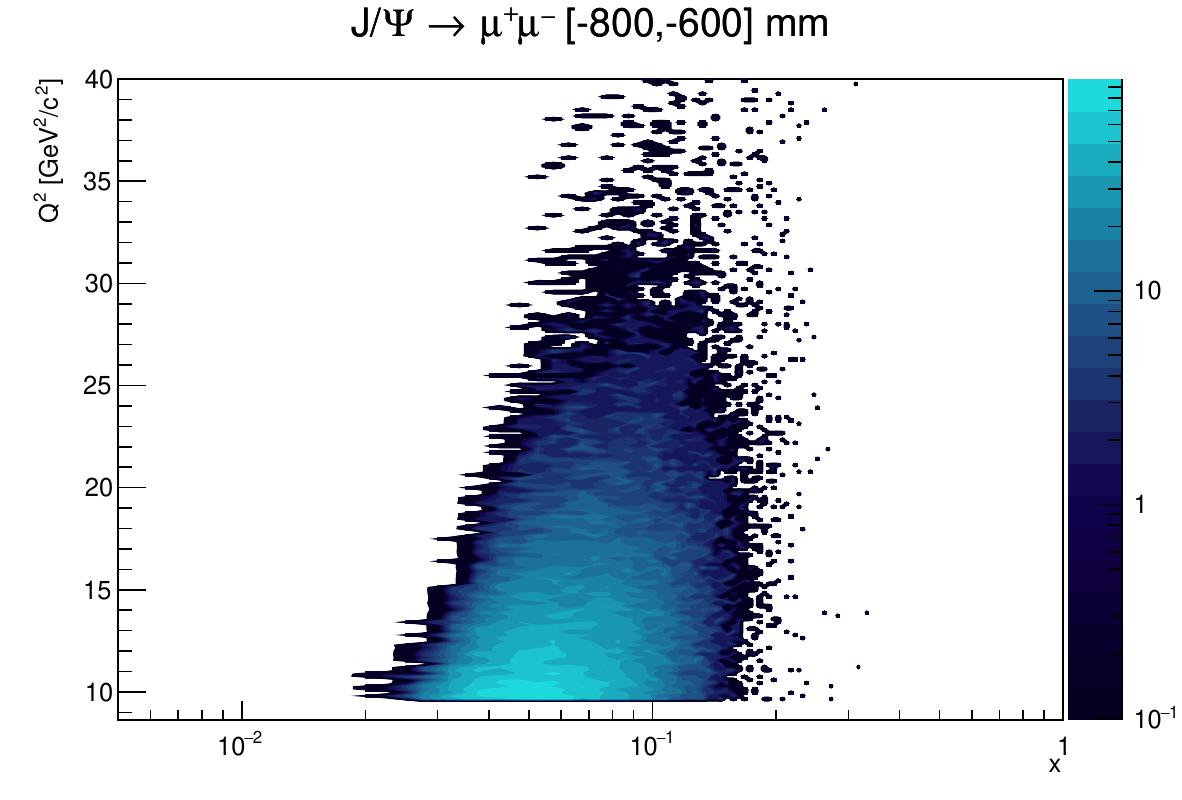}
\end{center}
\caption{Kinematic coverage in the $x-Q^2$ plane.}
\label{fig:kin2}
\end{figure}

The simulated $x-Q^2$ distribution of $J/\psi \to \mu^+\mu^-$ events is investigated in four $20~\rm{cm}$-long colliding regions, representing four possible $z$ positions of the LHCspin cell: $[-560,-360]~\rm{mm}$, which represents the SMOG2 cell position, i.e. the closest possible position to the VELO, together with three upstream configurations as indicated in Fig.~\ref{fig:kin2}.
These results indicate a broad kinematic coverage, which is enlarged as the gas cell is brought closer to the vertex locator.

\begin{figure}[ht!]
\begin{center}
\includegraphics[width=.7\textwidth]{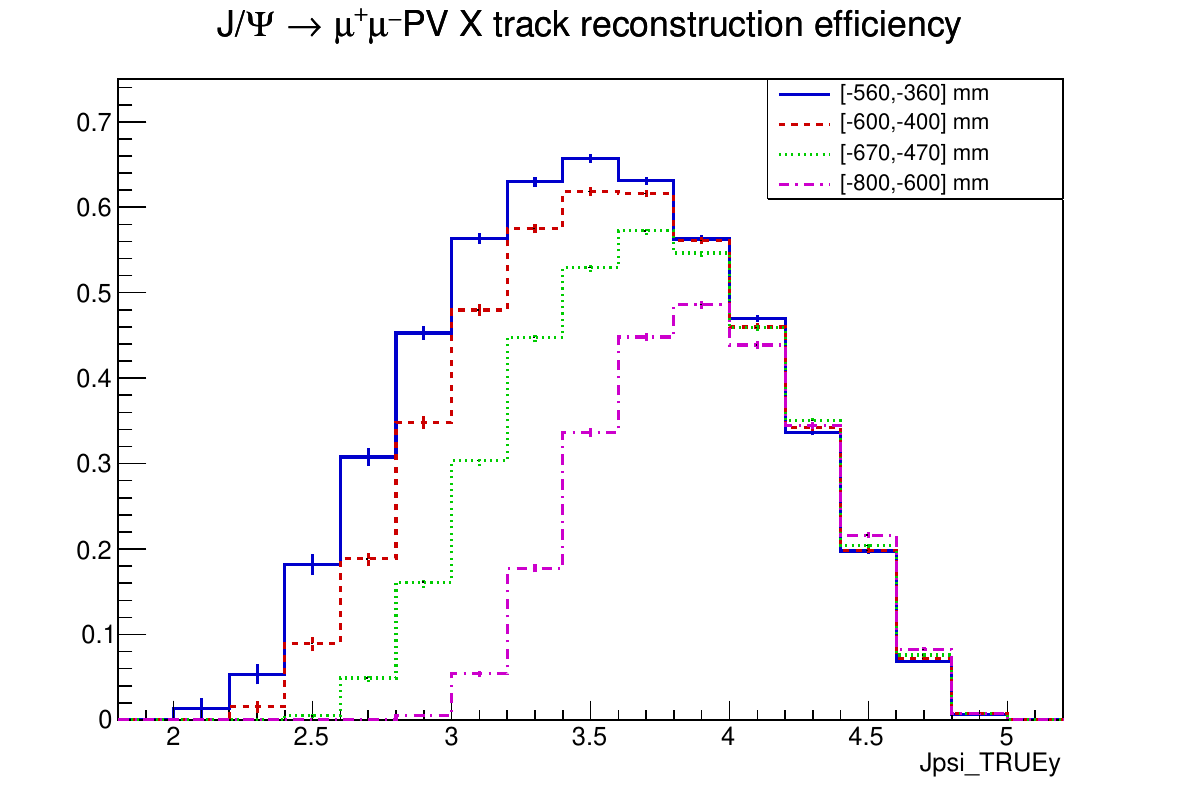}
\end{center}
\caption{Reconstruction efficiencies for $J/\psi \to \mu^+ \mu^-$ events.}
\label{fig:effs}
\end{figure}

Fig~\ref{fig:effs} shows the efficiency of reconstructing the PV and both tracks in $J/\psi \to \mu^+\mu^-$ decays as a function of the $J/\psi$ rapidity in the four considered cell positions.
Again, as the cell is placed further upstream, the rapidity coverage shrinks, albeit no dedicated vertexing and tracking algorithms have been used for this study.

\subsection{Analysis of pseudo-data}
\label{sec:tssa}
To create a pseudo-dataset for LHCspin, the polarization of the target gas is emulated by weighting events according to a given function~\cite{HERMES:2020ifk}. A variable called $\rho$ is computed based on the $J/\psi$ $x,p_T$ and $\phi$ angle values:
\begin{align}
\label{eq:model}
\rho =& 
    \frac{1}{2} \Bigg[ 1 + \left( a_1 + a_2\frac{x-\overline{x}}{x_{max}}
    + a_3\frac{p_T-\overline{p_T}}{p_{T~max}} \right) \sin \phi + \left( b_1 + b_2\frac{x-\overline{x}}{x_{max}}
    + b_3\frac{p_T-\overline{p_T}}{p_{T~max}} \right) \sin 2\phi
    \Bigg]
\end{align}

\noindent
where the overline denotes the average and $max$ indicates the largest value in the $p_T$ or $x$ spectrum. For each event, a random number between 0 and 1 is extracted according to a flat distribution: if the outcome is greater than $\rho$, a $-1$ tag is assigned to the event, and $+1$ otherwise, representing the polarization state. In particular, Eq.~\ref{eq:model} emulates a Sivers amplitude at the first order in the Taylor expansion of $p_T$ and $x$, with the second harmonic accounting for a possible higher-twist contribution.

\begin{figure}[ht!]
\begin{center}
\includegraphics[width=.49\textwidth]{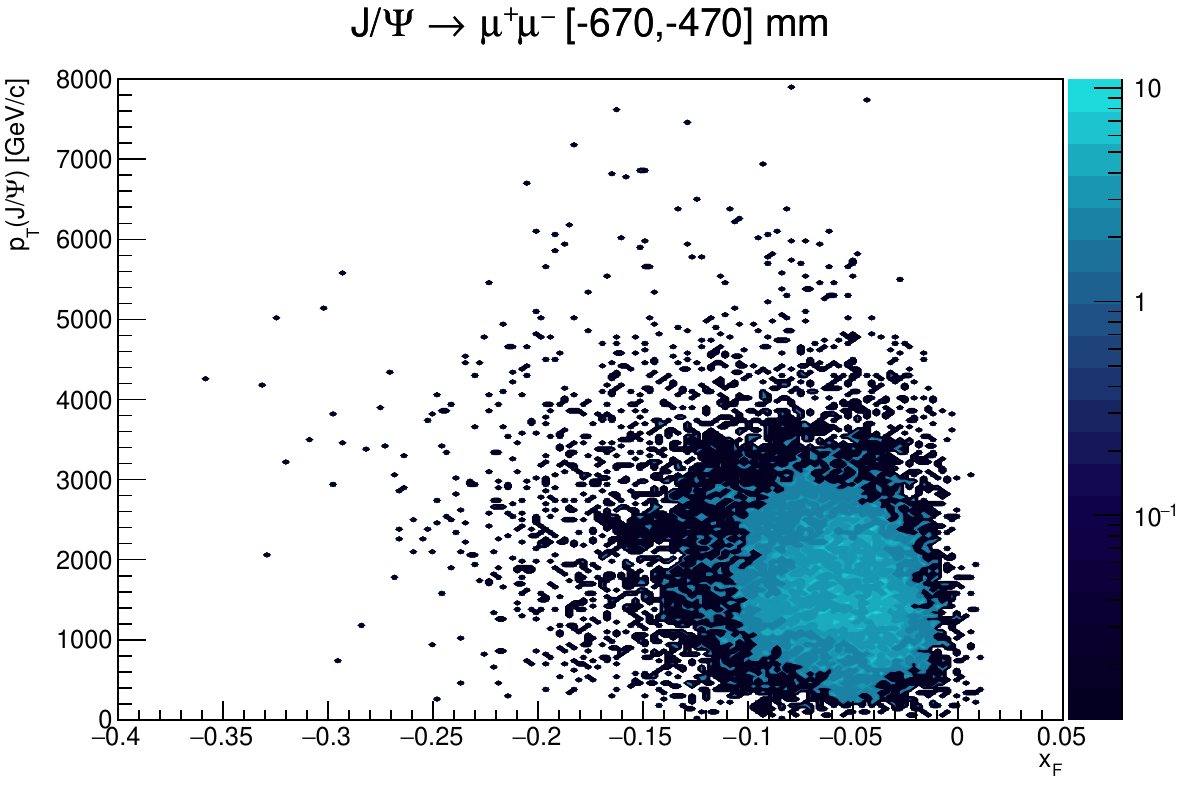}
\hfill
\includegraphics[width=.49\textwidth]{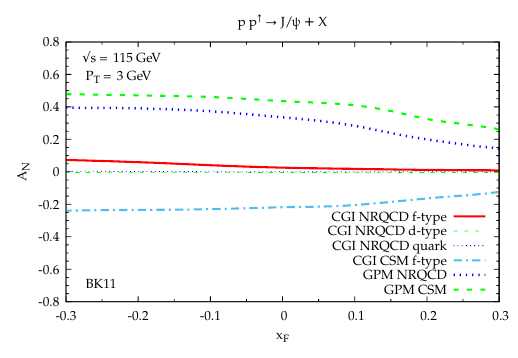}
\end{center}
\caption{Left: $x_F-p_T$ spectrum of simulated $J/\psi \to \mu^+\mu^-$ events.
Right: predicted asymmetry for polarized p-H collisions at $\sqrt{s}=115$ GeV~\cite{DAlesio:2020eqo}}
\label{fig:model}
\end{figure}

The distribution of simulated $J/\psi \to \mu^+\mu^-$ events in the $x_F-p_T$ plane is shown in Fig.~\ref{fig:model} (left). This sample is used in the following to emulate a measurement of the gluon Sivers function (GSF) with LHCspin.
The GSF is investigated in~\cite{DAlesio:2020}, where several models are shown to predict a sizeable asymmetry in the negative Feynman-x hemisphere (Fig.~\ref{fig:model}, right) in the kinematic range of LHCspin.
Qualitatively based on this prediction, the chosen parameters for Eq.~\ref{eq:model} are $a_1=0.1$, $a_2=a_3=0.05$ and $b_1=0.02$, $b_2=b_3=0.01$, i.e. a $10\%$ amplitude with a mild dependence on the kinematics.
\\
The TSSA can now be computed via Eq.~\ref{eq:tssa} by counting the events having a given polarization state as described in the following.

\begin{figure}[ht!]
\begin{center}
\includegraphics[width=.49\textwidth]{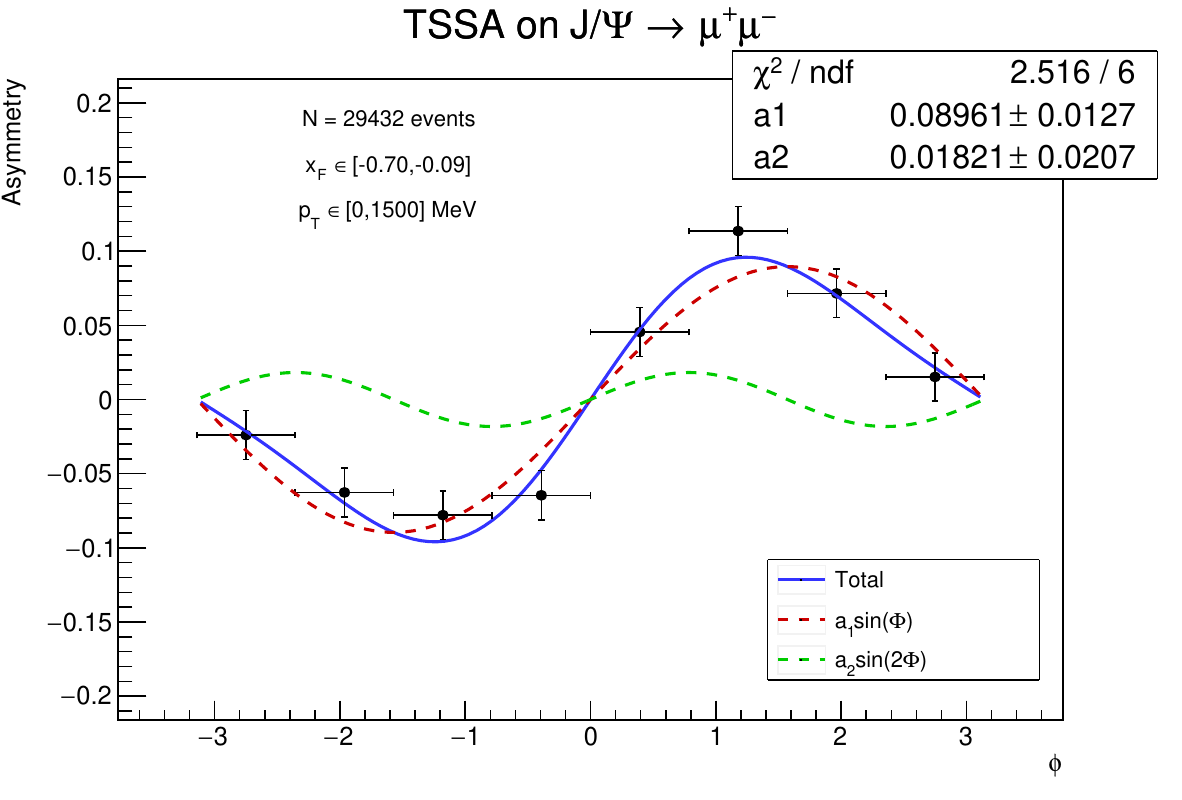}
\hfill
\includegraphics[width=.49\textwidth]{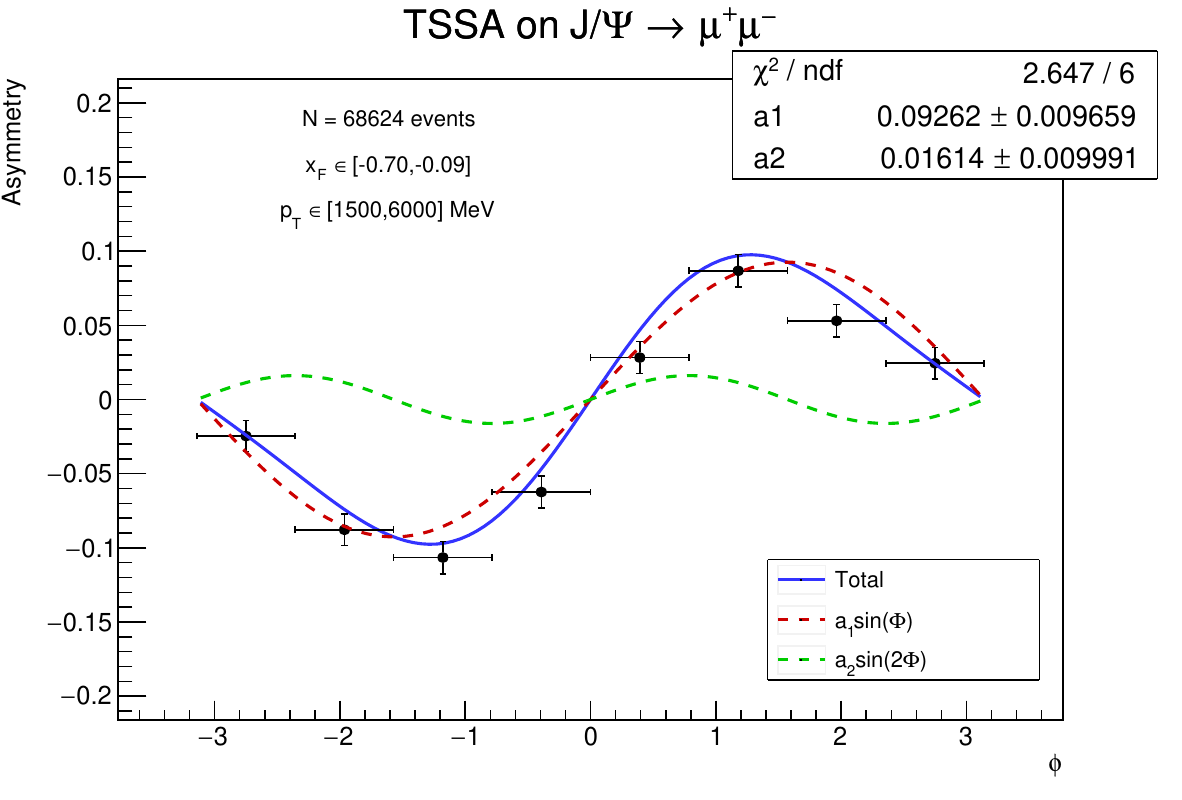}
\newline
\includegraphics[width=.49\textwidth]{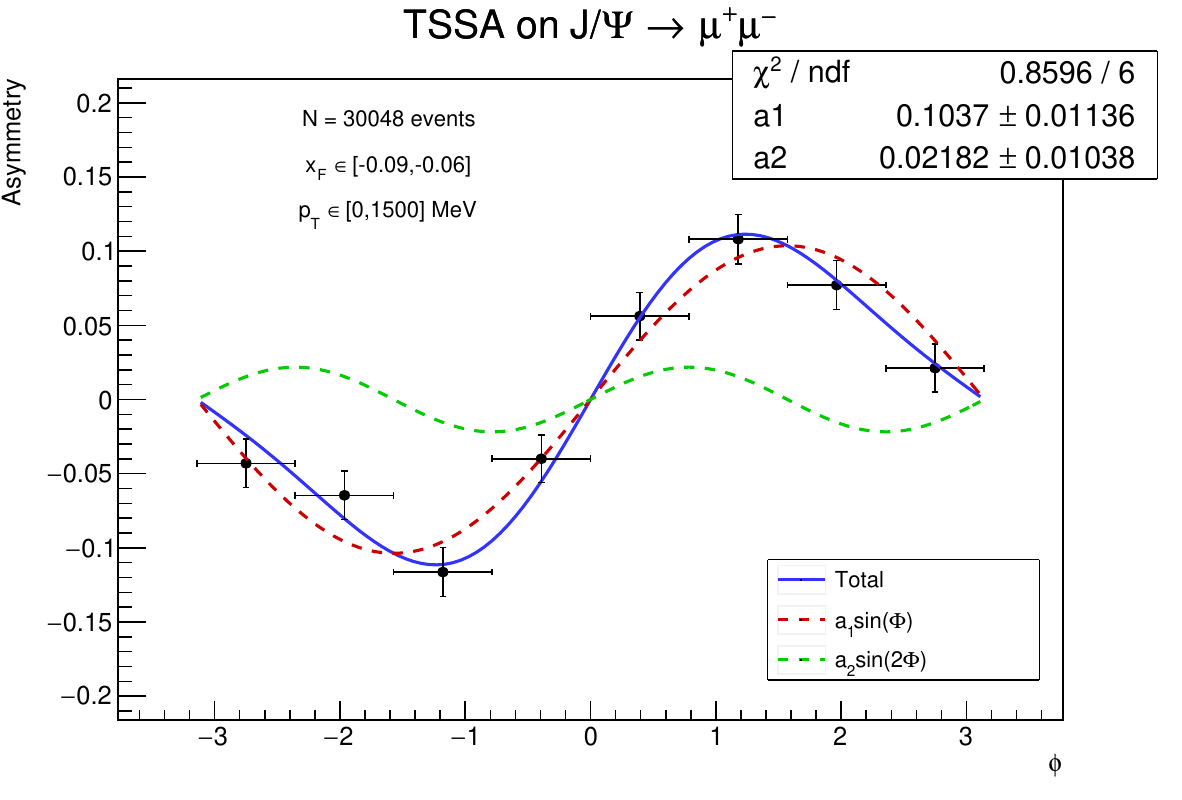}
\hfill
\includegraphics[width=.49\textwidth]{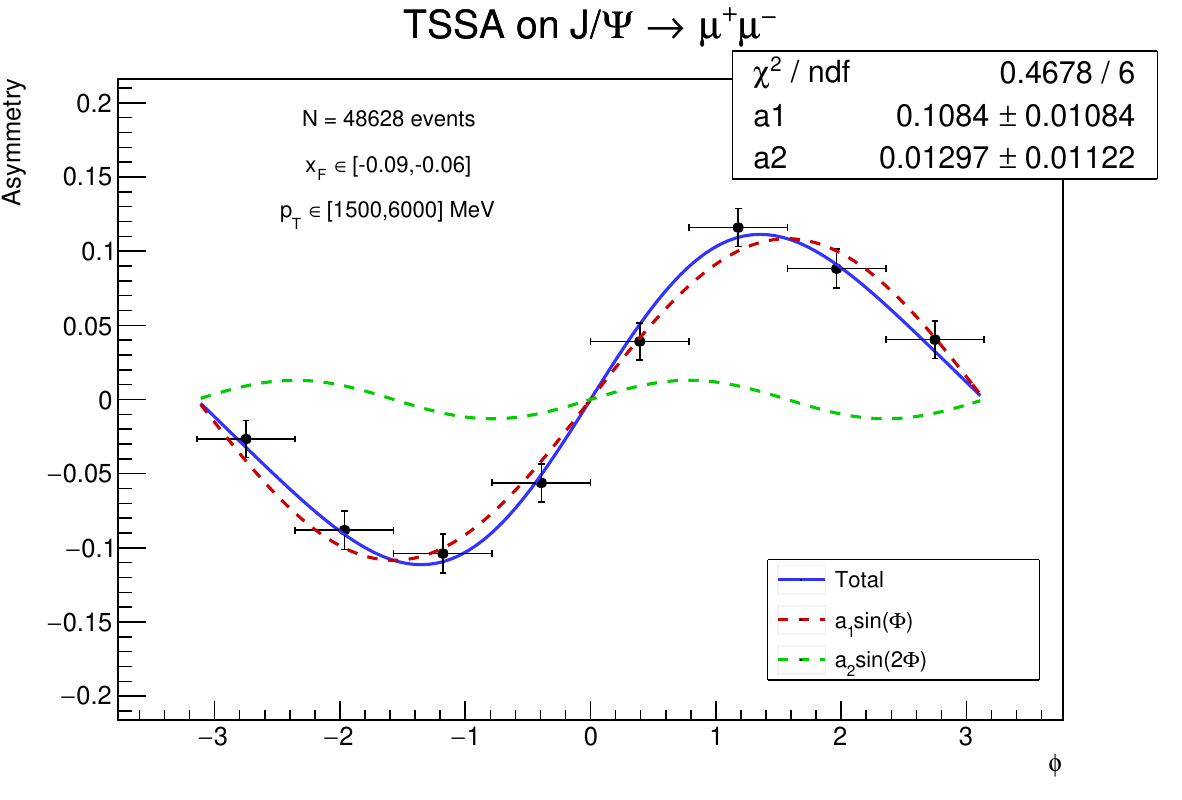}
\newline
\includegraphics[width=.49\textwidth]{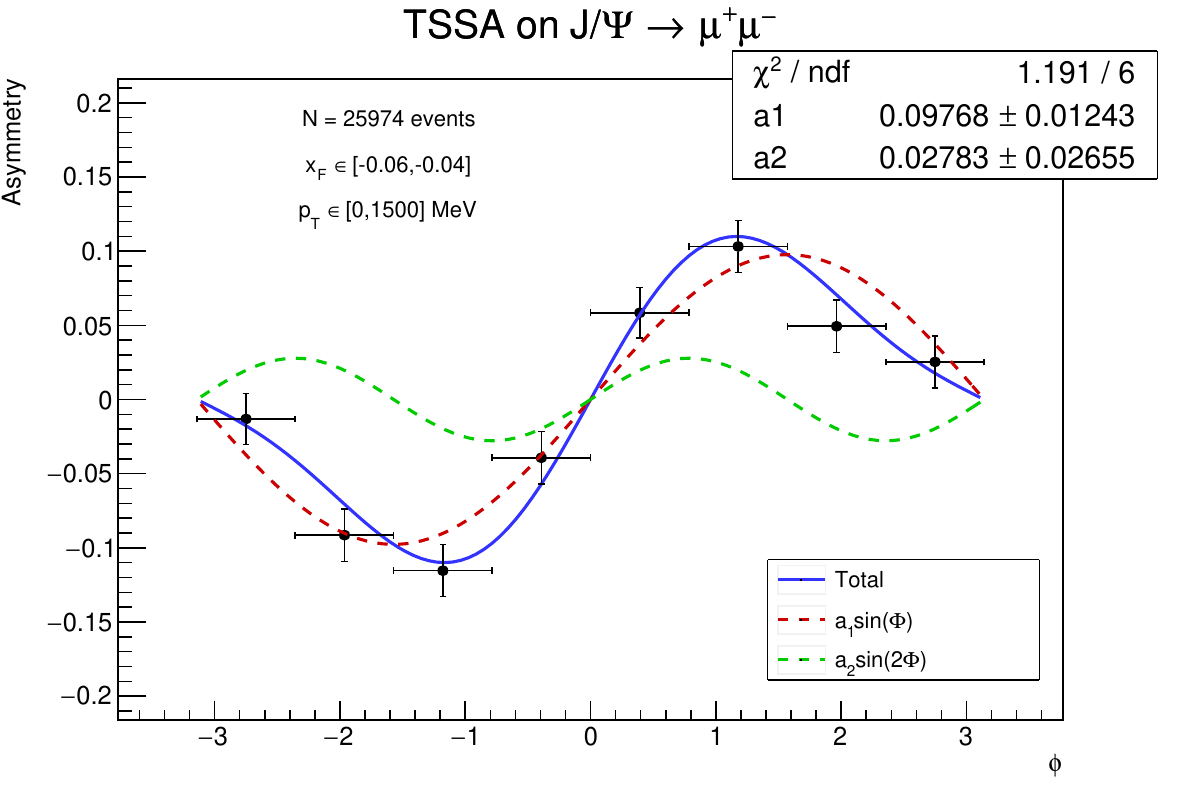}
\hfill
\includegraphics[width=.49\textwidth]{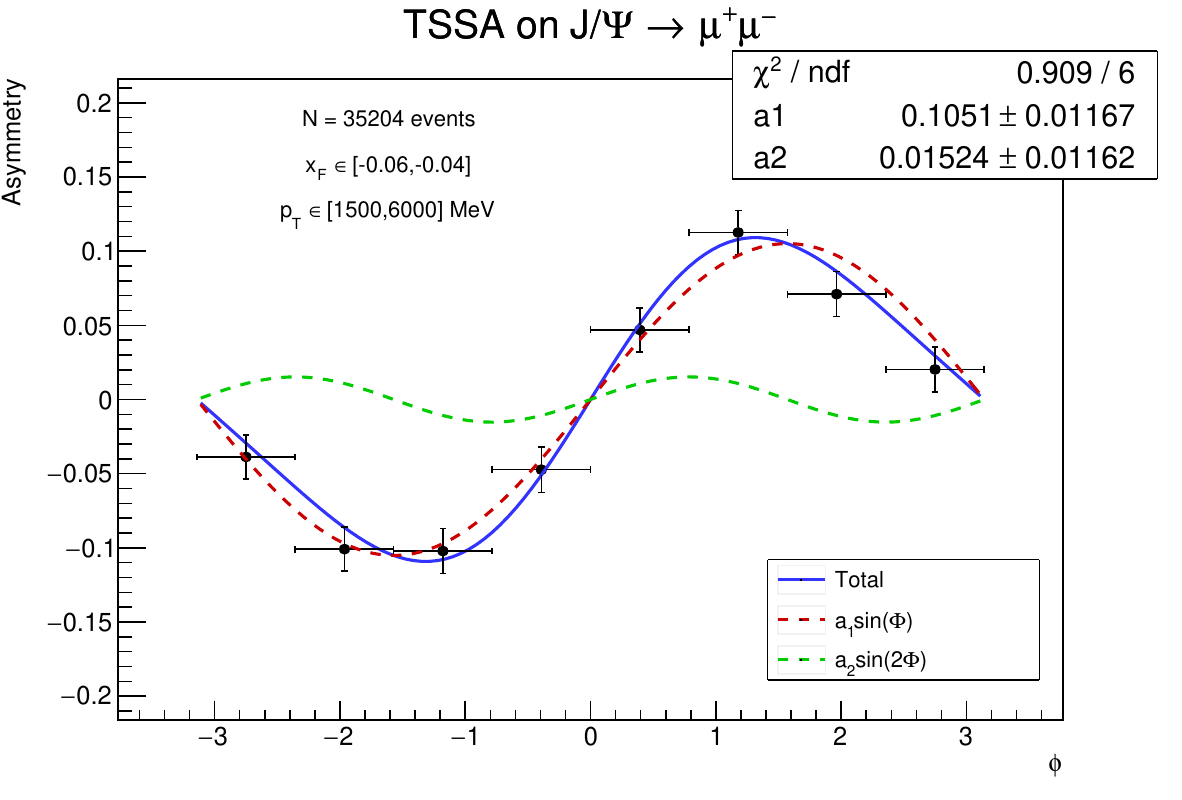}
\newline
\includegraphics[width=.49\textwidth]{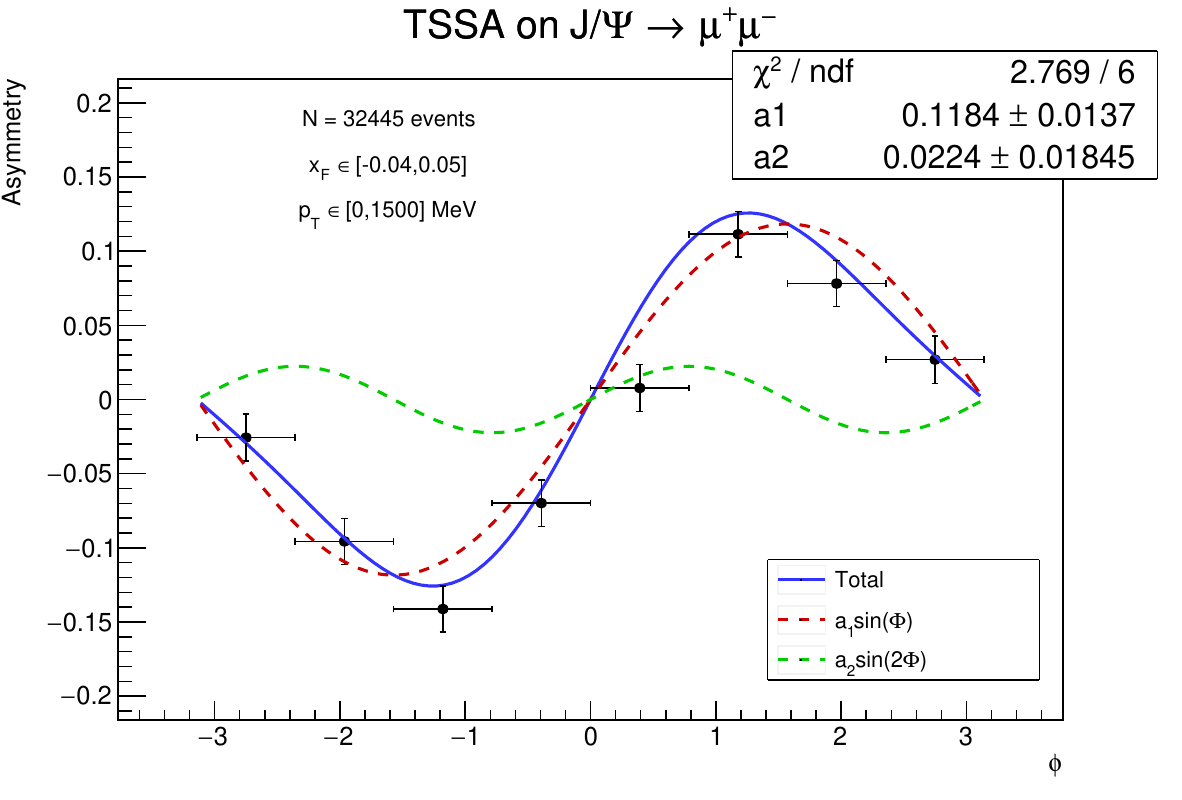}
\hfill
\includegraphics[width=.49\textwidth]{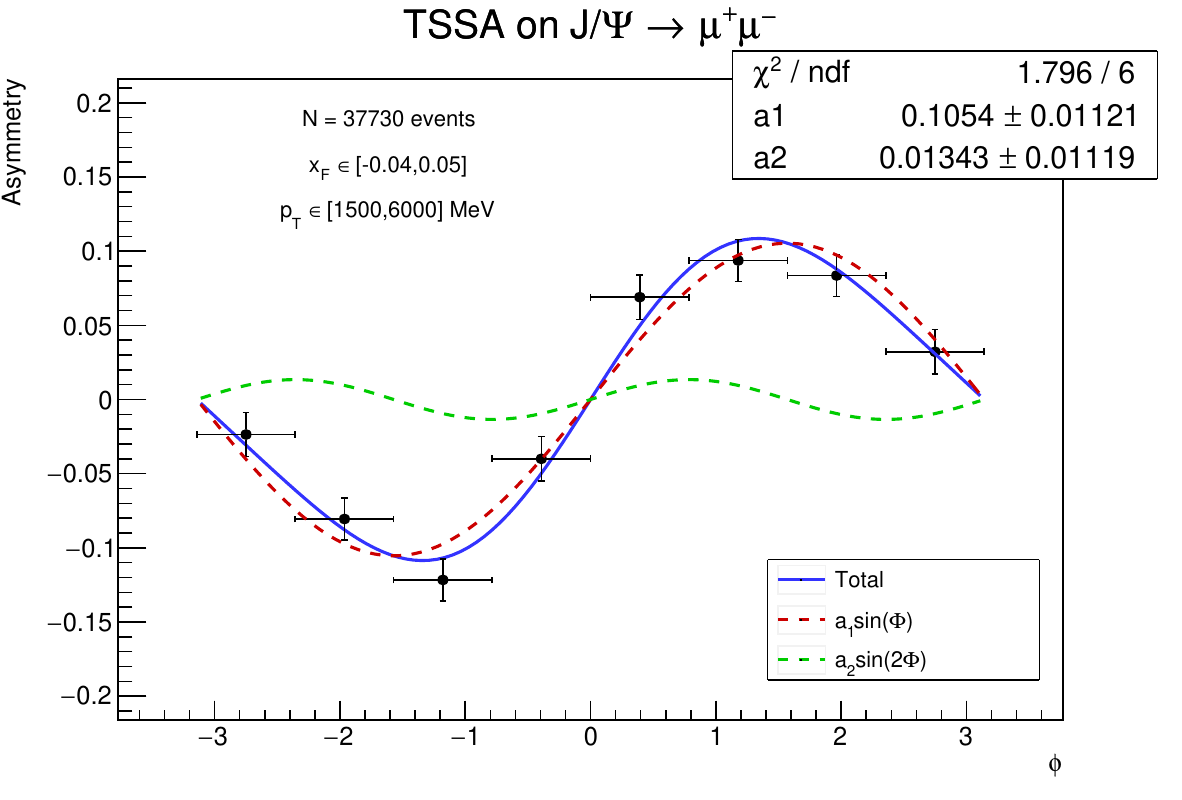}
\end{center}
\caption{Fits to azimuthal modulations in eight different $p_T - x_F$ bins.}
\label{fig:res_phi}
\end{figure}

The pseudo data are split into 2D $x_F-p_T$ bins, and further divided into $\phi$ bins, where the spin asymmetry is computed according to Eq.~\ref{eq:tssa}. The uncertainty is evaluated by propagating the statistical uncertainties on $N^{\uparrow}$ and $N^{\downarrow}$ and a $100\%$ polarization degree is assumed at this stage. 
For each $x-p_T$ bin, the $\phi$ modulation is fitted with the function:
\begin{equation}
    f = a_1 \sin \phi + a_2 \sin 2 \phi.
\end{equation}
The results are shown in Fig.~\ref{fig:res_phi}. The fitted amplitudes are compatible with the parameters used in the generated model (Eq.~\ref{eq:model}), i.e. no bias is observed.
Within the available statistics, which correspond to about two days of data-taking, a $1\%$ relative precision can be attained whereas the fit is not sensitive to a second harmonic with the chosen binning scheme. 

The impact of the uncertainty in the polarization degree on this measurement is evaluated by repeating the fits, incorporating the polarization degree uncertainty at each data point. Within the available statistics, the precision on the $a_1$ extraction is limited by the statistics if the error on the polarization degree is $5\%$. The uncertainties on the $a_1$ values are less than $10\%$ bigger with respect infinite precision on $P$. However, if the error on $P$ grows to $20\%$, the results show a systematic effect amounting to $30-40\%$ of the statistical error. At $50\%$, the statistical and systematic errors become comparable.

\section{Atomic recombination and cell coating}

\label{sec:coating}

The areal density of a polarized hydrogen jet target produced by an ABS is limited to about $<10^{12}$ atoms/cm$^2$. This limitation can be overcome using T-shaped tubular storage cells (SC), which have been employed at several storage rings \cite{Haeberli1, article-steffens, IUCF5:1994, HERMES, ANKE, BINP}.
The straight T-beam tube serves to compress the gas injected via a side tube, or via a capillary 
as in the case of the unpolarized SMOG2 target.
The incoming polarized atoms from the ABS beam collide with the inner wall of the cell a number of times (typically 100) before exiting the cell, boosting the target density by approximately two orders of magnitude with respect to a free particle jet.

As for the jet target, a magnetic holding field is required inside the storage cell to define the polarization direction and prevent depolarization due to beam-induced fields. When injecting
"pure" hyperfine substates ($m_F=+1$ or $-1$)\footnote{$m_F$ is the projection of the total angular momentum along the quantization axis (the direction of the holding field).} into the cell, a relatively small magnetic field of a few mT is sufficient. In this configuration, the atomic flux from the ABS is reduced to 50\% of its maximum value compared to using two substates. 
Alternatively, injecting hydrogen atoms in two substates with a defined nuclear spin $m_I$ but different electron spin projection $m_J= \pm 1/2$, maximizes the intensity. However, a magnetic field several times the critical field $B_c(H) = 50.7$~mT is then required to decouple electron and proton spins and achieve the highest polarization level. For deuterium, the situation is more favorable, with a critical field of only $B_c(D)=11.4$~mT. 

Maintaining nuclear polarization during frequent collisions of polarized hydrogen or deuterium atoms with the wall of the storage cell is crucial. Many studies on storage cell coatings have been performed in the last decades \cite{article-steffens, NIKEF, WiseIUCF, HermesPL, PRL1, PRL2, PRICE, HERMES2, PAX, ANKE}. 
At the 30 GeV HERA electron storage ring 
(HERA-e), a polarized H and D gas target has been operated from 1996 to 2002 by the HERMES collaboration. The 40 cm long, thin-walled storage cell, with Drifilm coating, was operated at about 100 K under conditions that led to the formation of a stable water-ice layer. Polarization measurements performed by using a Breit-Rabi polarimeter, consistently showed values exceeding 80\% of the maximum during long periods, with minimal recombination.

Unfortunately, Drifilm coating, a Fluorine compound, is not allowed at the LHC near beamline surfaces. Currently, the only relevant coating complying with CERN rules is amorphous carbon (a-C). This led to the idea to investigate the residual proton polarization in molecules after recombination of polarized H on an a-C surface.

\subsection{Measurements with a Amorphous Carbon Coating}

Coatings for storage cells were studied using a dedicated setup at FJZ J\"ulich
\footnote{In collaboration with the University of Cologne, the PNPI, Gatchina, Russia, and CERN.}.
The original task of this apparatus was to measure nuclear spin polarization in H$_2$, D$_2$, and HD molecules formed after the recombination of polarized atoms. 

\begin{figure}[h!]
  \begin{center}
       \includegraphics[width=34pc, angle=0]{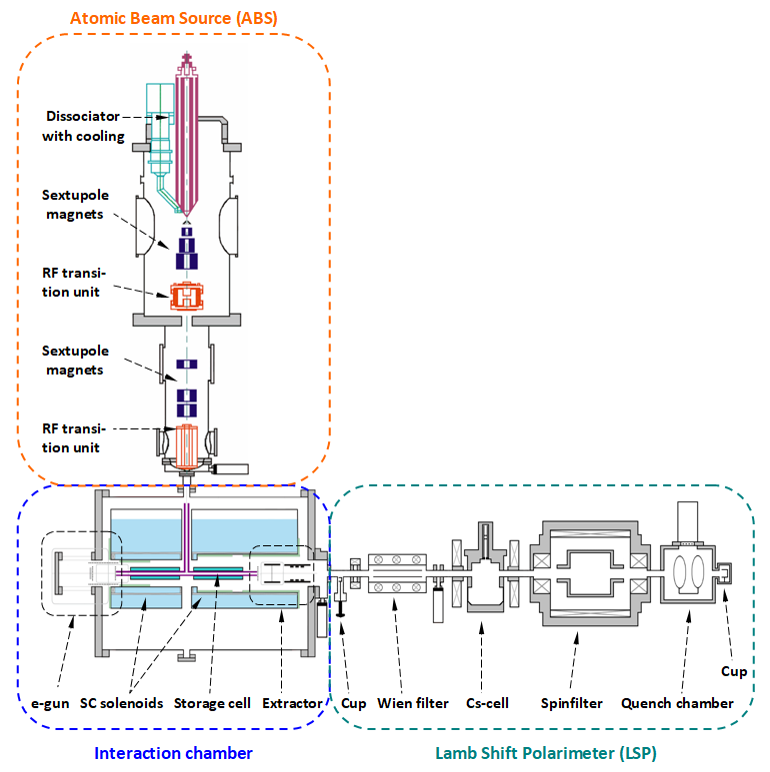}
    \caption{\label{SetUp} The design of the experimental setup at FZ Jülich: The ANKE-ABS is feeding a storage cell inside a superconducting solenoid (1~T) with polarized hydrogen/deuterium atoms. Inside the cell, the atoms can recombine on the wall surface into molecules. An electron beam from the left ionizes atoms and molecules, producing $p$ and H$_2^+$ ions, which are then accelerated into the Lamb-shift polarimeter where the nuclear polarization is measured.}
    \end{center}
\end{figure}

The apparatus (see fig.~\ref{SetUp}) consists of an ABS injecting atoms into a SC with the coating to be studied, and a reaction chamber with a storage cell inside a superconducting solenoid (1 T). Inside 
the cell, atoms can recombine at the walls into molecules. Atomic and molecular ions produced by electron bombardment within the SC are accelerated into a Lamb-shift polarimeter, where the nuclear polarization is measured. The T-shaped glass cell was a-C coated in collaboration with CERN. Further details of the procedure are described in the appendix.

\begin{figure}[h!]
  \begin{center}
       \includegraphics[width=30pc, angle=0]{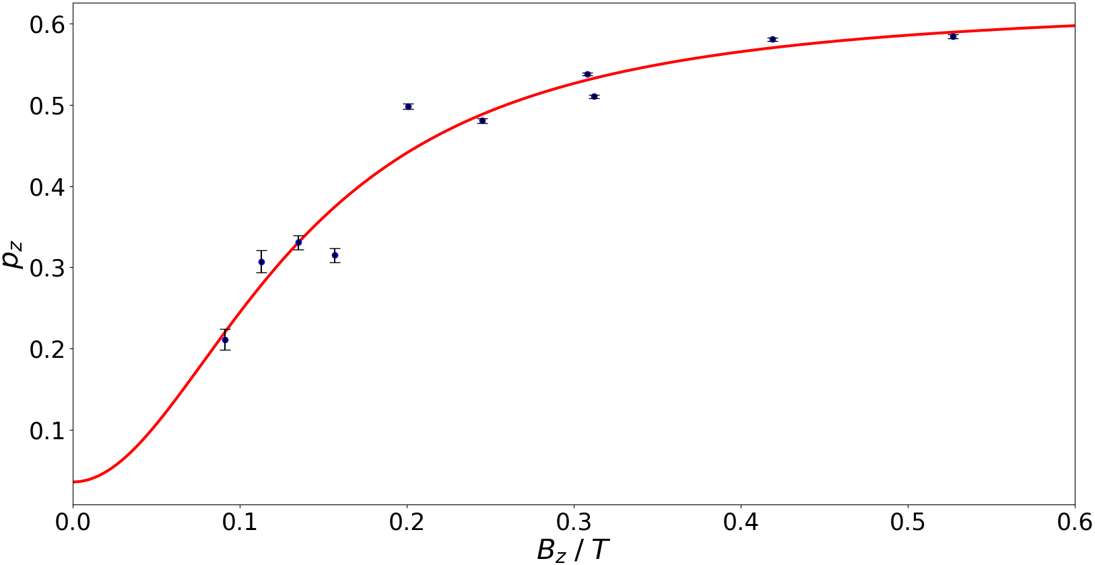}
    \caption{\label{Result} Measurement of the proton polarization $p_z$ as a function of the external magnetic field $B_z$ along the cell. Following Eq.~\ref{pol3} (in Appendix), the number of protons from atoms ($a$) and molecules ($b$) can be determined to calculate the recombination rate on the amorphous carbon coating. The overlaid fit function is $p_z(B) = p_{z0} \cdot e^{(-b/B)^2}$.}
    \end{center}
\end{figure}

The measurements with this cell revealed a high recombination rate of over $93\%$, along with a molecular polarization value in molecules of up to $P_m \sim 0.64$ (see Fig.\ref{Result}), which is the polarization of H$_2$ after $n$ wall collisions in an external magnetic field $B$. $P_m$ can be described by 
\begin{equation}
P_m(B,n)=P_{m0} \, e^{-n \left(\frac{B_{c,m}}{B} \right)^2}, \label{pol1}
\end{equation}
where $P_{m0}$ represents the molecular polarization induced by the recombination process, and $B_{c,m} = 5.4$~mT is the critical magnetic field of the hydrogen molecule. 
This indicates that about $74\% $ of the atomic polarization was preserved during the recombination process. Furthermore, during more than one week of measurement, no water film was built up on the cold cell surface ($\sim 100$~K) inside the superconducting solenoid, because this water film would suppress the recombination \cite{Tarek}. 

\subsection{A polarized molecular gas target}
Since a well-established surface coatings such as water ice, aluminum, or Teflon are not viable options for LHC, an amorphous carbon coating appears to be the only feasible solution so far. The recent investigations showed a nearly complete recombination of atoms into molecules on this surface, along with polarization values reaching up to $p_z\sim 0.64$. The statistical uncertainty in such experiments is proportional to $\sim 1/(\rho \cdot p_z^2)$, where $\rho$ is the gas target density. Since the target density of a molecular target, at the same ABS flux, increases with $\sqrt{2}$, the loss in polarization to about 50\% is partially compensated. Thus, this new type of molecular Polarized Gas Target will offer a comparable figure-of-merit as previously employed atomic storage-cell targets.

For nuclear-polarized molecules, where the proton spins are aligned in a parallel configuration, meaning the molecule is in an ortho-state with a symmetric spin wave function, the rotational magnetic moment $J$, must be odd. Thus, at room temperature, most of the molecules will have $J=1$, which then couples to the nuclear spins $I$. During wall interactions, the projections $m_J= +1$ or $-1$ of the rotational magnetic moment can flip, potentially altering the nuclear spins projection along the quantization axis, i.e.~the nuclear polarization is rapidly lost. To prevent this depolarization during wall collisions, a strong magnetic field is needed to decouple the nuclear spin from the rotational magnetic moment $J$. Depending on the number of wall collisions (Eq.~\ref{pol1}), which is mostly determined by the cell's diameter and length, a magnetic field between 0.2 and 0.3~T is needed along the cell. 

In conclusion, using a storage cell with an amorphous carbon coating represents a promising strategy to increase the effective target thickness of a polarized fixed target system for LHCb, although further 
measurements are needed to consolidate the results and better understand the conditions for producing a high density, high polarization molecular target.
This approach would represent a highly innovative target system. 
However, because standard method for polarization measurements used for atomic targets can not be applied, a novel target polarization measurement method must be established. We propose to perform that by calibrating a direct high energy proton-proton elastic scattering asymmetry, as described in Sec.~\ref{sec:abs_pol}.

\section{Development of an absolute polarimeter}

\label{sec:abs_pol}

As discussed in Sec.~\ref{sec:coating}, if the system will be equipped with a storage cell coated with amorphous carbon, 
a very high degree of atomic recombination is expected due to wall collisions. Given that the Breit-Rabi polarimeter cannot be used for measuring the nuclear polarization of molecules, a different approach is in order. 

\begin{figure}[h!]
    \centering
    \includegraphics[width=0.55\textwidth]{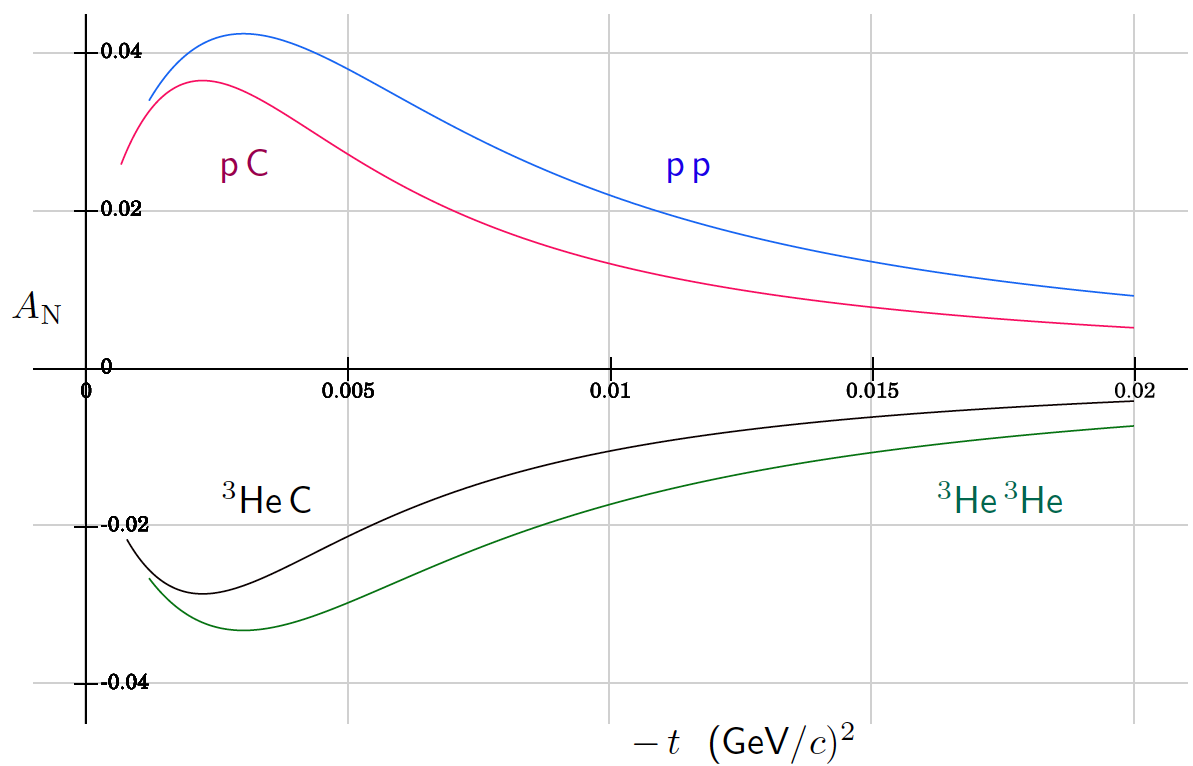} 
    \hspace{2mm}
       \includegraphics[width=0.35\textwidth]{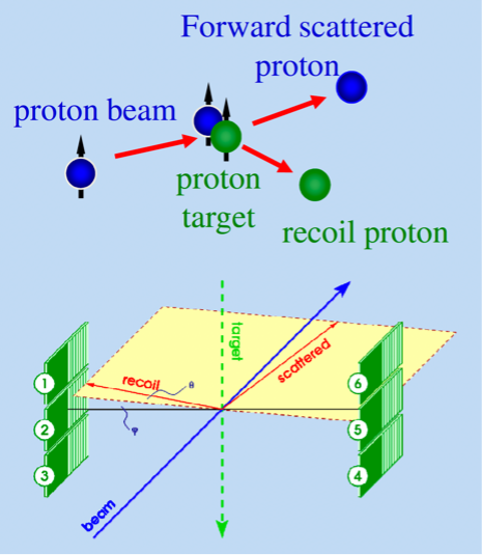} 
    \caption{Left: Theoretical estimations of the analyzing power $A_N$ as a function of the invariant momentum transfer $-t$ at LHC beam energies, for different collision systems. Right: Schematic representation of the recoil detector concept
     \cite{Buttimore_2001,Buttimoreslides2019,Buttimore2019}.}
   \label{fig:abspol}
\end{figure}

Absolute polarimetry, based on Coulomb-Nuclear Interference (CNI)~\cite{BETHE1958190}, is a viable option. This technique has been successfully adopted at RHIC, although with a beam energy more than one order of magnitude lower~\cite{ZELENSKI2005248,795735,Poblaguev:2019Eb} than the LHC top energy. Theoretical predictions of the expected analyzing power ($A_N$) for proton-proton elastic scattering at LHC energies driven by Coulomb-Nuclear Interference have been elaborated~\cite{Buttimore_2001,Buttimoreslides2019,Buttimore2019} (Fig.\ref{fig:abspol}, left), but need to be validated experimentally. 
This method requires the identification of elastically scattered protons produced in the collision of the 7 TeV proton beam on the polarized target gas. At the LHC, the emitted proton energy is expected to range between 1 and 6 MeV, with recoil angles close to 90 degrees (Fig.\ref{fig:abspol}, right).

It is worth noting that, to validate the estimates of analyzing power, the proton-proton left-right asymmetries measured by the detector system must be calibrated in parallel with the measurement of the target polarization obtained from the Breit-Rabi polarimeter using a polarized atomic beam generated by the ABS.

\subsection{The experimental setup}
The absolute polarimeter system to be developed (Fig.\ref{fig:abspol}, right) lays its foundations on a decade-long experience at the RHIC accelerator, whose setup consists of the following key components:
\begin{itemize}
    \item {\bf{Polarized Atomic Hydrogen Jet Target}} - provides the polarized atomic beam, which crosses the LHC proton beam perpendicularly.  
    \item {\bf{Scattering Chamber}} - the enclosed volume where the collisions between beam protons and target protons occur. The Recoil detectors are mounted directly on the flanges on both sides of the beamline.
    \item {\bf{Recoil Detector}} - utilize silicon detectors to identify recoil protons and measure their kinetic energy and time-of-flight (TOF) to select elastic scattering events, as described in Sec.\ref{rhic-recoil}. 
    \item {\bf{Breit-Rabi Polarimeter}} (BRP) - measures the nuclear polarization of the atomic hydrogen of the Atomic Jet.
    \item {\bf{Target Gas Analyzer}} (TGA) - measures the atomic-to-molecular fraction of the Atomic Jet.
\end{itemize}

\subsection{Recoil Detector}

\label{rhic-recoil}

The recoil detector has the function of detecting recoil protons from beam-target collisions and measuring their energy. The time-of-light (TOF) technique is then used to identify and select elastically scattered protons, whose left-right asymmetry will allow the determination of the analyzing power.      

In particular, the RHIC setup can be described as follows \cite{ZELENSKI2005248,795735,Poblaguev:2019Eb}:

\begin{itemize}
    \item The detectors select events within a specific TOF interval around the expected value for recoil protons of a given energy. The silicon detectors are positioned to the left and right with respect to the magnetic holding field of the jet target at a distance of approximately 70 cm. Each detector is 70.4 $\times$ 50 mm$^2$ in size with a 4.4 mm readout pitch, covering an azimuthal angle of 15$^\circ$ centered on the horizontal mid-plane.
    \item The silicon strip detectors are 400 $\mu m$ thick, such to fully absorb recoil protons with kinetic energies up to 7 MeV. The energy calibration of the silicon detectors can be performed using Am (5.486 MeV) and Gd (3.183 MeV) sources placed near the detectors. For the case of RHIC, a resolution of $\Delta T_R = 0.6$ MeV has been achieved for stopped protons. For protons with higher energies that punch through the detectors, the energy is corrected using the detector thickness and energy loss tables for silicon.
    \item The TOF of the recoil protons is measured relative to the bunch crossing time provided by the accelerator RF clock. The kinetic energy ($T_R$) of the recoil proton is related to its TOF by the non-relativistic relation:
    \begin{equation}
        T_R = 0.5 \cdot M_p \cdot \left(\frac{D}{TOF}\right)^2~,
    \end{equation}
    where $M_p$ is the mass of the proton, $D$ is the distance from the interaction point to the detector, and $TOF$ is the measured time-of-flight. At RHIC, the TOF resolution of approximately 3 ns has been achieved, accounting for the intrinsic time resolution of the detectors and the beam bunch length.
    \item The mass of the undetected forward-scattered system (missing mass) must be reconstructed to identify elastic scattering events. Events are selected based on the correlation between the recoil angle and kinetic energy, ensuring they match the expected values for elastic scattering.
    \item The selected event yield is sorted by kinetic energy bins and further categorized by spin states (target up-down) and detector side (left-right). This allows for the calculation of raw asymmetries and precise measurements of the analyzing power ($A_N$).
\end{itemize}

\subsection{Measurement of the Analyzing Power $A_N$}
    
    The selected event yield is binned according to recoil energy $T_R$. Within each $T_R$ bin, events are further categorized based on spin states (target up or down) and detector side relative to the target polarization axis (Left or Right). The raw asymmetries are then computed using the square-root formula, which effectively cancels out luminosity and acceptance effects \cite{Poblaguev_2020}.
 
  \begin{equation}
\epsilon_{target} = \frac{\sqrt{\mathrm{N}_{\uparrow }^{L}{\mathrm{N}_{\downarrow }^{R}}}-\sqrt{\mathrm{N}_{\uparrow }^{R}{\mathrm{N}_{\downarrow }^{L}}}}{\sqrt{\mathrm{N}_{\uparrow }^{L}{\mathrm{N}_{\downarrow }^{R}}}+\sqrt{\mathrm{N}_{\uparrow }^{R}{\mathrm{N}_{\downarrow }^{L}}}}~.
  \end{equation}

    The analyzing power $A_N$ can be derived from the measurement of the raw asymmetries using the formula:
    \begin{equation}
        A_N = -\frac{\epsilon_{target}}{P_T} \frac{1}{1-R_{BG}}~,
    \end{equation}
    where $P_T$ is the target polarization, as measured by the Breit-Rabi polarimeter, and $R_{BG}$ represents the background fraction for each recoil energy $T_R$ bin. The background contributions include: (a) particles from calibration sources, (b) beam scraping, and (c) beam scattering from the unpolarized residual target gas. At RHIC, the dominant component was (c), due to unfocused molecular hydrogen, which was accounted for as a dilution factor affecting the target polarization. Consequently, $R_{BG}$ was estimated to be between 0.02 and 0.03, based on contributions from components (a) and (b).

Based on the RHIC experience and depending on the desired level of accuracy, it may be beneficial to install a movable Target Gas Analyzer before the first sextupole magnet of the Breit-Rabi polarimeter. This would allow for periodic measurements of the molecular content during the beam time, reducing the aforementioned systematic error.
\section{The polarized jet target at the IR4}

\label{sec:ir4}

The amorphous carbon coating provides a unique opportunity to use a storage cell containing high-density polarized molecular hydrogen. This, however, requires to develop an absolute polarimeter which must be designed and calibrated directly at the LHC.
The most suitable location for this R\&D is the LHC Interaction Region 4 (IR4), where the absence of dense LHC equipment and the relatively low radiation levels (comparable to those at the main four experiment interaction points) make this area particularly well suited.

\subsection{Description}
\begin{figure}[h!]
\hspace{-1.4cm}
  \includegraphics [width=1.16\textwidth] {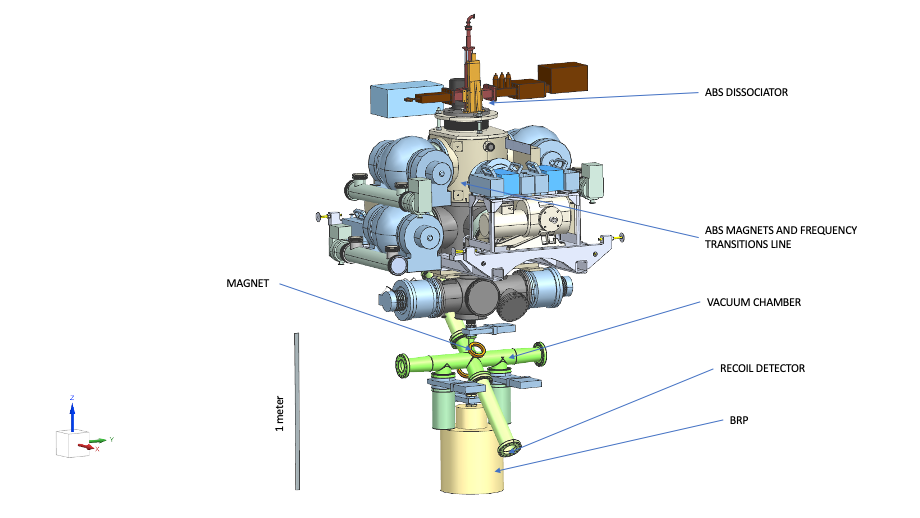} 
    \caption{\label{IP4-Setup} 
    CAD of the polarized gas target system, showing the main components and the dimensions of the apparatus.}
\end{figure}

\begin{figure}[h!]
\begin{center}
    \includegraphics [width=1.1\textwidth] {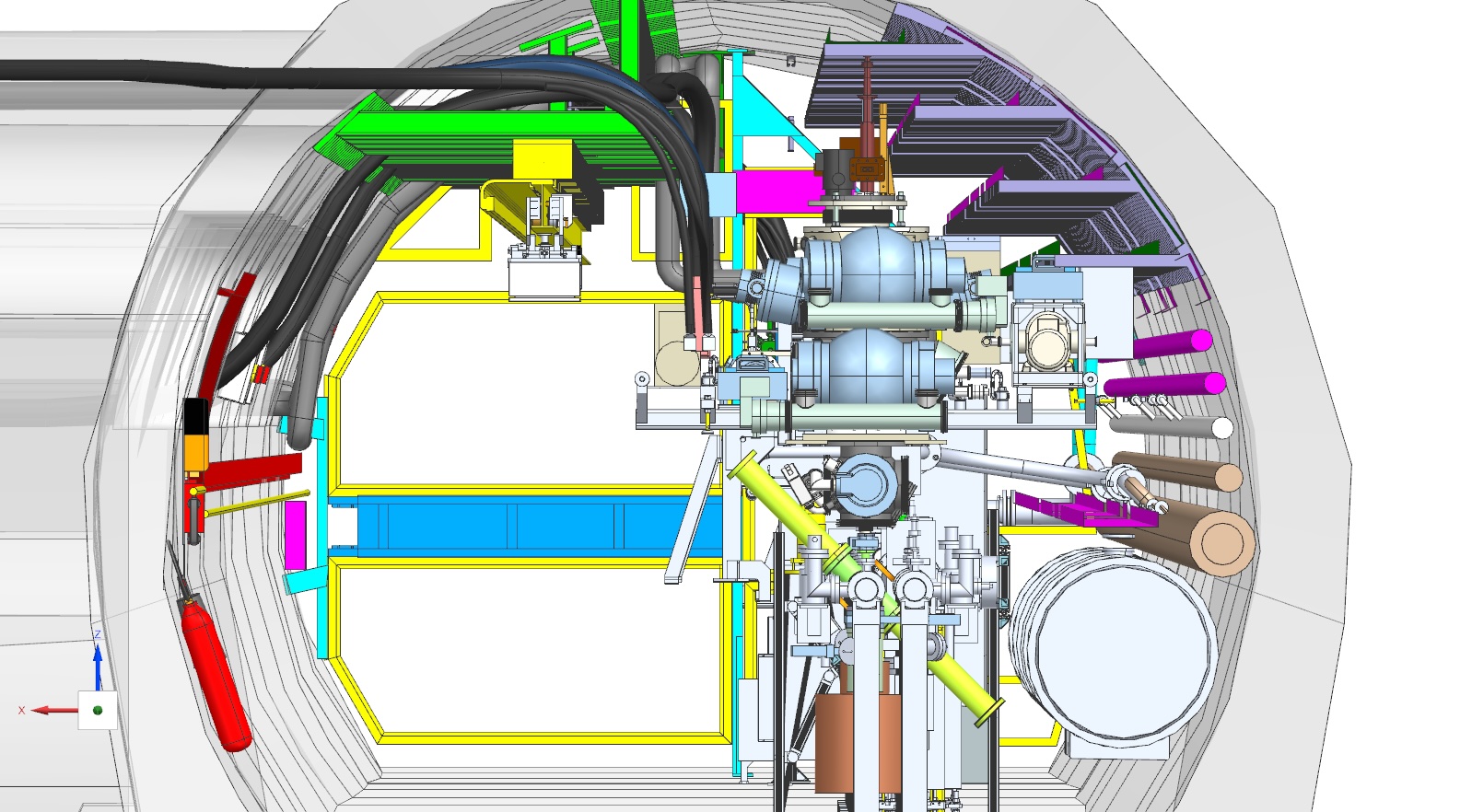} 
    \caption{\label{ir4tunnel} 
    CAD model of the target system implementation in the LHC tunnel at IR4.}
\end{center}
\end{figure}

A polarized jet target similar to the HJET-polarimeter\cite{Zelenski:2005mz} used at RHIC/BNL is proposed to be used at the IR4. It would consist of a polarized atomic beam source (ABS) to produce a jet of polarized hydrogen atoms, a target chamber with a holding field magnet, a Breit-Rabi polarimeter (BRP) to analyze the polarization of the hydrogen jet and an absolute polarimeter to measure the analyzing power of the elastic polarized p-p scattering\footnote{Similar arguments can be used in case of p-d scattering.}. The knowledge of the analyzing power in this energy range will be crucial for the later measurements at LHCb. The proposed setup is sketched in Figs.~\ref{IP4-Setup} (stand alone) and~\ref{ir4tunnel} (within the LHC tunnel).

The ABS of the HERMES/PAX-experiments\cite{Nass:2003mk} will be adapted for this purpose. It has been recently relocated to INFN-Ferrara, Italy in order to provide the necessary refurbishment and to make it compliant with a future installation in the LHC. The target chamber, along with a holding field magnet to define the spin axis of the hydrogen jet atoms, is highlighted in green in Fig.~\ref{Vacuumchamber}. However, to finalize the drawing, tracking calculations have to be conducted to optimize the beam path of the atoms through the ABS, target chamber, and BRP. This is currently ongoing by adapting the software described in Ref.\cite{Wise:2003aip}. 

\subsection{Polarized beam aperture}
The HERMES-ABS was optimized to inject a high intensity polarized atomic beam into the injection tube of a storage cell\cite{Nass:2003mk}. Therefore the size of the atomic beam had to be smaller than 10\mm, the diameter of that injection tube. For IR4 a high density polarized atomic beam is necessary. Therefore the ABS needs to be optimized to that using the aforementioned tracking programs. Fine tuning can be done using several parameters (hydrogen flux and nozzle temperature in the dissociator) to change the velocity and the velocity spread of the atomic beam within a certain range\cite{Nass:2008th}, which affect the trajectories through the sextupole components in the ABS. Using these techniques, the beam focus as well as the beam size can be adjusted. Diameters below 10\mm in a free jet have been achieved\cite{Zelenski:2005mz}. 

\subsection{The vacuum chamber}

\begin{figure}[h!]
    \includegraphics [width=0.9\textwidth] {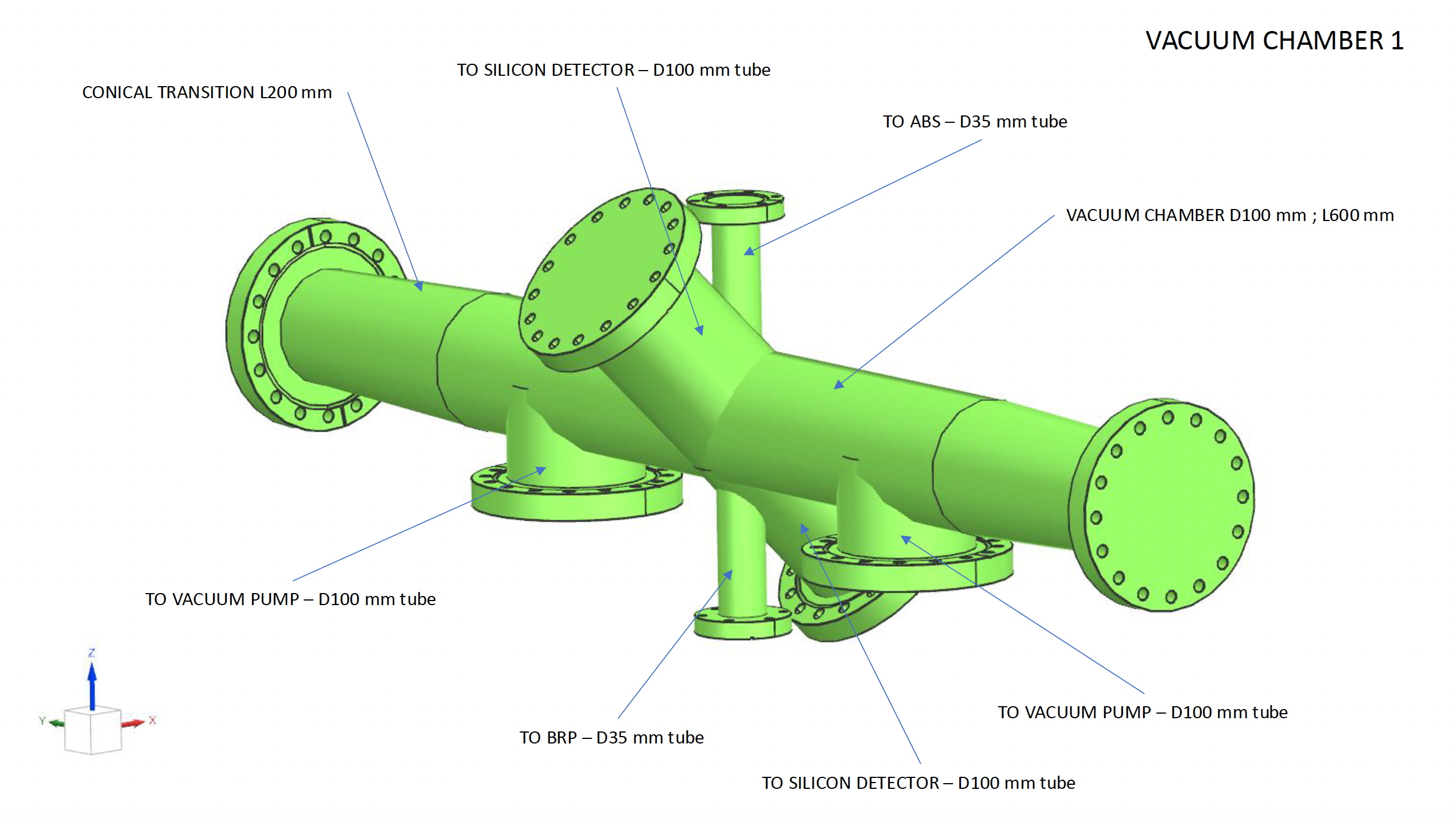} 
    \caption{\label{Vacuumchamber} 
    Detailed view of the vacuum chamber.}
\end{figure}

The vacuum chamber, made of 3-mm-thick stainless steel (AISI316L - low carbon), has been designed to closely resemble an extension of the beam pipe (see Fig.~\ref{Vacuumchamber}).
This helps maintaining the impedance at a normal level and makes simulations easier. Ongoing simulations \cite{benoit} will also determine whether an internal screen is recommended.
Apart from the tube connecting the ABS and the BRP, the two flanges for the upstream and downstream vacuum pumps, as well as the connection to the two cylinders (positioned at 45$^\circ$ to optimize the arm length), are visible  in Fig.~\ref{IP4-Setup}. These cylinders will contain the detectors for identifying the recoil protons produced in the CNI scattering process.

A holding field of up to approximately 300 mT will surround the vacuum chamber. The magnetic field can be produced by simple Helmholtz coils (yellow circles in Fig.~\ref{IP4-Setup}) in the case of a jet beam target. In its final configuration for LHCb, however, the presence of an extended storage cell requires the installation of a dipole magnet.

\subsection{The Beam Gas Vertex system}

The beam-gas vertex (BGV) detector \cite{PhysRevAccelBeams.22.042801} is an instrument designed to non-invasively measure the transverse beam size in the LHC by reconstructing tracks from beam-gas interactions. This system uses a gas target that creates local pressure bumps (nominally 1 $\times$ 10$^{-7}$ mbar) and detects tracks originating from the same beam-gas interaction vertex.
A BGV detector was installed at IR4 in 2016 as part of the R\&D for the High-Luminosity LHC project (see Fig.~\ref{fig:bgv}); however, the plan is to dismantle the system during LS3.
The LHCspin setup could not only take the place of the BGV, essentially substituting a simple gas target with a more complex one, but, if properly adapted for the purpose, also serve as a detector for beam size and emittance measurements.

\subsection{Physics opportunities}
\label{sec:output}

Due to the uniqueness of the collisions of a $7~\rm{TeV}$ proton beam against polarized hydrogen at rest, the possibility of physics measurements with a minimal experimental setup at IR4, ahead of the installation in LHCb, is under study. 

\begin{figure}
    \centering
    \includegraphics [width=0.9\textwidth] {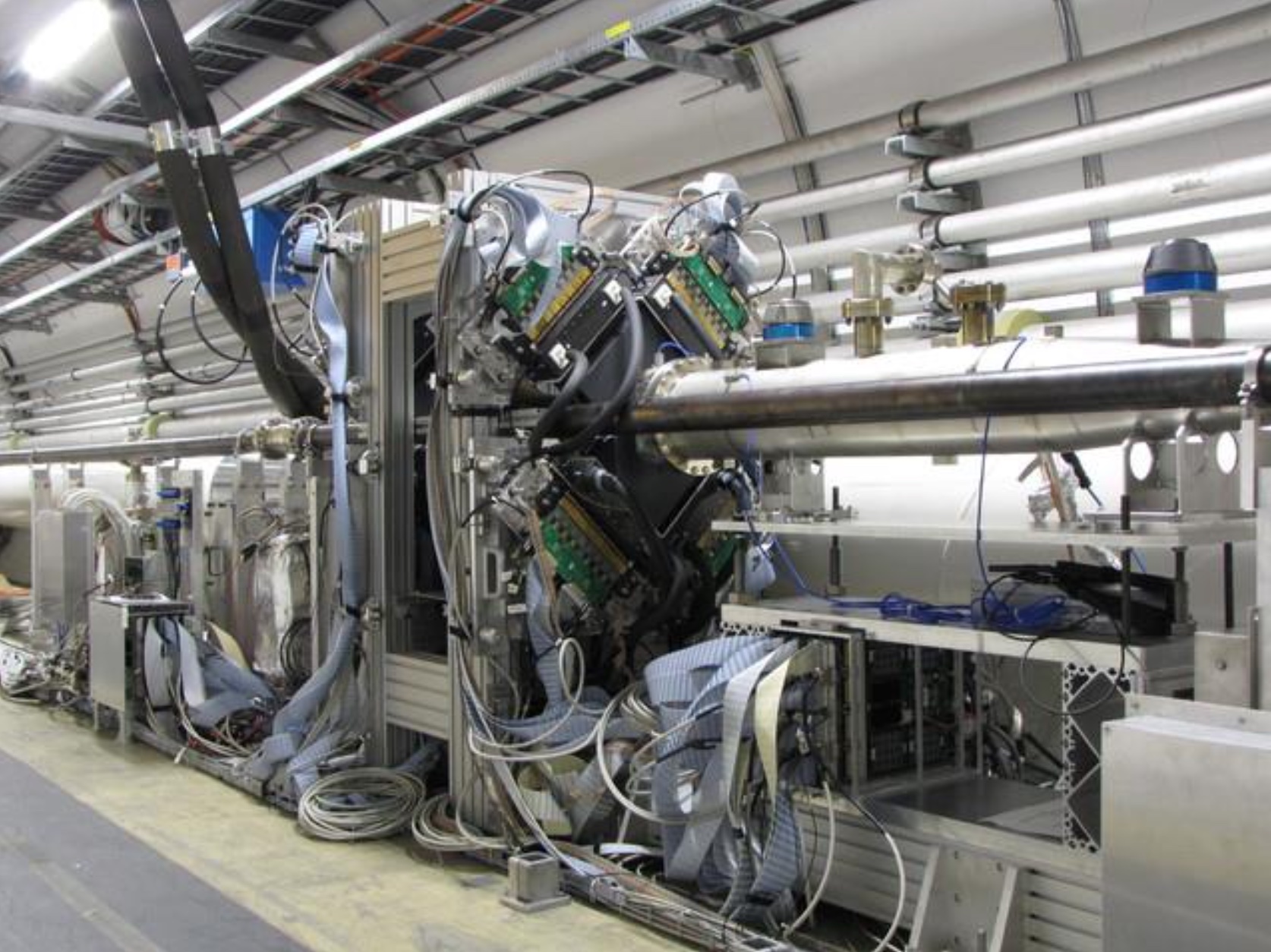} 
    \caption{The Beam Gas Vertex system at the IR4.}
    \label{fig:bgv}
\end{figure}

The scintillating fiber trackers, infrastructure and services of the existing BGV system~\cite{PhysRevAccelBeams.22.042801}, shown in Fig.~\ref{fig:bgv}, represent a solid starting point to further develop an experimental setup.

\begin{figure}
    \centering
    \includegraphics [width=0.99\textwidth] {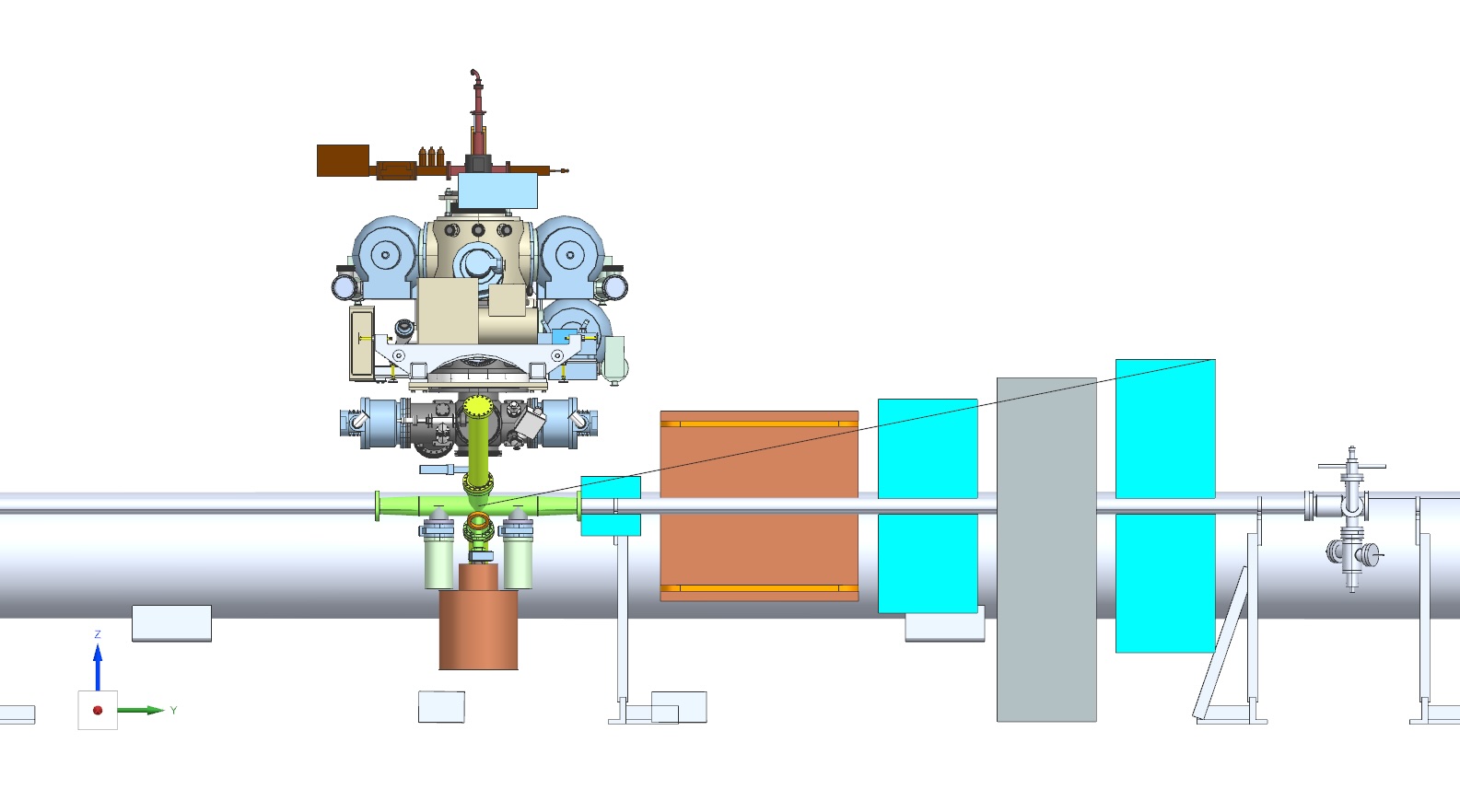} 
    \caption{An experimental setup at the IR4. From left to right: the polarized gas target, a first tracking detector, a dipole magnet, the second tracking detector, an absorber wall, and a muon detector.}
    \label{fig:ir4_detector}
\end{figure}

In Fig.~\ref{fig:ir4_detector}, a simple spectrometer concept is shown which includes a series of tracking stations, a dipole magnet, and muon stations placed behind an iron shielding.

Momentum resolution at the percent level can be attained for tracks with momentum up to $10$ GeV considering $10$ position measurements performed with $\mathcal{O}~(1~\rm{mm})$ resolution, a magnetic bending power of $\mathcal{O}~(1~\rm{T\cdot m})$ and a few meters of total lever arm.
Full simulation with \texttt{GEANT4}~\cite{GEANT4:2002zbu,Allison:2006ve,Allison:2016lfl} are under development to estimate more realistic performance and to determine the kinematic coverage whose main limitation is represented by the transverse size of the cavern.

\section{Working Group organization}
\label{sec:wg}

The R\&D work is organized into Working Groups that are
operating in parallel. The current division of the groups is as follows:

\begin{itemize}
    \item Polarized Atomic Beam Source;
    \item Breit-Rabi Polarimeter;
    \item Absolute Polarimeter;
    \item Mini Spectrometer at the IR4;
    \item Physics channels;
    \item Integration into LHCb.
\end{itemize}

\section{Conclusions}

\label{sec:Conclusions}

Over the last 20 years, the first measurements of azimuthal asymmetries in Semi-Inclusive Deep Inelastic Scattering have highlighted the critical role of the nucleon's non-collinear degrees of freedom in shaping its 3-D dynamical structure. This insight has been reinforced by theoretical advancement, including the development of new theory frameworks and formalisms, QCD-inspired models and phenomenological fits, and lattice QCD calculations. A wealth of significant data has been published across various kinematic domains, and more is expected in the near future. 

In this context, the unique opportunity to use LHC beams – both proton and heavy-ion – for polarized fixed-target measurements at a significantly reduced cost is particular compelling and unique. Moreover, LHCspin findings not only have the potential to anticipate parts of the upcoming US based Electron-Ion Collider program on polarized physics but also provide complementary measurements and approaches. 

A diverse research program that extends beyond the energy frontier is a fundamental component of the European Particle Physics Strategy. In its initial phase at the LHC Interaction Region 4, and later installed at LHCb, LHCspin will be able to deliver high-quality polarized data in the coming years, exploring a unique kinematical domain. These contributions will have a substantial impact on  advancing our understanding of the complexities of the strong interaction.

\section*{Acknowledgments}
\label{sec:acknow}

We sincerely appreciate the support of our colleagues from both the LHCb collaboration and the Physics Beyond Colliders group.

\section*{Appendix: Coating studies at FZJ}

\label{sec:Appendix}

\begin{figure}[h!]
  \begin{center}
       \includegraphics[width=30pc, angle=0]{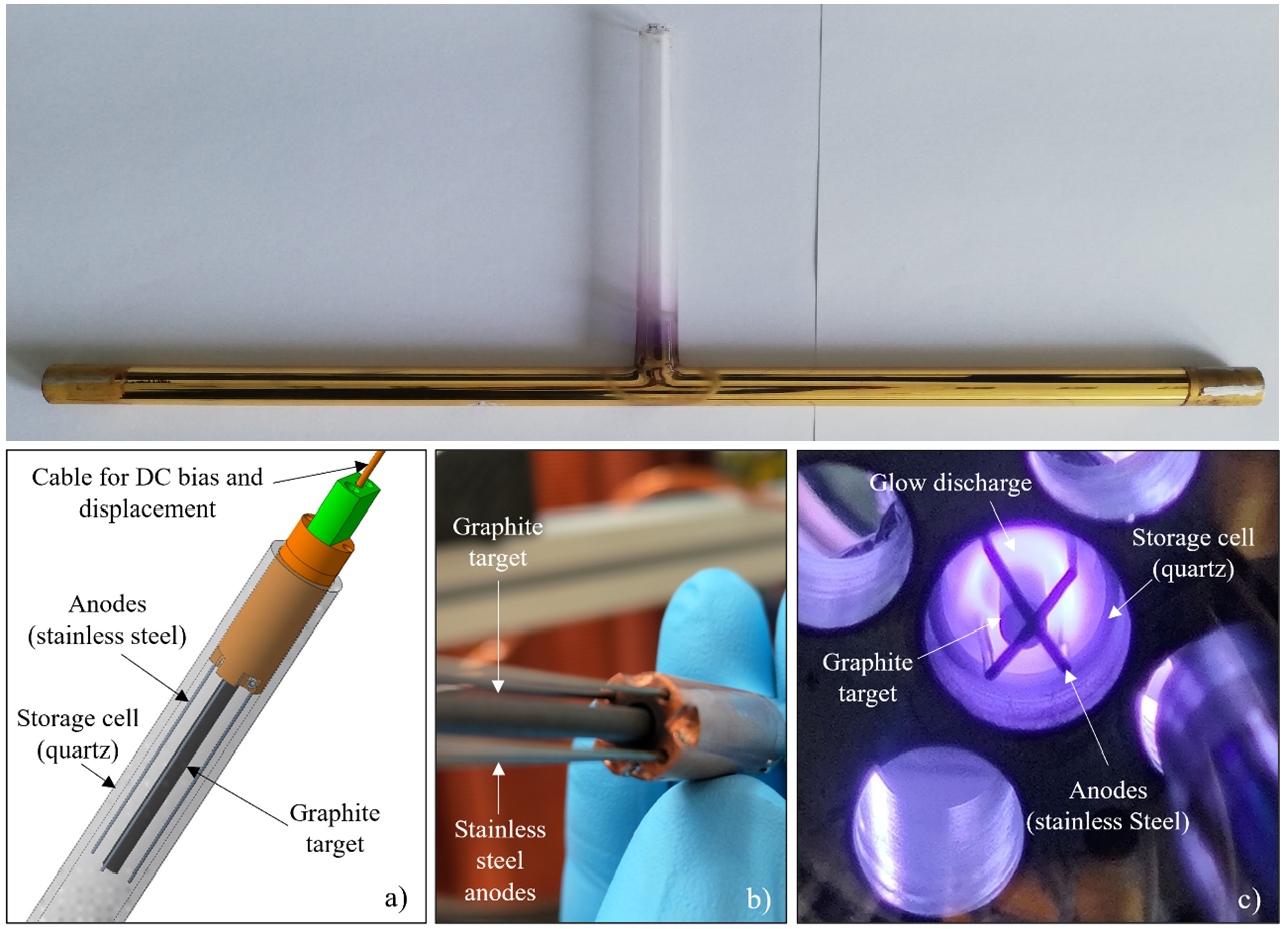}
    \caption{\label{Cell} A conventional storage cell and the production of a 200~nm amorphous Carbon coating by sputtering. }
    \end{center}
\end{figure}

As discussed in Section~\ref{sec:coating}, the apparatus used for the coating studies conducted at the Research Center FZJ in J\"ulich (Germany), consists of an ABS, which was also used for the polarized target of the ANKE@COSY experiment, a reaction chamber with a storage cell inside a superconducting solenoid and a Lamb-shift polarimeter to measure the nuclear spin polarization of the atoms and molecules (see Fig.~\ref{SetUp}).
When the polarized atoms enter the storage cell it is important that the magnetic field on their trajectory is always non-vanishing, otherwise the polarization might be lost. Changes of the field direction is not a problem if the precession of the electron spin is fast enough ($\sim$ GHz) that the spin will follow this field changes adiabatically due to the relatively slow velocity of the atoms ($\sim 1000$~m/s).
The nuclear spin is coupled to the electron spin and will just follow. Depending on the surface coating, the atoms might recombine into molecules or stay as atoms inside the cell. The cell itself is made from fused quartz with an inner diameter of 11~mm and a length of 400~mm. Inside, it is coated with a thin gold layer to induce a constant electric potential along the cell, that is isolated to the support system by the fused quartz itself. On top of the gold layer, 200~nm of amorphous Carbon was added\footnote{P. Costa Pinto, Technology Department, CERN.} by sputtering from a carbon electrode (see fig.~\ref{Cell}).

\begin{figure}[h!]
  \begin{center}
       \includegraphics[width=30pc, angle=0]{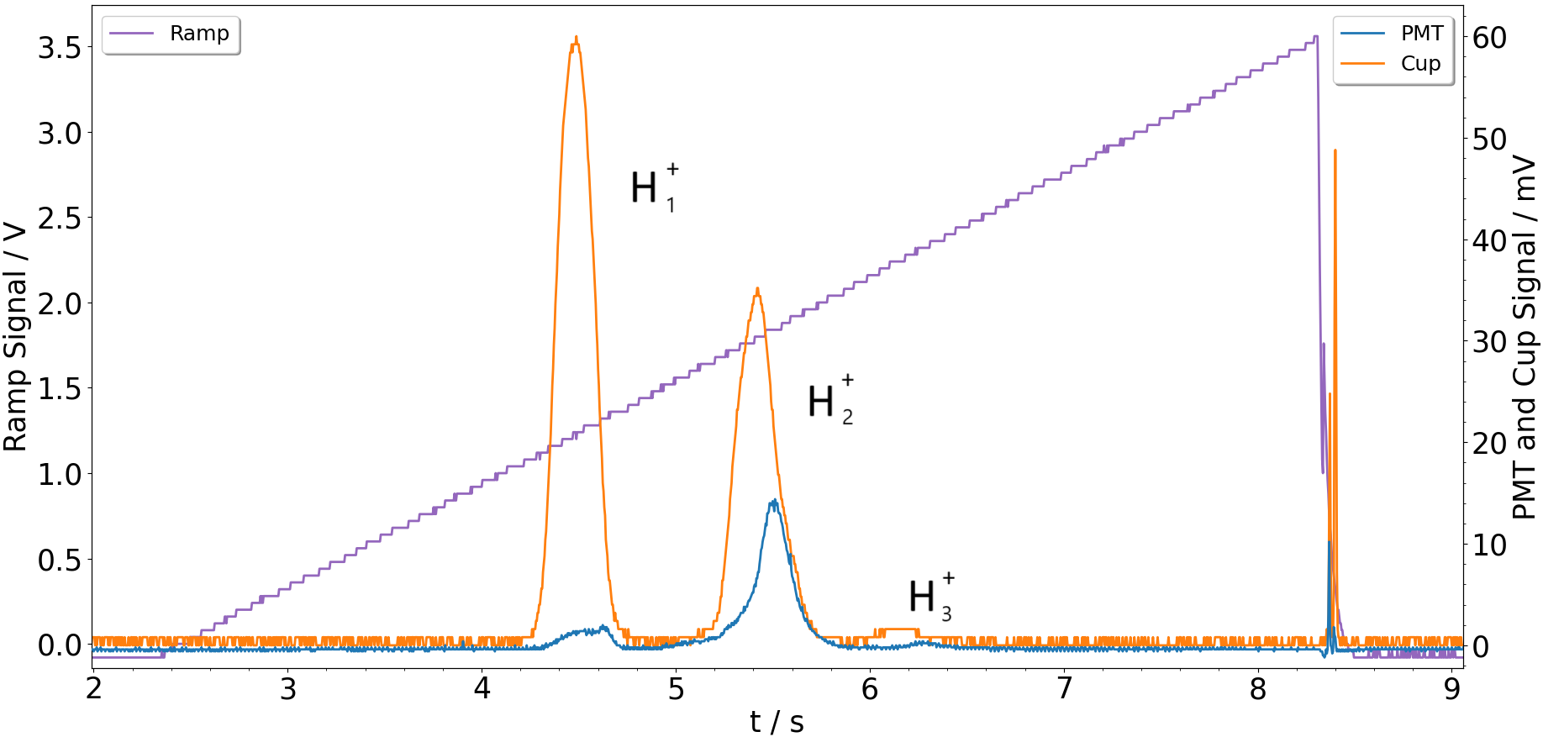}
    \caption{\label{WF} The mass spectra of the ion beam leaving the cell, produced by varying the electric field of a Wien filter as function of time. The intensity can be measured with a Faraday cup directly or by interactions with the vacuum chamber wall in front of a photomultiplier.}
    \end{center}
\end{figure}

An electron beam of about 150 eV is focused into the cell and ionizes atoms and molecules. The electric potential of the cell (0.3 - 2~keV) accelerates the produced $p$ and $H_2^+$ ions into the Lamb-shift polarimeter (LSP) where the nuclear polarization can be determined for both ions independently.\\
The first component of the LSP is a Wien filter that separates the ions due to their different velocities. Their intensities are either measured with a Faraday cup or with a photomultiplier at the end of the LSP and gives a first hint of the recombination rate in the storage cell (see fig.~\ref{WF}). 

\begin{figure}[h!]
  \begin{center}
       \includegraphics[width=32pc, angle=0]{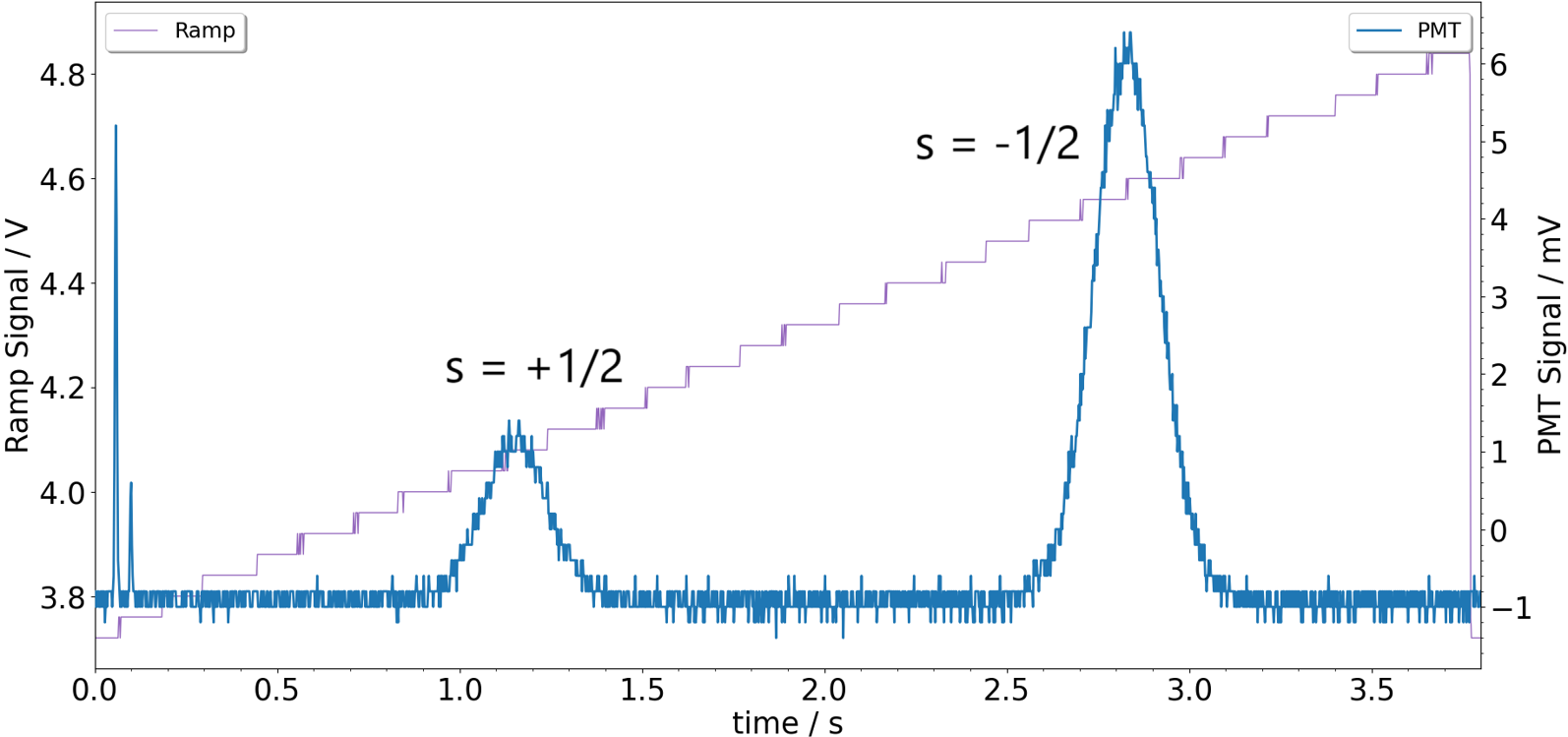}
    \caption{\label{Lyman} The intensity of the Lyman-$\alpha$ photons as function of time when the magnetic field in the spinfilter is ramped. In this example, protons of the primary $H_2^+$ ions with $s=+1/2$ will contribute to the first and with $s=-1/2$ to the second resonance. Thus, the polarization of the molecules in the storage cell is determined as $P_z=0.56 \pm 0.02$.  }
    \end{center}
\end{figure}

In the next step, the ions reach a Cesium cell where metastable hydrogen atoms in the $2S_{1/2}$ state are produced by charge exchange with Cs vapor. The efficiency for the $ p + Cs \rightarrow H_{2S} + Cs^+$ reaction is $10 - 15\%$, about 40 times higher compared to that of the production from $H_{2}^+$. As long as this charge exchange process occurs in a strong magnetic field the nuclear spin of the protons is conserved and, therefore, only dedicated hyperfine substates are populated. E.g., a proton with $m_I=+1/2$ can catch an electron with $m_J=+1/2$ or $-1/2$ to build the hyperfine substates $\alpha 1$ ($|m_J=+1/2, m_I=+1/2>$) or $\beta 3$ (($|m_J=-1/2, m_I=+1/2>$) only. Afterwards, the spinfilter will quench all metastable atoms into the ground state $1S_{1/2}$, but at resonant conditions of a longitudinal magnetic and a transversal electric field in combination with an induced radio-frequency of 1.60975~GHz, metastable atoms in a single hyperfine state can survive. These residual metastable atoms are than quenched into the ground state with a strong electric field (Stark effect) and the produced Lyman-$\alpha$ photons are registered with a photomultiplier as function of the magnetic field. By that (see Fig.~\ref{Lyman}) the amount of protons with spin up ($N_{+1/2}$) and down ($N_{-1/2}$) can be compared to measure the polarization $P_z=\frac{N_{+1/2} - N_{-1/2}}{N_{+1/2} + N_{-1/2}}$ of the primary beam directly (see Fig.\ref{Lyman}) \cite{LSP1, LSP2}. 

The superconducting solenoids encircling the storage cell generate a powerful magnetic field up to 1~T, ensuring that the electron and nuclear spin of hydrogen or deuterium atoms are decoupled and couple to the external magnetic field separately. Despite continuous collisions with the cell's surface coating, the polarization of these atoms persists as long as they remain within that magnetic field. \\
However, if the coating facilitates the recombination of atoms into their elementary molecules ($H_2$, $D_2$, or $HD$), these molecules may lose their nuclear polarization even in significantly stronger magnetic fields. This is attributed to wall collisions inducing random changes in the molecule's rotational angular momentum projection $m_J$ that is coupled with the nuclear spins of both protons $m_I$ within the molecule. This coupling is nearly 10 times less than the coupling to the electron spin, i.e. $B_c=5.4$~mT compared to $B_c=50.7$~mT, but it will lead to a transition of the nuclear spin state. Notably, the recombination process itself can result in polarization loss.\\
According to T. Wise et al.~\cite{wise1},
the molecular polarization $P_m$ of $H_2$ after $n$ wall collisions in an external magnetic field $B$, can be described by 
\begin{equation}
P_m(B,n)=P_{m0} \, e^{-n \left(\frac{B_{c,m}}{B} \right)^2}, \label{pol1_app}
\end{equation}
where $P_{m0}$ represents the molecular polarization induced by the recombination process and $B_{c,m} = 5.4$~mT denotes the critical magnetic field of the hydrogen molecule. The distribution of the number of wall collisions is characterized by the probability density function $W(n)=\alpha \, e^{-\alpha n}$, with $\alpha$ being a parameter determined by the surface material and storage cell geometry \cite{engels2015}. Thus, the mean value of such an exponential distribution is $\tilde{n}=\frac{ln(2)}{\alpha}$, which can be utilized to calculate the weighted average: 
\begin{equation}
        \bar{P}_m(B)=\frac{P_{m0}}{1+\frac{\tilde{n}}{ln(2)} \left( \frac{B_{c,m}}{B} \right)^2}. \label{pol1.2}
\end{equation}

 The electron beam produces protons by ionizing hydrogen atoms and can also interact with $H_2$ molecules, generating both $H_2^+$ ions and protons. Consequently, when the Wien filter is employed to filter out protons, the polarization of the corresponding ion beam $P_{p}(B)$ encompasses both, the polarization of the fraction $a$ of protons $P_a$ originating from the ionization of their respective atoms and the polarization of the fraction $b$ of those protons arising from the molecules:
\begin{equation}
    P_p(B)= a P_a+b \bar{P}_m = a P_a+ \frac{ b P_{m0}}{1+\frac{\tilde{n}}{ln(2)} \left( \frac{B_{c,m}}{B} \right)^2}. \label{pol3}
\end{equation}

Now, the nuclear polarization of the protons and the $H_2^+$ ions can be measured as a function of the applied magnetic field $B$ along the cell. A fit to the measurement for the $H_2^+$ ions delivers the original molecular polarization after the recombination $P_{m0}$ and the average number of wall collisions $\tilde{n}$ following equation~\ref{pol1.2}. The corresponding measurement for the protons (see Fig.~\ref{Result}) can be fitted with equation~\ref{pol3} and the known polarization of the hydrogen atoms from the ABS ($P_a\sim 0.9$), which does not depend on the magnetic field, to get the amount of protons from the atoms $a$ and the molecules $b$. Due to the know cross sections for the ionization reactions $H + e \rightarrow p + 2e$ and $H_2 + e \rightarrow p + H + 2e$ \cite{PRL1} the recombination rate can be directly determined. 

Accordingly, the measurements with the amorphous carbon coated cell delivered surprisingly a large recombination rate of $> 93\%$, but at the same time a polarization value up to $P_m \sim 0.64$. Thus, about $74\% $ of the atomic polarization was preserved in the recombination process. In addition, during more than one week of measurement no water film was built up on the cold cell surface ($\sim 100$~K) inside the superconducting solenoid, because this water film would suppress the recombination \cite{Tarek}. \\
Following the literature in astrophysics, the large recombination rate of the hydrogen atoms hints to photon-induced recombination on the amorphous carbon coated surface. The only possible photon source for this process is the dissociator of the ABS, which is a bright source for photons of the Balmer lines (transitions from $n=3 \rightarrow n=2$). Of course, the energy of these photons ($ < 3.4$~eV) is not enough to break the C-H bond (4.3~eV) on the amorphous carbon surface. But at the same time an even larger amount of Lyman photons (transitions from $n=2 \rightarrow n=1$) at energies $10.2 < E_{photon} < 13.6$~eV must be produced. These photons cannot be detected easily outside the dissociator, because they will be fully absorbed by the glass tube around the dissociator plasma, which must be cooled during operation. But they can leave the plasma via the nozzle and follow the trajectories of the hydrogen atoms through the ABS into the storage cell. To prove this thesis a small modification of the apparatus was made: Inside the storage cell an aluminum rod with a highly polished surface deflected the Lyman photons onto the beam line of the LSP. Another mirror inside the quenching chamber deflected them into the photomultiplier that is dedicated to these photon energies. These measurements allowed to estimate the number of photons reaching the storage cell: a number of about $10^{18}$/s was found, which is definitely enough to drive the recombination process \cite{Bilen}.

\clearpage
\setboolean{inbibliography}{true}
\bibliographystyle{LHCb}
\bibliography{main,gfs,LHCb-TDR,LHCb-DP,LHCb-PAPER,physics,coating}

\ifx\mcitethebibliography\mciteundefinedmacro
\PackageError{LHCb.bst}{mciteplus.sty has not been loaded}
{This bibstyle requires the use of the mciteplus package.}\fi
\providecommand{\href}[2]{#2}
\begin{mcitethebibliography}{100}
\mciteSetBstSublistMode{n}
\mciteSetBstMaxWidthForm{subitem}{\alph{mcitesubitemcount})}
\mciteSetBstSublistLabelBeginEnd{\mcitemaxwidthsubitemform\space}
{\relax}{\relax}

\bibitem{Kuhn_2009}
S.~E. Kuhn, J.-P. Chen, and E.~Leader, \ifthenelse{\boolean{articletitles}}{\emph{Spin structure of the nucleon—status and recent results}, }{}\href{https://doi.org/10.1016/j.ppnp.2009.02.001}{Progress in Particle and Nuclear Physics \textbf{63} (2009) 1–50}\relax
\mciteBstWouldAddEndPuncttrue
\mciteSetBstMidEndSepPunct{\mcitedefaultmidpunct}
{\mcitedefaultendpunct}{\mcitedefaultseppunct}\relax
\EndOfBibitem
\bibitem{Leader_2014}
E.~Leader and C.~Lorcé, \ifthenelse{\boolean{articletitles}}{\emph{The angular momentum controversy: What’s it all about and does it matter?}, }{}\href{https://doi.org/10.1016/j.physrep.2014.02.010}{Physics Reports \textbf{541} (2014) 163–248}\relax
\mciteBstWouldAddEndPuncttrue
\mciteSetBstMidEndSepPunct{\mcitedefaultmidpunct}
{\mcitedefaultendpunct}{\mcitedefaultseppunct}\relax
\EndOfBibitem
\bibitem{aidala2020probingnucleonsnucleihigh}
C.~A. Aidala {\em et~al.}, \ifthenelse{\boolean{articletitles}}{\emph{Probing nucleons and nuclei in high energy collisions}, }{} 2020\relax
\mciteBstWouldAddEndPuncttrue
\mciteSetBstMidEndSepPunct{\mcitedefaultmidpunct}
{\mcitedefaultendpunct}{\mcitedefaultseppunct}\relax
\EndOfBibitem
\bibitem{Anselmino:2020vlp}
M.~Anselmino, A.~Mukherjee, and A.~Vossen, \ifthenelse{\boolean{articletitles}}{\emph{{Transverse spin effects in hard semi-inclusive collisions}}, }{}\href{https://doi.org/10.1016/j.ppnp.2020.103806}{Prog.\ Part.\ Nucl.\ Phys.\  \textbf{114} (2020) 103806}, \href{http://arxiv.org/abs/2001.05415}{{\normalfont\ttfamily arXiv:2001.05415}}\relax
\mciteBstWouldAddEndPuncttrue
\mciteSetBstMidEndSepPunct{\mcitedefaultmidpunct}
{\mcitedefaultendpunct}{\mcitedefaultseppunct}\relax
\EndOfBibitem
\bibitem{Avakian:2019drf}
H.~Avakian, B.~Parsamyan, and A.~Prokudin, \ifthenelse{\boolean{articletitles}}{\emph{{Spin orbit correlations and the structure of the nucleon}}, }{}\href{https://doi.org/10.1393/ncr/i2019-10155-3}{Riv.\ Nuovo Cim.\  \textbf{42} (2019) 1}, \href{http://arxiv.org/abs/1909.13664}{{\normalfont\ttfamily arXiv:1909.13664}}\relax
\mciteBstWouldAddEndPuncttrue
\mciteSetBstMidEndSepPunct{\mcitedefaultmidpunct}
{\mcitedefaultendpunct}{\mcitedefaultseppunct}\relax
\EndOfBibitem
\bibitem{Abdul_Khalek_2022}
R.~Abdul~Khalek {\em et~al.}, \ifthenelse{\boolean{articletitles}}{\emph{Science requirements and detector concepts for the electron-ion collider}, }{}\href{https://doi.org/10.1016/j.nuclphysa.2022.122447}{Nuclear Physics A \textbf{1026} (2022) 122447}\relax
\mciteBstWouldAddEndPuncttrue
\mciteSetBstMidEndSepPunct{\mcitedefaultmidpunct}
{\mcitedefaultendpunct}{\mcitedefaultseppunct}\relax
\EndOfBibitem
\bibitem{osti_1177771}
J.~L. Abelleira~Fernandez, \ifthenelse{\boolean{articletitles}}{\emph{A large hadron electron collider at cern}, }{}\href{https://doi.org/10.1088/0954-3899/39/7/075001}{Journal of Physics.\ G, Nuclear and Particle Physics \textbf{39} (2015) }\relax
\mciteBstWouldAddEndPuncttrue
\mciteSetBstMidEndSepPunct{\mcitedefaultmidpunct}
{\mcitedefaultendpunct}{\mcitedefaultseppunct}\relax
\EndOfBibitem
\bibitem{accardi2023stronginteractionphysicsluminosity}
A.~Accardi {\em et~al.}, \ifthenelse{\boolean{articletitles}}{\emph{Strong interaction physics at the luminosity frontier with 22 gev electrons at jefferson lab}, }{} 2023\relax
\mciteBstWouldAddEndPuncttrue
\mciteSetBstMidEndSepPunct{\mcitedefaultmidpunct}
{\mcitedefaultendpunct}{\mcitedefaultseppunct}\relax
\EndOfBibitem
\bibitem{Collins_2011}
J.~Collins, {\em Foundations of Perturbative QCD}, Cambridge Monographs on Particle Physics, Nuclear Physics and Cosmology, Cambridge University Press, 2011\relax
\mciteBstWouldAddEndPuncttrue
\mciteSetBstMidEndSepPunct{\mcitedefaultmidpunct}
{\mcitedefaultendpunct}{\mcitedefaultseppunct}\relax
\EndOfBibitem
\bibitem{Boglione_2016}
M.~Boglione and A.~Prokudin, \ifthenelse{\boolean{articletitles}}{\emph{Phenomenology of transverse spin: Past, present and future}, }{}\href{https://doi.org/10.1140/epja/i2016-16154-6}{The European Physical Journal A \textbf{52} (2016) }\relax
\mciteBstWouldAddEndPuncttrue
\mciteSetBstMidEndSepPunct{\mcitedefaultmidpunct}
{\mcitedefaultendpunct}{\mcitedefaultseppunct}\relax
\EndOfBibitem
\bibitem{Bacchetta_2016}
A.~Bacchetta, \ifthenelse{\boolean{articletitles}}{\emph{Where do we stand with a 3-d picture of the proton?}, }{}\href{https://doi.org/10.1140/epja/i2016-16163-5}{The European Physical Journal A \textbf{52} (2016) }\relax
\mciteBstWouldAddEndPuncttrue
\mciteSetBstMidEndSepPunct{\mcitedefaultmidpunct}
{\mcitedefaultendpunct}{\mcitedefaultseppunct}\relax
\EndOfBibitem
\bibitem{Hadjidakis_2021}
C.~Hadjidakis {\em et~al.}, \ifthenelse{\boolean{articletitles}}{\emph{A fixed-target programme at the lhc: Physics case and projected performances for heavy-ion, hadron, spin and astroparticle studies}, }{}\href{https://doi.org/10.1016/j.physrep.2021.01.002}{Physics Reports \textbf{911} (2021) 1–83}\relax
\mciteBstWouldAddEndPuncttrue
\mciteSetBstMidEndSepPunct{\mcitedefaultmidpunct}
{\mcitedefaultendpunct}{\mcitedefaultseppunct}\relax
\EndOfBibitem
\bibitem{Diehl_2016}
M.~Diehl, \ifthenelse{\boolean{articletitles}}{\emph{Introduction to gpds and tmds}, }{}\href{https://doi.org/10.1140/epja/i2016-16149-3}{The European Physical Journal A \textbf{52} (2016) }\relax
\mciteBstWouldAddEndPuncttrue
\mciteSetBstMidEndSepPunct{\mcitedefaultmidpunct}
{\mcitedefaultendpunct}{\mcitedefaultseppunct}\relax
\EndOfBibitem
\bibitem{Angeles_Martinez_2015}
R.~Angeles-Martinez {\em et~al.}, \ifthenelse{\boolean{articletitles}}{\emph{Transverse momentum dependent (tmd) parton distribution functions: Status and prospects}, }{}\href{https://doi.org/10.5506/aphyspolb.46.2501}{Acta Physica Polonica B \textbf{46} (2015) 2501}\relax
\mciteBstWouldAddEndPuncttrue
\mciteSetBstMidEndSepPunct{\mcitedefaultmidpunct}
{\mcitedefaultendpunct}{\mcitedefaultseppunct}\relax
\EndOfBibitem
\bibitem{Sivers:1989cc}
D.~W. Sivers, \ifthenelse{\boolean{articletitles}}{\emph{{Single Spin Production Asymmetries from the Hard Scattering of Point-Like Constituents}}, }{}\href{https://doi.org/10.1103/PhysRevD.41.83}{Phys.\ Rev.\ D \textbf{41} (1990) 83}\relax
\mciteBstWouldAddEndPuncttrue
\mciteSetBstMidEndSepPunct{\mcitedefaultmidpunct}
{\mcitedefaultendpunct}{\mcitedefaultseppunct}\relax
\EndOfBibitem
\bibitem{ji2020protonspin30years}
X.~Ji, F.~Yuan, and Y.~Zhao, \ifthenelse{\boolean{articletitles}}{\emph{Proton spin after 30 years: what we know and what we don't?}, }{} 2020\relax
\mciteBstWouldAddEndPuncttrue
\mciteSetBstMidEndSepPunct{\mcitedefaultmidpunct}
{\mcitedefaultendpunct}{\mcitedefaultseppunct}\relax
\EndOfBibitem
\bibitem{LHCSpin_3}
M.~Santimaria {\em et~al.}, \ifthenelse{\boolean{articletitles}}{\emph{The {LHCspin} project}, }{}\href{https://doi.org/10.21468/scipostphysproc.8.050}{{SciPost} Physics Proceedings (2022) }\relax
\mciteBstWouldAddEndPuncttrue
\mciteSetBstMidEndSepPunct{\mcitedefaultmidpunct}
{\mcitedefaultendpunct}{\mcitedefaultseppunct}\relax
\EndOfBibitem
\bibitem{LHCSpin_4}
P.~Di~Nezza {\em et~al.}, \ifthenelse{\boolean{articletitles}}{\emph{{The LHCspin project}}, }{}\href{https://doi.org/10.22323/1.380.0347}{PoS \textbf{PANIC2021} (2022) 347}\relax
\mciteBstWouldAddEndPuncttrue
\mciteSetBstMidEndSepPunct{\mcitedefaultmidpunct}
{\mcitedefaultendpunct}{\mcitedefaultseppunct}\relax
\EndOfBibitem
\bibitem{LHCFT}
M.~A. Ross {\em et~al.}, \ifthenelse{\boolean{articletitles}}{\emph{Lhc fixed target experiments}, }{}\href{https://doi.org/https://doi.org/10.23731/CYRM-2020-004}{CERN Yellow Reports: Monographs \textbf{4} (2020) }\relax
\mciteBstWouldAddEndPuncttrue
\mciteSetBstMidEndSepPunct{\mcitedefaultmidpunct}
{\mcitedefaultendpunct}{\mcitedefaultseppunct}\relax
\EndOfBibitem
\bibitem{Nass:2003mk}
A.~Nass {\em et~al.}, \ifthenelse{\boolean{articletitles}}{\emph{{The HERMES polarized atomic beam source}}, }{}\href{https://doi.org/10.1016/S0168-9002(03)00986-0}{Nucl.\ Instrum.\ Meth.\ A \textbf{505} (2003) 633}\relax
\mciteBstWouldAddEndPuncttrue
\mciteSetBstMidEndSepPunct{\mcitedefaultmidpunct}
{\mcitedefaultendpunct}{\mcitedefaultseppunct}\relax
\EndOfBibitem
\bibitem{Garcia_2024}
O.~B. Garcia {\em et~al.}, \ifthenelse{\boolean{articletitles}}{\emph{High-density gas target at the lhcb experiment}, }{}\href{https://doi.org/10.1103/physrevaccelbeams.27.111001}{Physical Review Accelerators and Beams \textbf{27} (2024) }\relax
\mciteBstWouldAddEndPuncttrue
\mciteSetBstMidEndSepPunct{\mcitedefaultmidpunct}
{\mcitedefaultendpunct}{\mcitedefaultseppunct}\relax
\EndOfBibitem
\bibitem{Vlachos:2018lfe}
S.~Vlachos {\em et~al.}, \ifthenelse{\boolean{articletitles}}{\emph{{The LHC Beam Gas Vertex Detector - a Non-Invasive Profile Monitor for High Energy Machines}}, }{} in {\em {6th International Beam Instrumentation Conference}}, \href{https://doi.org/10.18429/JACoW-IBIC2017-WE3AB1}{ WE3AB1, 2018}\relax
\mciteBstWouldAddEndPuncttrue
\mciteSetBstMidEndSepPunct{\mcitedefaultmidpunct}
{\mcitedefaultendpunct}{\mcitedefaultseppunct}\relax
\EndOfBibitem
\bibitem{LHCbcollaboration:2903094}
L.~LHCb~collaboration, \ifthenelse{\boolean{articletitles}}{\emph{{LHCb Upgrade II Scoping Document}}, }{} tech. rep., CERN, Geneva, 2024.
\newblock doi:~\href{https://doi.org/10.17181/CERN.2RXP.HDK0}{10.17181/CERN.2RXP.HDK0}\relax
\mciteBstWouldAddEndPuncttrue
\mciteSetBstMidEndSepPunct{\mcitedefaultmidpunct}
{\mcitedefaultendpunct}{\mcitedefaultseppunct}\relax
\EndOfBibitem
\bibitem{DIEHL_2023}
S.~Diehl, \ifthenelse{\boolean{articletitles}}{\emph{Experimental exploration of the 3d nucleon structure}, }{}\href{https://doi.org/https://doi.org/10.1016/j.ppnp.2023.104069}{Progress in Particle and Nuclear Physics \textbf{133} (2023) 104069}\relax
\mciteBstWouldAddEndPuncttrue
\mciteSetBstMidEndSepPunct{\mcitedefaultmidpunct}
{\mcitedefaultendpunct}{\mcitedefaultseppunct}\relax
\EndOfBibitem
\bibitem{Pasquini}
{Pasquini, Barbara}, \ifthenelse{\boolean{articletitles}}{\emph{{RECENT ADVANCES ON SPIN AND 3D NUCLEON STRUCTURE}}, }{} 2021.
\newblock \url{https://indico.bnl.gov/event/9726/contributions/47605}\relax
\mciteBstWouldAddEndPuncttrue
\mciteSetBstMidEndSepPunct{\mcitedefaultmidpunct}
{\mcitedefaultendpunct}{\mcitedefaultseppunct}\relax
\EndOfBibitem
\bibitem{BHATTACHARYA_2017}
S.~Bhattacharya, A.~Metz, and J.~Zhou, \ifthenelse{\boolean{articletitles}}{\emph{Generalized tmds and the exclusive double drell–yan process}, }{}\href{https://doi.org/https://doi.org/10.1016/j.physletb.2017.05.081}{Physics Letters B \textbf{771} (2017) 396}\relax
\mciteBstWouldAddEndPuncttrue
\mciteSetBstMidEndSepPunct{\mcitedefaultmidpunct}
{\mcitedefaultendpunct}{\mcitedefaultseppunct}\relax
\EndOfBibitem
\bibitem{Boer:1997nt}
D.~Boer and P.~J. Mulders, \ifthenelse{\boolean{articletitles}}{\emph{{Time reversal odd distribution functions in leptoproduction}}, }{}\href{https://doi.org/10.1103/PhysRevD.57.5780}{Phys.\ Rev.\ D \textbf{57} (1998) 5780}, \href{http://arxiv.org/abs/hep-ph/9711485}{{\normalfont\ttfamily arXiv:hep-ph/9711485}}\relax
\mciteBstWouldAddEndPuncttrue
\mciteSetBstMidEndSepPunct{\mcitedefaultmidpunct}
{\mcitedefaultendpunct}{\mcitedefaultseppunct}\relax
\EndOfBibitem
\bibitem{K-M_96}
A.~M. Kotzinian and P.~J. Mulders, \ifthenelse{\boolean{articletitles}}{\emph{Longitudinal quark polarization in transversely polarized nucleons}, }{}\href{https://doi.org/10.1103/PhysRevD.54.1229}{Phys.\ Rev.\ D \textbf{54} (1996) 1229}\relax
\mciteBstWouldAddEndPuncttrue
\mciteSetBstMidEndSepPunct{\mcitedefaultmidpunct}
{\mcitedefaultendpunct}{\mcitedefaultseppunct}\relax
\EndOfBibitem
\bibitem{Bacchetta:2017gcc}
A.~Bacchetta {\em et~al.}, \ifthenelse{\boolean{articletitles}}{\emph{Extraction of partonic transverse momentum distributions from semi-inclusive deep-inelastic scattering, drell-yan and z-boson production}, }{}\href{https://doi.org/10.1007/JHEP06(2017)081}{Journal of High Energy Physics \textbf{6} (2017) 081}\relax
\mciteBstWouldAddEndPuncttrue
\mciteSetBstMidEndSepPunct{\mcitedefaultmidpunct}
{\mcitedefaultendpunct}{\mcitedefaultseppunct}\relax
\EndOfBibitem
\bibitem{saggiatore}
A.~Bacchetta and M.~Contalbrigo, \ifthenelse{\boolean{articletitles}}{\emph{The proton in 3d}, }{}Nuovo Saggiatore \textbf{28} (2012) 16\relax
\mciteBstWouldAddEndPuncttrue
\mciteSetBstMidEndSepPunct{\mcitedefaultmidpunct}
{\mcitedefaultendpunct}{\mcitedefaultseppunct}\relax
\EndOfBibitem
\bibitem{Arnold_2009}
S.~Arnold, A.~Metz, and M.~Schlegel, \ifthenelse{\boolean{articletitles}}{\emph{Dilepton production from polarized hadron hadron collisions}, }{}\href{https://doi.org/10.1103/physrevd.79.034005}{Physical Review D \textbf{79} (2009) }\relax
\mciteBstWouldAddEndPuncttrue
\mciteSetBstMidEndSepPunct{\mcitedefaultmidpunct}
{\mcitedefaultendpunct}{\mcitedefaultseppunct}\relax
\EndOfBibitem
\bibitem{Airapetian:2009ae}
HERMES collaboration, A.~Airapetian {\em et~al.}, \ifthenelse{\boolean{articletitles}}{\emph{{Observation of the Naive-T-odd Sivers Effect in Deep-Inelastic Scattering}}, }{}\href{https://doi.org/10.1103/PhysRevLett.103.152002}{Phys.\ Rev.\ Lett.\  \textbf{103} (2009) 152002}, \href{http://arxiv.org/abs/0906.3918}{{\normalfont\ttfamily arXiv:0906.3918}}\relax
\mciteBstWouldAddEndPuncttrue
\mciteSetBstMidEndSepPunct{\mcitedefaultmidpunct}
{\mcitedefaultendpunct}{\mcitedefaultseppunct}\relax
\EndOfBibitem
\bibitem{HERMES:2020ifk}
HERMES, A.~Airapetian {\em et~al.}, \ifthenelse{\boolean{articletitles}}{\emph{{Azimuthal single- and double-spin asymmetries in semi-inclusive deep-inelastic lepton scattering by transversely polarized protons}}, }{}\href{https://doi.org/10.1007/JHEP12(2020)010}{JHEP \textbf{12} (2020) 010}, \href{http://arxiv.org/abs/2007.07755}{{\normalfont\ttfamily arXiv:2007.07755}}\relax
\mciteBstWouldAddEndPuncttrue
\mciteSetBstMidEndSepPunct{\mcitedefaultmidpunct}
{\mcitedefaultendpunct}{\mcitedefaultseppunct}\relax
\EndOfBibitem
\bibitem{Alekseev:2008aa}
COMPASS collaboration, M.~Alekseev {\em et~al.}, \ifthenelse{\boolean{articletitles}}{\emph{{Collins and Sivers asymmetries for pions and kaons in muon-deuteron DIS}}, }{}\href{https://doi.org/10.1016/j.physletb.2009.01.060}{Phys.\ Lett.\  \textbf{B673} (2009) 127}, \href{http://arxiv.org/abs/0802.2160}{{\normalfont\ttfamily arXiv:0802.2160}}\relax
\mciteBstWouldAddEndPuncttrue
\mciteSetBstMidEndSepPunct{\mcitedefaultmidpunct}
{\mcitedefaultendpunct}{\mcitedefaultseppunct}\relax
\EndOfBibitem
\bibitem{Adolph:2012sp}
COMPASS collaboration, C.~Adolph {\em et~al.}, \ifthenelse{\boolean{articletitles}}{\emph{{II – Experimental investigation of transverse spin asymmetries in muon-proton SIDIS processes: Sivers asymmetries}}, }{}\href{https://doi.org/10.1016/j.physletb.2012.09.056}{Phys.\ Lett.\  \textbf{B717} (2012) 383}, \href{http://arxiv.org/abs/1205.5122}{{\normalfont\ttfamily arXiv:1205.5122}}\relax
\mciteBstWouldAddEndPuncttrue
\mciteSetBstMidEndSepPunct{\mcitedefaultmidpunct}
{\mcitedefaultendpunct}{\mcitedefaultseppunct}\relax
\EndOfBibitem
\bibitem{Adolph:2014fjw}
COMPASS collaboration, C.~Adolph {\em et~al.}, \ifthenelse{\boolean{articletitles}}{\emph{{A high-statistics measurement of transverse spin effects in dihadron production from muon–proton semi-inclusive deep-inelastic scattering}}, }{}\href{https://doi.org/10.1016/j.physletb.2014.06.080}{Phys.\ Lett.\  \textbf{B736} (2014) 124}, \href{http://arxiv.org/abs/1401.7873}{{\normalfont\ttfamily arXiv:1401.7873}}\relax
\mciteBstWouldAddEndPuncttrue
\mciteSetBstMidEndSepPunct{\mcitedefaultmidpunct}
{\mcitedefaultendpunct}{\mcitedefaultseppunct}\relax
\EndOfBibitem
\bibitem{Adolph:2014zba}
COMPASS collaboration, C.~Adolph {\em et~al.}, \ifthenelse{\boolean{articletitles}}{\emph{{Collins and Sivers asymmetries in muonproduction of pions and kaons off transversely polarised protons}}, }{}\href{https://doi.org/10.1016/j.physletb.2015.03.056}{Phys.\ Lett.\  \textbf{B744} (2015) 250}, \href{http://arxiv.org/abs/1408.4405}{{\normalfont\ttfamily arXiv:1408.4405}}\relax
\mciteBstWouldAddEndPuncttrue
\mciteSetBstMidEndSepPunct{\mcitedefaultmidpunct}
{\mcitedefaultendpunct}{\mcitedefaultseppunct}\relax
\EndOfBibitem
\bibitem{Adolph:2016dvl}
COMPASS collaboration, C.~Adolph {\em et~al.}, \ifthenelse{\boolean{articletitles}}{\emph{{Sivers asymmetry extracted in SIDIS at the hard scale of the Drell-Yan process at COMPASS}}, }{}\href{https://doi.org/10.1016/j.physletb.2017.04.042}{Phys.\ Lett.\  \textbf{B770} (2017) 138}, \href{http://arxiv.org/abs/1609.07374}{{\normalfont\ttfamily arXiv:1609.07374}}\relax
\mciteBstWouldAddEndPuncttrue
\mciteSetBstMidEndSepPunct{\mcitedefaultmidpunct}
{\mcitedefaultendpunct}{\mcitedefaultseppunct}\relax
\EndOfBibitem
\bibitem{Parsamyan:2013ug}
COMPASS collaboration, B.~Parsamyan, \ifthenelse{\boolean{articletitles}}{\emph{{Six ``beyond Collins and Sivers'' transverse spin asymmetries at COMPASS}}, }{}\href{https://doi.org/10.1134/S106377961401078X}{Phys.\ Part.\ Nucl.\  \textbf{45} (2014) 158}, \href{http://arxiv.org/abs/1301.6615}{{\normalfont\ttfamily arXiv:1301.6615}}\relax
\mciteBstWouldAddEndPuncttrue
\mciteSetBstMidEndSepPunct{\mcitedefaultmidpunct}
{\mcitedefaultendpunct}{\mcitedefaultseppunct}\relax
\EndOfBibitem
\bibitem{Courtoy:2015haa}
A.~Courtoy, S.~Bae\ss{}ler, M.~Gonz\'alez-Alonso, and S.~Liuti, \ifthenelse{\boolean{articletitles}}{\emph{Beyond-standard-model tensor interaction and hadron phenomenology}, }{}\href{https://doi.org/10.1103/PhysRevLett.115.162001}{Phys.\ Rev.\ Lett.\  \textbf{115} (2015) 162001}\relax
\mciteBstWouldAddEndPuncttrue
\mciteSetBstMidEndSepPunct{\mcitedefaultmidpunct}
{\mcitedefaultendpunct}{\mcitedefaultseppunct}\relax
\EndOfBibitem
\bibitem{Collins:2002kn}
J.~C. Collins, \ifthenelse{\boolean{articletitles}}{\emph{Leading-twist single-transverse-spin asymmetries: Drell–yan and deep-inelastic scattering}, }{}\href{https://doi.org/https://doi.org/10.1016/S0370-2693(02)01819-1}{Physics Letters B \textbf{536} (2002) 43}\relax
\mciteBstWouldAddEndPuncttrue
\mciteSetBstMidEndSepPunct{\mcitedefaultmidpunct}
{\mcitedefaultendpunct}{\mcitedefaultseppunct}\relax
\EndOfBibitem
\bibitem{COMPASS:2017jbv}
COMPASS, M.~Aghasyan {\em et~al.}, \ifthenelse{\boolean{articletitles}}{\emph{{First measurement of transverse-spin-dependent azimuthal asymmetries in the Drell-Yan process}}, }{}\href{https://doi.org/10.1103/PhysRevLett.119.112002}{Phys.\ Rev.\ Lett.\  \textbf{119} (2017) 112002}, \href{http://arxiv.org/abs/1704.00488}{{\normalfont\ttfamily arXiv:1704.00488}}\relax
\mciteBstWouldAddEndPuncttrue
\mciteSetBstMidEndSepPunct{\mcitedefaultmidpunct}
{\mcitedefaultendpunct}{\mcitedefaultseppunct}\relax
\EndOfBibitem
\bibitem{COMPASS:2023vqt}
COMPASS, G.~D. Alexeev {\em et~al.}, \ifthenelse{\boolean{articletitles}}{\emph{{Final COMPASS Results on the Transverse-Spin-Dependent Azimuthal Asymmetries in the Pion-Induced Drell-Yan Process}}, }{}\href{https://doi.org/10.1103/PhysRevLett.133.071902}{Phys.\ Rev.\ Lett.\  \textbf{133} (2024) 071902}, \href{http://arxiv.org/abs/2312.17379}{{\normalfont\ttfamily arXiv:2312.17379}}\relax
\mciteBstWouldAddEndPuncttrue
\mciteSetBstMidEndSepPunct{\mcitedefaultmidpunct}
{\mcitedefaultendpunct}{\mcitedefaultseppunct}\relax
\EndOfBibitem
\bibitem{Adamczyk:2015gyk}
STAR collaboration, L.~Adamczyk {\em et~al.}, \ifthenelse{\boolean{articletitles}}{\emph{{Measurement of the transverse single-spin asymmetry in $p^\uparrow+p \to W^{\pm}/Z^0$ at RHIC}}, }{}\href{https://doi.org/10.1103/PhysRevLett.116.132301}{Phys.\ Rev.\ Lett.\  \textbf{116} (2016) 132301}, \href{http://arxiv.org/abs/1511.06003}{{\normalfont\ttfamily arXiv:1511.06003}}\relax
\mciteBstWouldAddEndPuncttrue
\mciteSetBstMidEndSepPunct{\mcitedefaultmidpunct}
{\mcitedefaultendpunct}{\mcitedefaultseppunct}\relax
\EndOfBibitem
\bibitem{Keller:2022abm}
SpinQuest collaboration, D.~Keller, \ifthenelse{\boolean{articletitles}}{\emph{{The Transverse Structure of the Deuteron with Drell-Yan}}, }{}\href{http://arxiv.org/abs/2205.01249}{{\normalfont\ttfamily arXiv:2205.01249}}\relax
\mciteBstWouldAddEndPuncttrue
\mciteSetBstMidEndSepPunct{\mcitedefaultmidpunct}
{\mcitedefaultendpunct}{\mcitedefaultseppunct}\relax
\EndOfBibitem
\bibitem{Hadjidakis:2018ifr}
C.~Hadjidakis {\em et~al.}, \ifthenelse{\boolean{articletitles}}{\emph{A fixed-target programme at the lhc: Physics case and projected performances for heavy-ion, hadron, spin and astroparticle studies}, }{}\href{https://doi.org/https://doi.org/10.1016/j.physrep.2021.01.002}{Physics Reports \textbf{911} (2021) 1}, A Fixed-Target Programme at the LHC: Physics Case and Projected Performances for Heavy-Ion, Hadron, Spin and Astroparticle Studies\relax
\mciteBstWouldAddEndPuncttrue
\mciteSetBstMidEndSepPunct{\mcitedefaultmidpunct}
{\mcitedefaultendpunct}{\mcitedefaultseppunct}\relax
\EndOfBibitem
\bibitem{Fernando_Keller_2023}
I.~P. Fernando and D.~Keller, \ifthenelse{\boolean{articletitles}}{\emph{Extraction of the sivers function with deep neural networks}, }{}\href{https://doi.org/10.1103/PhysRevD.108.054007}{Phys.\ Rev.\ D \textbf{108} (2023) 054007}\relax
\mciteBstWouldAddEndPuncttrue
\mciteSetBstMidEndSepPunct{\mcitedefaultmidpunct}
{\mcitedefaultendpunct}{\mcitedefaultseppunct}\relax
\EndOfBibitem
\bibitem{Huang_2016}
J.~Huang, Z.-B. Kang, I.~Vitev, and H.~Xing, \ifthenelse{\boolean{articletitles}}{\emph{Spin asymmetries for vector boson production in polarized collisions}, }{}\href{https://doi.org/10.1103/physrevd.93.014036}{Physical Review D \textbf{93} (2016) }\relax
\mciteBstWouldAddEndPuncttrue
\mciteSetBstMidEndSepPunct{\mcitedefaultmidpunct}
{\mcitedefaultendpunct}{\mcitedefaultseppunct}\relax
\EndOfBibitem
\bibitem{Boer:2015}
D.~Boer, C.~Lorcé, C.~Pisano, and J.~Zhou, \ifthenelse{\boolean{articletitles}}{\emph{The gluon sivers distribution: status and future prospects}, }{}\href{https://doi.org/10.48550/arXiv.1504.04332}{arXiv:1504.\ 04332 [hep-ph] (2015) }\relax
\mciteBstWouldAddEndPuncttrue
\mciteSetBstMidEndSepPunct{\mcitedefaultmidpunct}
{\mcitedefaultendpunct}{\mcitedefaultseppunct}\relax
\EndOfBibitem
\bibitem{PhysRevD.86.094007}
D.~Boer and C.~Pisano, \ifthenelse{\boolean{articletitles}}{\emph{Polarized gluon studies with charmonium and bottomonium at lhcb and after}, }{}\href{https://doi.org/10.1103/PhysRevD.86.094007}{Phys.\ Rev.\ D \textbf{86} (2012) 094007}\relax
\mciteBstWouldAddEndPuncttrue
\mciteSetBstMidEndSepPunct{\mcitedefaultmidpunct}
{\mcitedefaultendpunct}{\mcitedefaultseppunct}\relax
\EndOfBibitem
\bibitem{PhysRevD.110.034038}
N.~Kato, L.~Maxia, and C.~Pisano, \ifthenelse{\boolean{articletitles}}{\emph{Spin asymmetries for $c$-even quarkonium production as a probe of gluon distributions}, }{}\href{https://doi.org/10.1103/PhysRevD.110.034038}{Phys.\ Rev.\ D \textbf{110} (2024) 034038}\relax
\mciteBstWouldAddEndPuncttrue
\mciteSetBstMidEndSepPunct{\mcitedefaultmidpunct}
{\mcitedefaultendpunct}{\mcitedefaultseppunct}\relax
\EndOfBibitem
\bibitem{DAlesio:2020}
U.~D'Alesio {\em et~al.}, \ifthenelse{\boolean{articletitles}}{\emph{Process dependence of the gluon sivers function in ${p}^{\ensuremath{\uparrow}}p\ensuremath{\rightarrow}j/\ensuremath{\psi}+x$ within a tmd scheme in nrqcd}, }{}\href{https://doi.org/10.1103/PhysRevD.102.094011}{Phys.\ Rev.\ D \textbf{102} (2020) 094011}\relax
\mciteBstWouldAddEndPuncttrue
\mciteSetBstMidEndSepPunct{\mcitedefaultmidpunct}
{\mcitedefaultendpunct}{\mcitedefaultseppunct}\relax
\EndOfBibitem
\bibitem{Scarpa:2020}
F.~Scarpa {\em et~al.}, \ifthenelse{\boolean{articletitles}}{\emph{Studies of gluon tmds and their evolution using quarkonium-pair production at the lhc}, }{}\href{https://doi.org/10.1140/epjc/s10052-020-7619-1}{The European Physical Journal C \textbf{80} (2020) 87}\relax
\mciteBstWouldAddEndPuncttrue
\mciteSetBstMidEndSepPunct{\mcitedefaultmidpunct}
{\mcitedefaultendpunct}{\mcitedefaultseppunct}\relax
\EndOfBibitem
\bibitem{Bacchetta:2020vty}
A.~Bacchetta, F.~G. Celiberto, M.~Radici, and P.~Taels, \ifthenelse{\boolean{articletitles}}{\emph{{Transverse-momentum-dependent gluon distribution functions in a spectator model}}, }{}\href{https://doi.org/10.1140/epjc/s10052-020-8327-6}{Eur.\ Phys.\ J.\ C \textbf{80} (2020) 733}, \href{http://arxiv.org/abs/2005.02288}{{\normalfont\ttfamily arXiv:2005.02288}}\relax
\mciteBstWouldAddEndPuncttrue
\mciteSetBstMidEndSepPunct{\mcitedefaultmidpunct}
{\mcitedefaultendpunct}{\mcitedefaultseppunct}\relax
\EndOfBibitem
\bibitem{Bacchetta:2024fci}
A.~Bacchetta, F.~G. Celiberto, and M.~Radici, \ifthenelse{\boolean{articletitles}}{\emph{{T-odd gluon distribution functions in a spectator model}}, }{}\href{https://doi.org/10.1140/epjc/s10052-024-12927-y}{Eur.\ Phys.\ J.\ C \textbf{84} (2024) 576}, \href{http://arxiv.org/abs/2402.17556}{{\normalfont\ttfamily arXiv:2402.17556}}\relax
\mciteBstWouldAddEndPuncttrue
\mciteSetBstMidEndSepPunct{\mcitedefaultmidpunct}
{\mcitedefaultendpunct}{\mcitedefaultseppunct}\relax
\EndOfBibitem
\bibitem{Ji:1996ek}
X.~Ji, \ifthenelse{\boolean{articletitles}}{\emph{Gauge-invariant decomposition of nucleon spin}, }{}\href{https://doi.org/10.1103/PhysRevLett.78.610}{Phys.\ Rev.\ Lett.\  \textbf{78} (1997) 610}\relax
\mciteBstWouldAddEndPuncttrue
\mciteSetBstMidEndSepPunct{\mcitedefaultmidpunct}
{\mcitedefaultendpunct}{\mcitedefaultseppunct}\relax
\EndOfBibitem
\bibitem{Koempel:2012}
J.~Koempel, P.~Kroll, A.~Metz, and J.~Zhou, \ifthenelse{\boolean{articletitles}}{\emph{Exclusive production of quarkonia as a probe of the generalized parton distribution for gluons}, }{}\href{https://doi.org/10.1103/PhysRevD.85.051502}{Phys.\ Rev.\ D \textbf{85} (2012) 051502}\relax
\mciteBstWouldAddEndPuncttrue
\mciteSetBstMidEndSepPunct{\mcitedefaultmidpunct}
{\mcitedefaultendpunct}{\mcitedefaultseppunct}\relax
\EndOfBibitem
\bibitem{Aaij:2022}
R.~Aaij {\em et~al.}, \ifthenelse{\boolean{articletitles}}{\emph{Study of coherent $j/\psi$ production in lead-lead collisions at $\sqrt{s_{NN}}=5$ tev}, }{}\href{https://doi.org/10.1007/JHEP07(2022)117}{Journal of High Energy Physics \textbf{7} (2022) 117}\relax
\mciteBstWouldAddEndPuncttrue
\mciteSetBstMidEndSepPunct{\mcitedefaultmidpunct}
{\mcitedefaultendpunct}{\mcitedefaultseppunct}\relax
\EndOfBibitem
\bibitem{Aaij:2022_2}
LHCb Collaboration, R.~Aaij {\em et~al.}, \ifthenelse{\boolean{articletitles}}{\emph{$j/\ensuremath{\psi}$ photoproduction in pb-pb peripheral collisions at $\sqrt{{s}_{NN}}=5$ tev}, }{}\href{https://doi.org/10.1103/PhysRevC.105.L032201}{Phys.\ Rev.\ C \textbf{105} (2022) L032201}\relax
\mciteBstWouldAddEndPuncttrue
\mciteSetBstMidEndSepPunct{\mcitedefaultmidpunct}
{\mcitedefaultendpunct}{\mcitedefaultseppunct}\relax
\EndOfBibitem
\bibitem{Aaij:2023}
R.~Aaij {\em et~al.}, \ifthenelse{\boolean{articletitles}}{\emph{Study of exclusive photoproduction of charmonium in ultra-peripheral lead-lead collisions}, }{}\href{https://doi.org/10.1007/JHEP06(2023)146}{Journal of High Energy Physics \textbf{6} (2023) 146}\relax
\mciteBstWouldAddEndPuncttrue
\mciteSetBstMidEndSepPunct{\mcitedefaultmidpunct}
{\mcitedefaultendpunct}{\mcitedefaultseppunct}\relax
\EndOfBibitem
\bibitem{Koempel_2012}
J.~Koempel, P.~Kroll, A.~Metz, and J.~Zhou, \ifthenelse{\boolean{articletitles}}{\emph{Exclusive production of quarkonia as a probe of the generalized parton distribution for gluons}, }{}\href{https://doi.org/10.1103/PhysRevD.85.051502}{Phys.\ Rev.\ D \textbf{85} (2012) 051502}\relax
\mciteBstWouldAddEndPuncttrue
\mciteSetBstMidEndSepPunct{\mcitedefaultmidpunct}
{\mcitedefaultendpunct}{\mcitedefaultseppunct}\relax
\EndOfBibitem
\bibitem{PhysRevLett.121.202301}
P.~Bozek and W.~Broniowski, \ifthenelse{\boolean{articletitles}}{\emph{Elliptic flow in ultrarelativistic collisions with polarized deuterons}, }{}\href{https://doi.org/10.1103/PhysRevLett.121.202301}{Phys.\ Rev.\ Lett.\  \textbf{121} (2018) 202301}\relax
\mciteBstWouldAddEndPuncttrue
\mciteSetBstMidEndSepPunct{\mcitedefaultmidpunct}
{\mcitedefaultendpunct}{\mcitedefaultseppunct}\relax
\EndOfBibitem
\bibitem{Broniowski:2019kjo}
W.~Broniowski and P.~Bozek, \ifthenelse{\boolean{articletitles}}{\emph{{Elliptic flow in ultrarelativistic collisions with light polarized nuclei}}, }{}\href{https://doi.org/10.1103/PhysRevC.101.024901}{Phys.\ Rev.\ C \textbf{101} (2020) 024901}, \href{http://arxiv.org/abs/1906.09045}{{\normalfont\ttfamily arXiv:1906.09045}}\relax
\mciteBstWouldAddEndPuncttrue
\mciteSetBstMidEndSepPunct{\mcitedefaultmidpunct}
{\mcitedefaultendpunct}{\mcitedefaultseppunct}\relax
\EndOfBibitem
\bibitem{Beni__2018}
S.~Benić and Y.~Hatta, \ifthenelse{\boolean{articletitles}}{\emph{Single spin asymmetries in ultraperipheral collisions}, }{}\href{https://doi.org/10.1103/physrevd.98.094025}{Physical Review D \textbf{98} (2018) }\relax
\mciteBstWouldAddEndPuncttrue
\mciteSetBstMidEndSepPunct{\mcitedefaultmidpunct}
{\mcitedefaultendpunct}{\mcitedefaultseppunct}\relax
\EndOfBibitem
\bibitem{Barshel:2014}
C.~Barschel, \ifthenelse{\boolean{articletitles}}{\emph{Precision luminosity measurement at lhcb with beam-gas imaging}, }{}PhD Thesis, RWTH Aachen Unuversity (2014)\relax
\mciteBstWouldAddEndPuncttrue
\mciteSetBstMidEndSepPunct{\mcitedefaultmidpunct}
{\mcitedefaultendpunct}{\mcitedefaultseppunct}\relax
\EndOfBibitem
\bibitem{SMOG2_paper}
O.~B. Garcia {\em et~al.}, \ifthenelse{\boolean{articletitles}}{\emph{High-density gas target at the lhcb experiment}, }{}\href{https://doi.org/10.1103/PhysRevAccelBeams.27.111001}{Phys.\ Rev.\ Accel.\ Beams \textbf{27} (2024) 111001}\relax
\mciteBstWouldAddEndPuncttrue
\mciteSetBstMidEndSepPunct{\mcitedefaultmidpunct}
{\mcitedefaultendpunct}{\mcitedefaultseppunct}\relax
\EndOfBibitem
\bibitem{Haeberli1}
W.~Haeberli, \ifthenelse{\boolean{articletitles}}{\emph{Storage cell target for polarized hydrogen and deuterium}, }{} in {\em High Energy Spin Physics} (W.~Meyer, E.~Steffens, and W.~Thiel, eds.), \href{https://doi.org/10.1007/978-3-642-76661-9_37}{ (Berlin, Heidelberg), 194--198, Springer Berlin Heidelberg, 1991}\relax
\mciteBstWouldAddEndPuncttrue
\mciteSetBstMidEndSepPunct{\mcitedefaultmidpunct}
{\mcitedefaultendpunct}{\mcitedefaultseppunct}\relax
\EndOfBibitem
\bibitem{article-steffens}
E.~Steffens and W.~Haeberli, \ifthenelse{\boolean{articletitles}}{\emph{Polarized gas targets}, }{}\href{https://doi.org/10.1088/0034-4885/66/11/R02}{Rep.\ Prog.\ Phys \textbf{66} (2003) 1887}\relax
\mciteBstWouldAddEndPuncttrue
\mciteSetBstMidEndSepPunct{\mcitedefaultmidpunct}
{\mcitedefaultendpunct}{\mcitedefaultseppunct}\relax
\EndOfBibitem
\bibitem{Knudsen}
J.~J. Shea, \ifthenelse{\boolean{articletitles}}{\emph{Foundations of vacuum science and technology}, }{}\href{https://doi.org/10.1109/MEI.1998.689278}{IEEE Electrical Insulation Magazine \textbf{14} (1998) 42}\relax
\mciteBstWouldAddEndPuncttrue
\mciteSetBstMidEndSepPunct{\mcitedefaultmidpunct}
{\mcitedefaultendpunct}{\mcitedefaultseppunct}\relax
\EndOfBibitem
\bibitem{MolFlow}
M.~Ady and R.~Kersevan, \ifthenelse{\boolean{articletitles}}{\emph{{Introduction to the latest version of the test-particle Monte Carlo code Molflow+}}, }{} \href{http://cdsweb.cern.ch/search?p=CERN-ACC-2014-0249&f=reportnumber&action_search=Search&c=LHCb+Reports} {CERN-ACC-2014-0249}, 2014\relax
\mciteBstWouldAddEndPuncttrue
\mciteSetBstMidEndSepPunct{\mcitedefaultmidpunct}
{\mcitedefaultendpunct}{\mcitedefaultseppunct}\relax
\EndOfBibitem
\bibitem{PCollins_VELO}
P.~Collins {\em et~al.}, \ifthenelse{\boolean{articletitles}}{\emph{The {LHCb VELO} upgrade}, }{}\href{https://doi.org/https://doi.org/10.1016/j.nima.2010.04.108}{Nuclear Instruments and Methods in Physics Research Section A: Accelerators, Spectrometers, Detectors and Associated Equipment \textbf{636} (2011) S185}, 7th International "Hiroshima"" Symposium on the Development and Application of Semiconductor Tracking Detectors\relax
\mciteBstWouldAddEndPuncttrue
\mciteSetBstMidEndSepPunct{\mcitedefaultmidpunct}
{\mcitedefaultendpunct}{\mcitedefaultseppunct}\relax
\EndOfBibitem
\bibitem{LHCb-PUB-2012-018}
R.~B. Appleby {\em et~al.}, \ifthenelse{\boolean{articletitles}}{\emph{{VELO aperture considerations for the LHCb Upgrade}}, }{} \href{http://cdsweb.cern.ch/search?p=LHCb-PUB-2012-018&f=reportnumber&action_search=Search&c=LHCb+Notes} {LHCb-PUB-2012-018}, 2012\relax
\mciteBstWouldAddEndPuncttrue
\mciteSetBstMidEndSepPunct{\mcitedefaultmidpunct}
{\mcitedefaultendpunct}{\mcitedefaultseppunct}\relax
\EndOfBibitem
\bibitem{Aperture_studies}
C.~B.~M. et~al, \ifthenelse{\boolean{articletitles}}{\emph{Calculation of the allowed aperture for a gas storage cell in ip8}, }{}Tech.\ Rep.\ CERN-PBC-Notes-2018-008 (2018)\relax
\mciteBstWouldAddEndPuncttrue
\mciteSetBstMidEndSepPunct{\mcitedefaultmidpunct}
{\mcitedefaultendpunct}{\mcitedefaultseppunct}\relax
\EndOfBibitem
\bibitem{CERN-PBC-Notes-2018-008}
C.~Boscolo~Meneguolo {\em et~al.}, \ifthenelse{\boolean{articletitles}}{\emph{{Calculation of the allowed aperture for a gas storage cell in IP8}}, }{} \href{http://cdsweb.cern.ch/search?p=CERN-PBC-Notes-2018-008&f=reportnumber&action_search=Search&c=LHCb+Reports} {CERN-PBC-Notes-2018-008}, 2018\relax
\mciteBstWouldAddEndPuncttrue
\mciteSetBstMidEndSepPunct{\mcitedefaultmidpunct}
{\mcitedefaultendpunct}{\mcitedefaultseppunct}\relax
\EndOfBibitem
\bibitem{CHIGGIATO2006382}
P.~Chiggiato and P.~{Costa Pinto}, \ifthenelse{\boolean{articletitles}}{\emph{{Ti–Zr–V} non-evaporable getter films: from development to large scale production for the {Large Hadron Collider}}, }{}\href{https://doi.org/https://doi.org/10.1016/j.tsf.2005.12.218}{Thin Solid Films \textbf{515} (2006) 382}, Proceedings of the Eighth International Conference on Atomically Controlled Surfaces, Interfaces and Nanostructures and the Thirteenth International Congress on Thin Films\relax
\mciteBstWouldAddEndPuncttrue
\mciteSetBstMidEndSepPunct{\mcitedefaultmidpunct}
{\mcitedefaultendpunct}{\mcitedefaultseppunct}\relax
\EndOfBibitem
\bibitem{HENRIST200195}
B.~Henrist, N.~Hilleret, C.~Scheuerlein, and M.~Taborelli, \ifthenelse{\boolean{articletitles}}{\emph{The secondary electron yield of {TiZr} and {TiZrV} non-evaporable getter thin film coatings}, }{}\href{https://doi.org/https://doi.org/10.1016/S0169-4332(00)00838-2}{Applied Surface Science \textbf{172} (2001) 95}\relax
\mciteBstWouldAddEndPuncttrue
\mciteSetBstMidEndSepPunct{\mcitedefaultmidpunct}
{\mcitedefaultendpunct}{\mcitedefaultseppunct}\relax
\EndOfBibitem
\bibitem{PhysRevSTAB.14.071001}
C.~Yin~Vallgren {\em et~al.}, \ifthenelse{\boolean{articletitles}}{\emph{Amorphous carbon coatings for the mitigation of electron cloud in the {CERN Super Proton Synchrotron}}, }{}\href{https://doi.org/10.1103/PhysRevSTAB.14.071001}{Phys.\ Rev.\ ST Accel.\ Beams \textbf{14} (2011) 071001}\relax
\mciteBstWouldAddEndPuncttrue
\mciteSetBstMidEndSepPunct{\mcitedefaultmidpunct}
{\mcitedefaultendpunct}{\mcitedefaultseppunct}\relax
\EndOfBibitem
\bibitem{vollenberg:ipac2021-wepab338}
W.~Vollenberg {\em et~al.}, \ifthenelse{\boolean{articletitles}}{\emph{{Amorphous Carbon Coating in SPS}}, }{} in {\em Proc. IPAC'21}, \href{https://doi.org/10.18429/JACoW-IPAC2021-WEPAB338}{ No.~12 in International Particle Accelerator Conference, 3475--3478, JACoW Publishing, Geneva, Switzerland, 2021}\relax
\mciteBstWouldAddEndPuncttrue
\mciteSetBstMidEndSepPunct{\mcitedefaultmidpunct}
{\mcitedefaultendpunct}{\mcitedefaultseppunct}\relax
\EndOfBibitem
\bibitem{Nass:2008th}
A.~Nass and E.~Steffens, \ifthenelse{\boolean{articletitles}}{\emph{{Direct Simulation of Low-Pressure Supersonic Gas Expansions and its Experimental Verification}}, }{}\href{https://doi.org/10.1016/j.nima.2008.10.002}{Nucl.\ Instrum.\ Meth.\ A \textbf{598} (2009) 653}, \href{http://arxiv.org/abs/0810.0393}{{\normalfont\ttfamily arXiv:0810.0393}}\relax
\mciteBstWouldAddEndPuncttrue
\mciteSetBstMidEndSepPunct{\mcitedefaultmidpunct}
{\mcitedefaultendpunct}{\mcitedefaultseppunct}\relax
\EndOfBibitem
\bibitem{Steffens:2019kgb}
E.~Steffens {\em et~al.}, \ifthenelse{\boolean{articletitles}}{\emph{{Design Consideration on a Polarized Gas Target for the LHC}}, }{}\href{https://doi.org/10.22323/1.346.0098}{PoS \textbf{SPIN2018} (2019) 098}\relax
\mciteBstWouldAddEndPuncttrue
\mciteSetBstMidEndSepPunct{\mcitedefaultmidpunct}
{\mcitedefaultendpunct}{\mcitedefaultseppunct}\relax
\EndOfBibitem
\bibitem{HERMES:BRP}
HERMES collaboration, C.~Baumgarten {\em et~al.}, \ifthenelse{\boolean{articletitles}}{\emph{{An atomic beam polarimeter to measure the nuclear polarization in the HERMES gaseous polarized hydrogen and deuterium target}}, }{}Nuc.\ Instrum.\ and Meth.\ A \textbf{82} (2006) 606\relax
\mciteBstWouldAddEndPuncttrue
\mciteSetBstMidEndSepPunct{\mcitedefaultmidpunct}
{\mcitedefaultendpunct}{\mcitedefaultseppunct}\relax
\EndOfBibitem
\bibitem{CERN-LHCC-2021-012}
LHCb, \ifthenelse{\boolean{articletitles}}{\emph{{Framework TDR for the LHCb Upgrade II: Opportunities in flavour physics, and beyond, in the HL-LHC era}}, }{} 2021\relax
\mciteBstWouldAddEndPuncttrue
\mciteSetBstMidEndSepPunct{\mcitedefaultmidpunct}
{\mcitedefaultendpunct}{\mcitedefaultseppunct}\relax
\EndOfBibitem
\bibitem{LHCB-FIGURE-2023-008}
LHCb, \ifthenelse{\boolean{articletitles}}{\emph{{Invariant mass spectra from SMOG2 $\;p\mathrm{Ar}\;$ and $\;p\mathrm{H}\;$ collisions from 2022 data}}, }{} 2023.
\newblock \url{https://cds.cern.ch/record/2859158}\relax
\mciteBstWouldAddEndPuncttrue
\mciteSetBstMidEndSepPunct{\mcitedefaultmidpunct}
{\mcitedefaultendpunct}{\mcitedefaultseppunct}\relax
\EndOfBibitem
\bibitem{BoenteGarcia:2024kba}
O.~Boente~Garcia {\em et~al.}, \ifthenelse{\boolean{articletitles}}{\emph{{High-density gas target at the LHCb experiment}}, }{}\href{https://doi.org/10.1103/PhysRevAccelBeams.27.111001}{Phys.\ Rev.\ Accel.\ Beams \textbf{27} (2024) 111001}, \href{http://arxiv.org/abs/2407.14200}{{\normalfont\ttfamily arXiv:2407.14200}}\relax
\mciteBstWouldAddEndPuncttrue
\mciteSetBstMidEndSepPunct{\mcitedefaultmidpunct}
{\mcitedefaultendpunct}{\mcitedefaultseppunct}\relax
\EndOfBibitem
\bibitem{conductance}
S.~Dushman, J.~M. Lafferty, R.~S. of~General Electric Research~Laboratory, and S.~C. Brown, \ifthenelse{\boolean{articletitles}}{\emph{Scientific foundations of vacuum technique}, }{}\href{https://doi.org/10.1119/1.1942137}{American Journal of Physics \textbf{30} (1962) 612}\relax
\mciteBstWouldAddEndPuncttrue
\mciteSetBstMidEndSepPunct{\mcitedefaultmidpunct}
{\mcitedefaultendpunct}{\mcitedefaultseppunct}\relax
\EndOfBibitem
\bibitem{Citron:2018lsq}
Z.~Citron {\em et~al.}, \ifthenelse{\boolean{articletitles}}{\emph{{Report from Working Group 5}: {Future physics opportunities for high-density QCD at the LHC with heavy-ion and proton beams}}, }{}\href{https://doi.org/10.23731/CYRM-2019-007.1159}{CERN Yellow Rep.\ Monogr.\  \textbf{7} (2019) 1159}, \href{http://arxiv.org/abs/1812.06772}{{\normalfont\ttfamily arXiv:1812.06772}}\relax
\mciteBstWouldAddEndPuncttrue
\mciteSetBstMidEndSepPunct{\mcitedefaultmidpunct}
{\mcitedefaultendpunct}{\mcitedefaultseppunct}\relax
\EndOfBibitem
\bibitem{Bursche:2649878}
A.~Bursche {\em et~al.}, \ifthenelse{\boolean{articletitles}}{\emph{{Physics opportunities with the ﬁxed-target program of the LHCb experiment using an unpolarized gas target}}, }{} tech. rep., CERN, Geneva, 2018\relax
\mciteBstWouldAddEndPuncttrue
\mciteSetBstMidEndSepPunct{\mcitedefaultmidpunct}
{\mcitedefaultendpunct}{\mcitedefaultseppunct}\relax
\EndOfBibitem
\bibitem{Santimaria:2023xdy}
M.~Santimaria {\em et~al.}, \ifthenelse{\boolean{articletitles}}{\emph{{The LHCspin project - A polarised gas target at the Large Hadron Collider}}, }{}\href{https://doi.org/10.1051/epjconf/202327605007}{EPJ Web Conf.\  \textbf{276} (2023) 05007}\relax
\mciteBstWouldAddEndPuncttrue
\mciteSetBstMidEndSepPunct{\mcitedefaultmidpunct}
{\mcitedefaultendpunct}{\mcitedefaultseppunct}\relax
\EndOfBibitem
\bibitem{Belyaev:2017sws}
I.~Belyaev and D.~Savrina, \ifthenelse{\boolean{articletitles}}{\emph{{Study of double parton scattering processes with heavy quarks}}, }{}\href{https://doi.org/10.1142/9789813227767_0008}{Adv.\ Ser.\ Direct.\ High Energy Phys.\  \textbf{29} (2018) 141}, \href{http://arxiv.org/abs/1711.10877}{{\normalfont\ttfamily arXiv:1711.10877}}\relax
\mciteBstWouldAddEndPuncttrue
\mciteSetBstMidEndSepPunct{\mcitedefaultmidpunct}
{\mcitedefaultendpunct}{\mcitedefaultseppunct}\relax
\EndOfBibitem
\bibitem{Lansberg:2015lva}
J.-P. Lansberg and H.-S. Shao, \ifthenelse{\boolean{articletitles}}{\emph{{Double-quarkonium production at a fixed-target experiment at the LHC (AFTER@LHC)}}, }{}\href{https://doi.org/10.1016/j.nuclphysb.2015.09.005}{Nucl.\ Phys.\ B \textbf{900} (2015) 273}, \href{http://arxiv.org/abs/1504.06531}{{\normalfont\ttfamily arXiv:1504.06531}}\relax
\mciteBstWouldAddEndPuncttrue
\mciteSetBstMidEndSepPunct{\mcitedefaultmidpunct}
{\mcitedefaultendpunct}{\mcitedefaultseppunct}\relax
\EndOfBibitem
\bibitem{LHCb:2018jry}
LHCb, R.~Aaij {\em et~al.}, \ifthenelse{\boolean{articletitles}}{\emph{{First Measurement of Charm Production in its Fixed-Target Configuration at the LHC}}, }{}\href{https://doi.org/10.1103/PhysRevLett.122.132002}{Phys.\ Rev.\ Lett.\  \textbf{122} (2019) 132002}, \href{http://arxiv.org/abs/1810.07907}{{\normalfont\ttfamily arXiv:1810.07907}}\relax
\mciteBstWouldAddEndPuncttrue
\mciteSetBstMidEndSepPunct{\mcitedefaultmidpunct}
{\mcitedefaultendpunct}{\mcitedefaultseppunct}\relax
\EndOfBibitem
\bibitem{Sjostrand:2007gs}
T.~Sjostrand, S.~Mrenna, and P.~Z. Skands, \ifthenelse{\boolean{articletitles}}{\emph{{A Brief Introduction to PYTHIA 8.1}}, }{}\href{https://doi.org/10.1016/j.cpc.2008.01.036}{Comput.\ Phys.\ Commun.\  \textbf{178} (2008) 852}, \href{http://arxiv.org/abs/0710.3820}{{\normalfont\ttfamily arXiv:0710.3820}}\relax
\mciteBstWouldAddEndPuncttrue
\mciteSetBstMidEndSepPunct{\mcitedefaultmidpunct}
{\mcitedefaultendpunct}{\mcitedefaultseppunct}\relax
\EndOfBibitem
\bibitem{LHCb-PROC-2010-056}
I.~Belyaev {\em et~al.}, \ifthenelse{\boolean{articletitles}}{\emph{{Handling of the generation of primary events in Gauss, the LHCb simulation framework}}, }{} tech. rep., CERN, Geneva, 2010\relax
\mciteBstWouldAddEndPuncttrue
\mciteSetBstMidEndSepPunct{\mcitedefaultmidpunct}
{\mcitedefaultendpunct}{\mcitedefaultseppunct}\relax
\EndOfBibitem
\bibitem{Pierog_2015}
T.~Pierog {\em et~al.}, \ifthenelse{\boolean{articletitles}}{\emph{{EPOS} {LHC}: Test of collective hadronization with data measured at the {CERN} {L}arge {H}adron {C}ollider}, }{}\href{https://doi.org/10.1103/physrevc.92.034906}{Phys.\ Rev.\  \textbf{C92} (2015) 034906}\relax
\mciteBstWouldAddEndPuncttrue
\mciteSetBstMidEndSepPunct{\mcitedefaultmidpunct}
{\mcitedefaultendpunct}{\mcitedefaultseppunct}\relax
\EndOfBibitem
\bibitem{GEANT4:2002zbu}
GEANT4, S.~Agostinelli {\em et~al.}, \ifthenelse{\boolean{articletitles}}{\emph{{GEANT4 - A Simulation Toolkit}}, }{}\href{https://doi.org/10.1016/S0168-9002(03)01368-8}{Nucl.\ Instrum.\ Meth.\ A \textbf{506} (2003) 250}\relax
\mciteBstWouldAddEndPuncttrue
\mciteSetBstMidEndSepPunct{\mcitedefaultmidpunct}
{\mcitedefaultendpunct}{\mcitedefaultseppunct}\relax
\EndOfBibitem
\bibitem{Allison:2006ve}
J.~Allison {\em et~al.}, \ifthenelse{\boolean{articletitles}}{\emph{{Geant4 developments and applications}}, }{}\href{https://doi.org/10.1109/TNS.2006.869826}{IEEE Trans.\ Nucl.\ Sci.\  \textbf{53} (2006) 270}\relax
\mciteBstWouldAddEndPuncttrue
\mciteSetBstMidEndSepPunct{\mcitedefaultmidpunct}
{\mcitedefaultendpunct}{\mcitedefaultseppunct}\relax
\EndOfBibitem
\bibitem{Clemencic:2011zza}
LHCb, M.~Clemencic {\em et~al.}, \ifthenelse{\boolean{articletitles}}{\emph{{The LHCb simulation application, Gauss: Design, evolution and experience}}, }{}\href{https://doi.org/10.1088/1742-6596/331/3/032023}{J.\ Phys.\ Conf.\ Ser.\  \textbf{331} (2011) 032023}\relax
\mciteBstWouldAddEndPuncttrue
\mciteSetBstMidEndSepPunct{\mcitedefaultmidpunct}
{\mcitedefaultendpunct}{\mcitedefaultseppunct}\relax
\EndOfBibitem
\bibitem{DAlesio:2020eqo}
U.~D'Alesio {\em et~al.}, \ifthenelse{\boolean{articletitles}}{\emph{{Process dependence of the gluon Sivers function in $p^\uparrow p \to J/\psi + X$ within a TMD scheme in NRQCD}}, }{}\href{https://doi.org/10.1103/PhysRevD.102.094011}{Phys.\ Rev.\ D \textbf{102} (2020) 094011}, \href{http://arxiv.org/abs/2007.03353}{{\normalfont\ttfamily arXiv:2007.03353}}\relax
\mciteBstWouldAddEndPuncttrue
\mciteSetBstMidEndSepPunct{\mcitedefaultmidpunct}
{\mcitedefaultendpunct}{\mcitedefaultseppunct}\relax
\EndOfBibitem
\bibitem{IUCF5:1994}
M.~A. Ross {\em et~al.}, \ifthenelse{\boolean{articletitles}}{\emph{Performance of a polarized-hydrogen storage cell target}, }{}\href{https://doi.org/10.1016/0168-9002(94)90079-5}{Nuc.\ Phys.\ Instrum.\ Meth.\ A \textbf{344} (1994) 307}\relax
\mciteBstWouldAddEndPuncttrue
\mciteSetBstMidEndSepPunct{\mcitedefaultmidpunct}
{\mcitedefaultendpunct}{\mcitedefaultseppunct}\relax
\EndOfBibitem
\bibitem{HERMES}
C.~Baumgarten {\em et~al.}, \ifthenelse{\boolean{articletitles}}{\emph{The storage cell of the polarized {H/D} internal gas target of the {HERMES} experiment at {HERA}}, }{}\href{https://doi.org/10.1016/S0168-9002(02)01752-7}{Nuc.\ Instrum.\ Meth.\ A \textbf{496} (2003) 277}\relax
\mciteBstWouldAddEndPuncttrue
\mciteSetBstMidEndSepPunct{\mcitedefaultmidpunct}
{\mcitedefaultendpunct}{\mcitedefaultseppunct}\relax
\EndOfBibitem
\bibitem{ANKE}
K.~Grigoryev {\em et~al.}, \ifthenelse{\boolean{articletitles}}{\emph{The polarized internal target at {ANKE}: First results}, }{}\href{https://doi.org/10.1063/1.2750938}{AIP Conf.\ Proc.\  \textbf{915} (2007) }\relax
\mciteBstWouldAddEndPuncttrue
\mciteSetBstMidEndSepPunct{\mcitedefaultmidpunct}
{\mcitedefaultendpunct}{\mcitedefaultseppunct}\relax
\EndOfBibitem
\bibitem{BINP}
R.~Gilman {\em et~al.}, \ifthenelse{\boolean{articletitles}}{\emph{A polarized gas internal target using a storage cell in an electron storage ring}, }{}\href{https://doi.org/https://doi.org/10.1016/0168-9002(93)90693-C}{Nucl.\ Instrum.\ Meth.\ A \textbf{327} (1993) 277}\relax
\mciteBstWouldAddEndPuncttrue
\mciteSetBstMidEndSepPunct{\mcitedefaultmidpunct}
{\mcitedefaultendpunct}{\mcitedefaultseppunct}\relax
\EndOfBibitem
\bibitem{NIKEF}
J.~F.~J. van~den Brand {\em et~al.}, \ifthenelse{\boolean{articletitles}}{\emph{Evidence for nuclear tensor polarization of deuterium molecules in storage cells}, }{}\href{https://doi.org/10.1103/PhysRevLett.78.1235}{Phys.\ Rev.\ Lett.\  \textbf{78} (1997) 1235}\relax
\mciteBstWouldAddEndPuncttrue
\mciteSetBstMidEndSepPunct{\mcitedefaultmidpunct}
{\mcitedefaultendpunct}{\mcitedefaultseppunct}\relax
\EndOfBibitem
\bibitem{WiseIUCF}
T.~Wise {\em et~al.}, \ifthenelse{\boolean{articletitles}}{\emph{Nuclear polarization of hydrogen molecules from recombination of polarized atoms}, }{}\href{https://doi.org/10.1103/PhysRevLett.87.042701}{Phys.\ Rev.\ Lett.\  \textbf{87} (2001) 042701}\relax
\mciteBstWouldAddEndPuncttrue
\mciteSetBstMidEndSepPunct{\mcitedefaultmidpunct}
{\mcitedefaultendpunct}{\mcitedefaultseppunct}\relax
\EndOfBibitem
\bibitem{HermesPL}
{The HERMES collaboration}, \ifthenelse{\boolean{articletitles}}{\emph{Nuclear polarization of molecular hydrogen recombined on a non-metallic surface}, }{}\href{https://doi.org/10.1140/epjd/e2004-00023-5}{Eur.\ Phys.\ J.\ D \textbf{29} (2004) 21}\relax
\mciteBstWouldAddEndPuncttrue
\mciteSetBstMidEndSepPunct{\mcitedefaultmidpunct}
{\mcitedefaultendpunct}{\mcitedefaultseppunct}\relax
\EndOfBibitem
\bibitem{PRL1}
R.~Engels {\em et~al.}, \ifthenelse{\boolean{articletitles}}{\emph{Production of hyperpolarized {${\mathrm{H}}_{2}$} molecules from {$\stackrel{\ensuremath{\rightarrow}}{\mathrm{H}}$} atoms in gas-storage cells}, }{}\href{https://doi.org/10.1103/PhysRevLett.115.113007}{Phys.\ Rev.\ Lett.\  \textbf{115} (2015) 113007}\relax
\mciteBstWouldAddEndPuncttrue
\mciteSetBstMidEndSepPunct{\mcitedefaultmidpunct}
{\mcitedefaultendpunct}{\mcitedefaultseppunct}\relax
\EndOfBibitem
\bibitem{PRL2}
R.~Engels {\em et~al.}, \ifthenelse{\boolean{articletitles}}{\emph{Production of {HD} molecules in definite hyperfine substates}, }{}\href{https://doi.org/10.1103/PhysRevLett.124.113003}{Phys.\ Rev.\ Lett.\  \textbf{124} (2020) 113003}\relax
\mciteBstWouldAddEndPuncttrue
\mciteSetBstMidEndSepPunct{\mcitedefaultmidpunct}
{\mcitedefaultendpunct}{\mcitedefaultseppunct}\relax
\EndOfBibitem
\bibitem{PRICE}
J.~S. Price and W.~Haeberli, \ifthenelse{\boolean{articletitles}}{\emph{Polarization measurement for polarized gas targets}, }{}\href{https://doi.org/https://doi.org/10.1016/0168-9002(93)90844-8}{Nucl.\ Instrum.\ Meth.\ A \textbf{326} (1993) 416}\relax
\mciteBstWouldAddEndPuncttrue
\mciteSetBstMidEndSepPunct{\mcitedefaultmidpunct}
{\mcitedefaultendpunct}{\mcitedefaultseppunct}\relax
\EndOfBibitem
\bibitem{HERMES2}
{J.\ Stewart for the HERMES collaboration}, \ifthenelse{\boolean{articletitles}}{\emph{{The {HERMES} polarized hydrogen internal gas target}}, }{}\href{https://doi.org/10.1063/1.55016}{AIP Conf.\ Proc.\  \textbf{421} (1998) 69}\relax
\mciteBstWouldAddEndPuncttrue
\mciteSetBstMidEndSepPunct{\mcitedefaultmidpunct}
{\mcitedefaultendpunct}{\mcitedefaultseppunct}\relax
\EndOfBibitem
\bibitem{PAX}
G.~Ciullo {\em et~al.}, \ifthenelse{\boolean{articletitles}}{\emph{The polarised internal target for the {PAX} experiment}, }{}\href{https://doi.org/10.1088/1742-6596/295/1/012150}{J.\ Phys.\ : Conf.\ Series \textbf{295} (2011) 012150}\relax
\mciteBstWouldAddEndPuncttrue
\mciteSetBstMidEndSepPunct{\mcitedefaultmidpunct}
{\mcitedefaultendpunct}{\mcitedefaultseppunct}\relax
\EndOfBibitem
\bibitem{Tarek}
T.~El-Kordy {\em et~al.}, \ifthenelse{\boolean{articletitles}}{\emph{Amorphous carbon-coated storage cell tests for the polarized gas target at {LHCb}}, }{}\href{https://doi.org/https://doi.org/10.1016/j.nima.2024.169707}{Nucl.\ Instrum.\ Meth.\ A \textbf{1068} (2024) 169707}\relax
\mciteBstWouldAddEndPuncttrue
\mciteSetBstMidEndSepPunct{\mcitedefaultmidpunct}
{\mcitedefaultendpunct}{\mcitedefaultseppunct}\relax
\EndOfBibitem
\bibitem{BETHE1958190}
H.~A. Bethe, \ifthenelse{\boolean{articletitles}}{\emph{Scattering and polarization of protons by nuclei}, }{}\href{https://doi.org/https://doi.org/10.1016/0003-4916(58)90017-4}{Annals of Physics \textbf{3} (1958) 190}\relax
\mciteBstWouldAddEndPuncttrue
\mciteSetBstMidEndSepPunct{\mcitedefaultmidpunct}
{\mcitedefaultendpunct}{\mcitedefaultseppunct}\relax
\EndOfBibitem
\bibitem{ZELENSKI2005248}
A.~Zelenski {\em et~al.}, \ifthenelse{\boolean{articletitles}}{\emph{Absolute polarized h-jet polarimeter development, for rhic}, }{}\href{https://doi.org/https://doi.org/10.1016/j.nima.2004.08.080}{Nuclear Instruments and Methods in Physics Research Section A: Accelerators, Spectrometers, Detectors and Associated Equipment \textbf{536} (2005) 248}, Polarized Sources and Targets for the 21st Century. Proceedings o f the 10th International Workshop on Polarized Sources and Targets\relax
\mciteBstWouldAddEndPuncttrue
\mciteSetBstMidEndSepPunct{\mcitedefaultmidpunct}
{\mcitedefaultendpunct}{\mcitedefaultseppunct}\relax
\EndOfBibitem
\bibitem{795735}
H.~Huang {\em et~al.}, \ifthenelse{\boolean{articletitles}}{\emph{A p-carbon cni polarimeter for rhic}, }{} in {\em Proceedings of the 1999 Particle Accelerator Conference (Cat. No.99CH36366)}, \href{https://doi.org/10.1109/PAC.1999.795735}{\textbf{1} 471--473 vol.1, 1999}\relax
\mciteBstWouldAddEndPuncttrue
\mciteSetBstMidEndSepPunct{\mcitedefaultmidpunct}
{\mcitedefaultendpunct}{\mcitedefaultseppunct}\relax
\EndOfBibitem
\bibitem{Poblaguev:2019Eb}
A.~Poblaguev {\em et~al.}, \ifthenelse{\boolean{articletitles}}{\emph{{Study of elastic proton-proton single and double spin analyzing powers at RHIC HJET polarimeter}}, }{}\href{https://doi.org/10.22323/1.346.0143}{PoS \textbf{SPIN2018} (2019) 143}\relax
\mciteBstWouldAddEndPuncttrue
\mciteSetBstMidEndSepPunct{\mcitedefaultmidpunct}
{\mcitedefaultendpunct}{\mcitedefaultseppunct}\relax
\EndOfBibitem
\bibitem{Buttimore_2001}
N.~H. Buttimore, E.~Leader, and T.~L. Trueman, \ifthenelse{\boolean{articletitles}}{\emph{An absolute polarimeter for high energy protons}, }{}\href{https://doi.org/10.1103/physrevd.64.094021}{Physical Review D \textbf{64} (2001) }\relax
\mciteBstWouldAddEndPuncttrue
\mciteSetBstMidEndSepPunct{\mcitedefaultmidpunct}
{\mcitedefaultendpunct}{\mcitedefaultseppunct}\relax
\EndOfBibitem
\bibitem{Buttimoreslides2019}
N.~Buttimore, \ifthenelse{\boolean{articletitles}}{\emph{\textsc{LHC}spin and polarimetry}, }{} Workshop presentation slides, 2019.
\newblock Presented at LHCspin kick-off meeting, Ferrara, July 2019\relax
\mciteBstWouldAddEndPuncttrue
\mciteSetBstMidEndSepPunct{\mcitedefaultmidpunct}
{\mcitedefaultendpunct}{\mcitedefaultseppunct}\relax
\EndOfBibitem
\bibitem{Buttimore2019}
N.~Buttimore, \ifthenelse{\boolean{articletitles}}{\emph{Private communication}, }{} 2019\relax
\mciteBstWouldAddEndPuncttrue
\mciteSetBstMidEndSepPunct{\mcitedefaultmidpunct}
{\mcitedefaultendpunct}{\mcitedefaultseppunct}\relax
\EndOfBibitem
\bibitem{Poblaguev_2020}
A.~A. Poblaguev {\em et~al.}, \ifthenelse{\boolean{articletitles}}{\emph{Systematic error analysis in the absolute hydrogen gas jet polarimeter at rhic}, }{}\href{https://doi.org/10.1016/j.nima.2020.164261}{Nuclear Instruments and Methods in Physics Research Section A: Accelerators, Spectrometers, Detectors and Associated Equipment \textbf{976} (2020) 164261}\relax
\mciteBstWouldAddEndPuncttrue
\mciteSetBstMidEndSepPunct{\mcitedefaultmidpunct}
{\mcitedefaultendpunct}{\mcitedefaultseppunct}\relax
\EndOfBibitem
\bibitem{Zelenski:2005mz}
A.~Zelenski {\em et~al.}, \ifthenelse{\boolean{articletitles}}{\emph{{Absolute polarized H-jet polarimeter development, for RHIC}}, }{}\href{https://doi.org/10.1016/j.nima.2004.08.080}{Nucl.\ Instrum.\ Meth.\ A \textbf{536} (2005) 248}\relax
\mciteBstWouldAddEndPuncttrue
\mciteSetBstMidEndSepPunct{\mcitedefaultmidpunct}
{\mcitedefaultendpunct}{\mcitedefaultseppunct}\relax
\EndOfBibitem
\bibitem{Wise:2003aip}
T.~Wise {\em et~al.}, \ifthenelse{\boolean{articletitles}}{\emph{{Design of a polarized atomic H source for a jet target at RHIC}}, }{}\href{https://doi.org/10.1063/1.1607272}{AIP Conf.\ Proc.\  \textbf{675} (2003) 934}\relax
\mciteBstWouldAddEndPuncttrue
\mciteSetBstMidEndSepPunct{\mcitedefaultmidpunct}
{\mcitedefaultendpunct}{\mcitedefaultseppunct}\relax
\EndOfBibitem
\bibitem{benoit}
B.~Salvant, \ifthenelse{\boolean{articletitles}}{\emph{Private communication}, }{} 2025\relax
\mciteBstWouldAddEndPuncttrue
\mciteSetBstMidEndSepPunct{\mcitedefaultmidpunct}
{\mcitedefaultendpunct}{\mcitedefaultseppunct}\relax
\EndOfBibitem
\bibitem{PhysRevAccelBeams.22.042801}
The BGV Collaboration, A.~Alexopoulos {\em et~al.}, \ifthenelse{\boolean{articletitles}}{\emph{Noninvasive lhc transverse beam size measurement using inelastic beam-gas interactions}, }{}\href{https://doi.org/10.1103/PhysRevAccelBeams.22.042801}{Phys.\ Rev.\ Accel.\ Beams \textbf{22} (2019) 042801}\relax
\mciteBstWouldAddEndPuncttrue
\mciteSetBstMidEndSepPunct{\mcitedefaultmidpunct}
{\mcitedefaultendpunct}{\mcitedefaultseppunct}\relax
\EndOfBibitem
\bibitem{Allison:2016lfl}
J.~Allison {\em et~al.}, \ifthenelse{\boolean{articletitles}}{\emph{{Recent developments in Geant4}}, }{}\href{https://doi.org/10.1016/j.nima.2016.06.125}{Nucl.\ Instrum.\ Meth.\ A \textbf{835} (2016) 186}\relax
\mciteBstWouldAddEndPuncttrue
\mciteSetBstMidEndSepPunct{\mcitedefaultmidpunct}
{\mcitedefaultendpunct}{\mcitedefaultseppunct}\relax
\EndOfBibitem
\bibitem{LSP1}
R.~Engels {\em et~al.}, \ifthenelse{\boolean{articletitles}}{\emph{Precision lamb-shift polarimeter for polarized atomic and ion beams}, }{}\href{https://doi.org/10.1063/1.1619550}{Rev.\ Sci.\ Instrum.\  \textbf{74} (2003) 4607}\relax
\mciteBstWouldAddEndPuncttrue
\mciteSetBstMidEndSepPunct{\mcitedefaultmidpunct}
{\mcitedefaultendpunct}{\mcitedefaultseppunct}\relax
\EndOfBibitem
\bibitem{LSP2}
R.~Engels {\em et~al.}, \ifthenelse{\boolean{articletitles}}{\emph{Background reduction by a getter pump around the ionization volume of a lamb-shift polarimeter and possible improvements of polarized ion sources}, }{}\href{https://doi.org/10.1063/1.1898923}{Rev.\ Sci.\ Instrum.\  \textbf{76} (2005) 053305}\relax
\mciteBstWouldAddEndPuncttrue
\mciteSetBstMidEndSepPunct{\mcitedefaultmidpunct}
{\mcitedefaultendpunct}{\mcitedefaultseppunct}\relax
\EndOfBibitem
\bibitem{wise1}
T.~Wise {\em et~al.}, \ifthenelse{\boolean{articletitles}}{\emph{Nuclear polarization of hydrogen molecules from recombination of polarized atoms}, }{}\href{https://doi.org/10.1103/PhysRevLett.87.042701}{Phys.\ Rev.\ Lett.\  \textbf{87} (2001) 042701}\relax
\mciteBstWouldAddEndPuncttrue
\mciteSetBstMidEndSepPunct{\mcitedefaultmidpunct}
{\mcitedefaultendpunct}{\mcitedefaultseppunct}\relax
\EndOfBibitem
\bibitem{engels2015}
R.~Engels, \ifthenelse{\boolean{articletitles}}{\emph{{Polarized Molecules: A new Option for Internal Storage-Cell Targets?}}, }{} in {\em Proceedings of XVIth International Workshop in Polarized Sources, Targets, and Polarimetry {\textemdash} PoS(PSTP2015)}, \href{https://doi.org/10.22323/1.243.0008}{\textbf{243} 008, 2016}\relax
\mciteBstWouldAddEndPuncttrue
\mciteSetBstMidEndSepPunct{\mcitedefaultmidpunct}
{\mcitedefaultendpunct}{\mcitedefaultseppunct}\relax
\EndOfBibitem
\bibitem{Bilen}
O.~Bilen, \ifthenelse{\boolean{articletitles}}{\emph{{Preparation and Test of Carbon coated Storage Cells for the LHCspin Project}}, }{} Bachelor Thesis, Heinrich-Heine-University Düsseldorf, 2022.
\newblock doi:~\href{https://doi.org/10.13140/RG.2.2.30248.55045}{10.13140/RG.2.2.30248.55045}\relax
\mciteBstWouldAddEndPuncttrue
\mciteSetBstMidEndSepPunct{\mcitedefaultmidpunct}
{\mcitedefaultendpunct}{\mcitedefaultseppunct}\relax
\EndOfBibitem
\end{mcitethebibliography}
\end{document}